\documentclass[12pt]{article}
\PassOptionsToPackage{natbib=true}{biblatex}
\usepackage[utf8]{inputenc}
\usepackage{array}
\usepackage{verbatim}
\usepackage{booktabs}
\usepackage{mathtools}
\usepackage{url}
\usepackage{amsmath}
\usepackage{amsthm}
\usepackage{amssymb}
\usepackage{graphicx}
\usepackage{setspace}
\usepackage{nameref}

\makeatletter

\providecommand{\tabularnewline}{\\}

\theoremstyle{plain}
\newtheorem{thm}{\protect\theoremname}
\theoremstyle{plain}
\newtheorem{assumption}[thm]{\protect\assumptionname}
\theoremstyle{plain}
\newtheorem{lem}[thm]{\protect\lemmaname}

\usepackage{array}
\usepackage{url}

\usepackage{nameref}

\providecommand{\tabularnewline}{\\}
\usepackage{cprotect}
\usepackage{ragged2e}
\usepackage{appendix}
\usepackage{changepage}
\usepackage{pdflscape}
\usepackage{rotating}
\usepackage[normalem]{ulem}
\usepackage{enumerate}

\newcommand{\blind}{1}

\providecommand{\assumptionname}{Assumption}
\providecommand{\lemmaname}{Lemma}

\providecommand{\theoremname}{Theorem}

\ifdefined\showcaptionsetup
\fi

\ifdefined\showcaptionsetup
 \PassOptionsToPackage{caption=false}{subfig}
\fi
\usepackage{subfig}
\makeatother

\usepackage[style=authoryear,backend=biber,doi=false,isbn=false,url=false,eprint=false]{biblatex}
\providecommand{\assumptionname}{Assumption}
\providecommand{\lemmaname}{Lemma}
\providecommand{\theoremname}{Theorem}

\begin{document}
\def\spacingset#1{\renewcommand{\baselinestretch}
{#1}\small\normalsize} \spacingset{1}
\if1\blind{

\title{Forecasting dementia incidence\thanks{We thank Andrew Harvey and numerous seminar participants for useful
comments. Funding from the the Keynes Fund, the Social Security Administration
through the Michigan Retirement Research Center (MRDRC grant UM25-02)
and the Economic and Social Research Council (``Centre for Microeconomic
Analysis of Public Policy at the Institute for Fiscal Studies'' (RES-544-28-50001)
for this work is gratefully acknowledged. }}
\author{Jérôme R. Simons\thanks{Faculty of Economics, University of Cambridge},
Yuntao Chen\thanks{ Division of Psychiatry, University College London},
Eric Brunner\thanks{ Department of Epidemiology and Public Health, University College
London}, Eric French\thanks{ Faculty of Economics, University of Cambridge and Institute for Fiscal
Studies}}
\date{\today}

\maketitle
} \fi

\if0\blind
{ \bigskip{}
\bigskip{}
\bigskip{}

\begin{center}
{\LARGE\textbf{Forecasting dementia incidence}}\textbf{ }
\par\end{center}

\medskip{}
} \fi

\bigskip{}

\begin{abstract}
This paper estimates the stochastic process of how dementia incidence
evolves over time. We proceed in two steps: first, we estimate a time
trend for dementia using a multi-state Cox model. The multi-state
model addresses problems of both interval censoring arising from infrequent
measurement and also measurement error in dementia. Second, we feed
the estimated mean and variance of the time trend into a Kalman filter
to infer the population level dementia process. Using data from the
English Longitudinal Study of Aging (ELSA), we find that dementia
incidence is no longer declining in England. Furthermore, our forecast
is that future incidence remains constant, although there is considerable
uncertainty in this forecast. Our two-step estimation procedure has
significant computational advantages by combining a multi-state model
with a time series method. To account for the short sample that is
available for dementia, we derive expressions for the Kalman filter's
convergence speed, size, and power to detect changes and conclude
our estimator performs well even in short samples. 
\end{abstract}
\noindent\textit{Keywords:} dementia incidence, time trends, forecasting
\vfill{}

\newpage\spacingset{1.8} 


\global\long\def\uwrite#1#2{\underset{#2}{\underbrace{#1}} }%
\global\long\def\blw#1{\ensuremath{\underline{#1}}}%
\global\long\def\abv#1{\ensuremath{\overline{#1}}}%
\global\long\def\vect#1{\mathbf{#1}}%

\global\long\def\smlseq#1{\{#1\} }%
\global\long\def\seq#1{\left\{  #1\right\}  }%
\global\long\def\smlsetof#1#2{\{#1\mid#2\} }%
\global\long\def\setof#1#2{\left\{  #1\mid#2\right\}  }%

\global\long\def\goesto{\ensuremath{\rightarrow}}%
\global\long\def\ngoesto{\ensuremath{\nrightarrow}}%
\global\long\def\uto{\ensuremath{\uparrow}}%
\global\long\def\dto{\ensuremath{\downarrow}}%
\global\long\def\uuto{\ensuremath{\upuparrows}}%
\global\long\def\ddto{\ensuremath{\downdownarrows}}%
\global\long\def\ulrto{\ensuremath{\nearrow}}%
\global\long\def\dlrto{\ensuremath{\searrow}}%

\global\long\def\setmap{\ensuremath{\rightarrow}}%
\global\long\def\elmap{\ensuremath{\mapsto}}%
\global\long\def\compose{\ensuremath{\circ}}%
\global\long\def\cont{C}%
\global\long\def\cadlag{D}%
\global\long\def\Ellp#1{\ensuremath{\mathcal{L}^{#1}}}%

\global\long\def\naturals{\ensuremath{\mathbb{N}}}%
\global\long\def\reals{\mathbb{R}}%
\global\long\def\complex{\mathbb{C}}%
\global\long\def\rationals{\mathbb{Q}}%
\global\long\def\integers{\mathbb{Z}}%

\global\long\def\abs#1{\ensuremath{\left|#1\right|}}%
\global\long\def\smlabs#1{\ensuremath{\lvert#1\rvert}}%
\global\long\def\bigabs#1{\ensuremath{\bigl|#1\bigr|}}%
\global\long\def\Bigabs#1{\ensuremath{\Bigl|#1\Bigr|}}%
\global\long\def\biggabs#1{\ensuremath{\biggl|#1\biggr|}}%
\global\long\def\norm#1{\ensuremath{\left\Vert #1\right\Vert }}%
\global\long\def\smlnorm#1{\ensuremath{\lVert#1\rVert}}%
\global\long\def\bignorm#1{\ensuremath{\bigl\|#1\bigr\|}}%
\global\long\def\Bignorm#1{\ensuremath{\Bigl\|#1\Bigr\|}}%
\global\long\def\biggnorm#1{\ensuremath{\biggl\|#1\biggr\|}}%

\global\long\def\Union{\ensuremath{\bigcup}}%
\global\long\def\Intsect{\ensuremath{\bigcap}}%
\global\long\def\union{\ensuremath{\cup}}%
\global\long\def\intsect{\ensuremath{\cap}}%
\global\long\def\pset{\ensuremath{\mathcal{P}}}%
\global\long\def\clsr#1{\ensuremath{\overline{#1}}}%
\global\long\def\symd{\ensuremath{\Delta}}%
\global\long\def\intr{\operatorname{int}}%
\global\long\def\cprod{\otimes}%
\global\long\def\Cprod{\bigotimes}%

\global\long\def\smlinprd#1#2{\ensuremath{\langle#1,#2\rangle}}%
\global\long\def\inprd#1#2{\ensuremath{\left\langle #1,#2\right\rangle }}%
\global\long\def\orthog{\ensuremath{\perp}}%
\global\long\def\dirsum{\ensuremath{\oplus}}%

\global\long\def\spn{\operatorname{sp}}%
\global\long\def\rank{\operatorname{rk}}%
\global\long\def\proj{\operatorname{proj}}%
\global\long\def\tr{\operatorname{tr}}%

\global\long\def\smpl{\ensuremath{\Omega}}%
\global\long\def\elsmp{\ensuremath{\omega}}%
\global\long\def\sigf#1{\mathcal{#1}}%
\global\long\def\sigfield{\ensuremath{\mathcal{F}}}%
\global\long\def\sigfieldg{\ensuremath{\mathcal{G}}}%
\global\long\def\flt#1{\mathcal{#1}}%
\global\long\def\filt{\mathcal{F}}%
\global\long\def\filtg{\mathcal{G}}%
\global\long\def\Borel{\ensuremath{\mathcal{B}}}%
\global\long\def\cyl{\ensuremath{\mathcal{C}}}%
\global\long\def\nulls{\ensuremath{\mathcal{N}}}%
\global\long\def\gauss{\mathfrak{g}}%
\global\long\def\leb{\mathfrak{m}}%

\global\long\def\prob{P}%
\global\long\def\Prob{\ensuremath{\mathbb{P}}}%
\global\long\def\Probs{\mathcal{P}}%
\global\long\def\PROBS{\mathcal{M}}%
\global\long\def\expect{\ensuremath{\mathbb{E}}}%
\global\long\def\probspc{\ensuremath{(\smpl,\filt,\Prob)}}%

\global\long\def\iid{\ensuremath{\textnormal{i.i.d.}}}%
\global\long\def\as{\ensuremath{\textnormal{a.s.}}}%
\global\long\def\asp{\ensuremath{\textnormal{a.s.p.}}}%
\global\long\def\io{\ensuremath{\ensuremath{\textnormal{i.o.}}}}%
\global\long\def\independent{\mathpalette{\independenT}{\perp}}%
\global\long\def\independenT#1#2{\mathrel{\rlap{$#1#2$}\mkern2mu  {#1#2}}}%
\global\long\def\indep{\independent}%
\global\long\def\distrib{\ensuremath{\sim}}%
\global\long\def\distiid{\ensuremath{\sim_{\iid}}}%
\global\long\def\asydist{\ensuremath{\overset{a}{\distrib}}}%
\global\long\def\inprob{\ensuremath{\overset{p}{\goesto}}}%
\global\long\def\inprobu#1{\ensuremath{\overset{#1}{\goesto}}}%
\global\long\def\inas{\ensuremath{\overset{\as}{\goesto}}}%
\global\long\def\eqas{=_{\as}}%
\global\long\def\inLp#1{\ensuremath{\overset{\Ellp{#1}}{\goesto}}}%
\global\long\def\indist{\ensuremath{\overset{d}{\goesto}}}%
\global\long\def\eqdist{=_{d}}%
\global\long\def\wkc{\ensuremath{\rightsquigarrow}}%
\global\long\def\wkcu#1{\overset{#1}{\ensuremath{\rightsquigarrow}}}%
\global\long\def\plim{\operatorname*{plim}}%

\global\long\def\var{\operatorname{var}}%
\global\long\def\lrvar{\operatorname{lrvar}}%
\global\long\def\cov{\operatorname{cov}}%
\global\long\def\corr{\operatorname{corr}}%
\global\long\def\bias{\operatorname{bias}}%
\global\long\def\MSE{\operatorname{MSE}}%
\global\long\def\med{\operatorname{med}}%

\global\long\def\simple{\mathcal{R}}%
\global\long\def\sring{\mathcal{A}}%
\global\long\def\sproc{\mathcal{H}}%
\global\long\def\Wiener{\ensuremath{\mathbb{W}}}%
\global\long\def\sint{\bullet}%
\global\long\def\cv#1{\left\langle #1\right\rangle }%
\global\long\def\smlcv#1{\langle#1\rangle}%
\global\long\def\qv#1{\left[#1\right]}%
\global\long\def\smlqv#1{[#1]}%

\global\long\def\trans{\ensuremath{\prime}}%
\global\long\def\indic{\ensuremath{\mathbf{1}}}%
\global\long\def\Lagr{\mathcal{L}}%
\global\long\def\grad{\nabla}%
\global\long\def\pmin{\ensuremath{\wedge}}%
\global\long\def\Pmin{\ensuremath{\bigwedge}}%
\global\long\def\pmax{\ensuremath{\vee}}%
\global\long\def\Pmax{\ensuremath{\bigvee}}%
\global\long\def\sgn{\operatorname{sgn}}%
\global\long\def\argmin{\operatorname*{argmin}}%
\global\long\def\argmax{\operatorname*{argmax}}%
\global\long\def\Rp{\operatorname{Re}}%
\global\long\def\Ip{\operatorname{Im}}%
\global\long\def\deriv{\ensuremath{\mathrm{d}}}%
\global\long\def\diffnspc{\ensuremath{\deriv}}%
\global\long\def\diff{\ensuremath{\,\deriv}}%
\global\long\def\i{\ensuremath{\mathrm{i}}}%
\global\long\def\e{\mathrm{e}}%
\global\long\def\sep{,\ }%
\global\long\def\defeq{\coloneqq}%
\global\long\def\eqdef{\eqqcolon}%

\section{Introduction}

\label{sec:intro}More than 55 million people worldwide live with
dementia, and this number is projected to increase over time (\citealt{nichols2022estimation}).
Dementia is a multidimensional challenge with major implications for
affected individuals, their families, social policy, and national
economies. In England and Wales, the number of people living with
dementia is predicted to increase substantially in the near future
\citep{ahmadi2017temporal}, causing significant increases in health
and social care costs \citep{collins2022will,banks2025long}. These
forecasts, however, are sensitive to future trends in dementia incidence
\citep{wolters2020twenty}. Moreover, forecasts of trends in dementia
incidence are sensitive to modeling assumptions \citep{CHEN2023e859}.
If the dementia incidence trend changes, the future burden of dementia
and associated costs might differ substantially from current forecasts.
Therefore, credible estimates of the dementia incidence trend and
forecasts of its likely trajectory are important in shaping social
care policy.

Nationally representative data that use consistent case definitions
of dementia over time only exist since about 2000, affording only
short sample periods to conduct inference. Consequently, epidemiological
studies to date have extrapolated the observed dementia incidence
rate trend without considering time-series uncertainty \citep{ahmadi2017temporal,CHEN2023e859,wolters2020twenty}.
We address this gap in methodology by formulating statistical methods
that account for samples that are short in the time dimension and
large in the number of cross sectional observations.

We estimate the model in two steps, using recent advances in estimation
of multi-state models with many parameters and contribute our own
time series model. Via the multi-state model, we estimate a time series
of dementia incidence based on \citet{CHEN2023e859} which accounts
for (i) individuals' dropping out (censoring) due to death which can
lead to underestimating dementia incidence if censoring is not addressed,
and (ii) potential misclassification of dementia. In a second step,
we use the estimated incidence rate and its sampling uncertainty as
inputs into a Kalman filter to estimate mean and variance of a dementia
process free of cross-sectional noise.

This procedure allows us to recover and model a stochastic process
for population level dementia incidence, while also addressing the
aforementioned problems of dementia measurement and censoring. We
incorporate this sampling uncertainty in dementia incidence by treating
it as observational noise in a Kalman filter. Using this setup, we
can compare different time series models of dementia incidence and
use our preferred model for forecasting.

Our two-step estimation methodology extends to all contexts where
some parameters in the main model are a time series that is estimated,
rather than known with certainty, which arises frequently in models
for panel data.

We estimate these trends in dementia incidence in England using data
from the English Longitudinal Study of Aging over the 2002-2018 period.
While we find some evidence that incidence fell in the early part
of our sample period, there is little evidence of a long-term trend
in dementia incidence. Our preferred model is a random walk with zero
drift, meaning that while dementia incidence changes over time, the
optimal forecast is its most recent value. We also produce confidence
intervals for these dementia forecasts and find that we cannot reject
significant increases or declines in incidence over the next decade.

To account for the short time dimension, we present two lines of argument
to establish that the proposed Kalman filter performs well in small
samples. First, we study the convergence properties in terms of the
Kalman gain factor. Second, we study the sensitivity or power of the
filter to detect changes in the dementia trend and its false positive
rate as a function of the signal-to-noise ratio and the number of
available time periods. We learn that even with our short sample,
the filter's properties stabilize rapidly enough to warrant inference
on the direction of the dementia trend.

The advantage of our procedure is that we are able to use the most
recent developments in multi-state modeling, including misclassification
models and interval-censoring, and add a time series method to this
framework by modeling the underlying time trend that shifts the dementia
hazard. The proposed methodology is very easy to implement and can
be applied in any context where units move between different states
over time.

\citet[Eq. 11.13]{aalen2008survival} suggest joint modeling and estimation
of the stochastic time trend along with demographic variables. However,
they note that ``{[}a{]} lot of work remains to be done on how best
to implement the models.'' Some studies engage in this joint modeling
\citep{gjessinghort,timedepfrailty,unkel2014time,dynamicfrailty,ragni2025timedepfrail},
but use simpler empirical frameworks than what we use. In contrast,
our two-step method trades statistical efficiency for computational
feasibility. Our methodology is deliberately kept simple and has potential
for wide-spread adoption.

\citet{davis2016handbook,davis2021count} suggest joint modeling and
estimation of the stochastic time trend along with demographic variables
using count time series models. These models perform well when the
time series is long. In contrast, our approach works well when the
number of cross-sectional units is large but there is only a modest
time series dimension.

\section{Data}

\label{sec:Data}

We used data from the English Longitudinal Study of Ageing (ELSA),
a longitudinal panel study of a representative sample of people aged
50 years or more living in private households in England. We use the
ELSA data spanning 17 years across wave 1 (2002-03) to wave 9 (2018-19),
allowing us to measure transitions into dementia and death over eight
two-year intervals and a total of 70,806 individual health transitions.
Mortality data were linked to participants who had provided written
consent for linkage to official records from the National Health Service
central register.

We use the measure of dementia from \citet{ahmadi2017temporal}, which
uses an algorithmic case definition based on coexistence of both cognitive
and functional impairment, or a report of a doctor's diagnosis of
dementia by the participant or caregiver. Cognitive impairment is
defined as impairment in two or more domains of cognitive function
(orientation to time, immediate and delayed memory and verbal fluency).
These measures are available for all nine waves except verbal fluency
at wave six, for which we impute using information from waves five
and seven. For individuals unable to take the cognitive function tests,
the Informant Questionnaire on Cognitive Decline was administered
to a proxy informant (usually an immediate family member), and a score
higher than 3·6 is used to identify cognitive impairment. Functional
impairment is defined as an inability to carry out one or more activities
of daily living independently, which includes getting into or out
of bed, walking across a room, bathing or showering, using the toilet,
dressing, cutting food, and eating. This case definition is less likely
to be affected by changes in diagnostic criteria and clinical practice
over time. Appendix \ref{app:thedata} provides further details of
the measurement or dementia and the sample we use.

\section{Modeling Dementia Incidence}

\label{sec:main-model}

We estimate the time series process of dementia incidence using a
two step estimator. In the first step, we estimate dementia incidence
as a function of demographics and time indicator variables. In the
second step, we develop models of the underlying time series process
given by the coefficients of the time indicators and their variance
estimates.

\subsection{Multi-state Markov model}

\label{sec:msm}
\begin{figure}
\caption{Illustration of the state space.}
\label{fig:multi-state} \centering{}\resizebox{0.4\textwidth}{!}{ \includegraphics{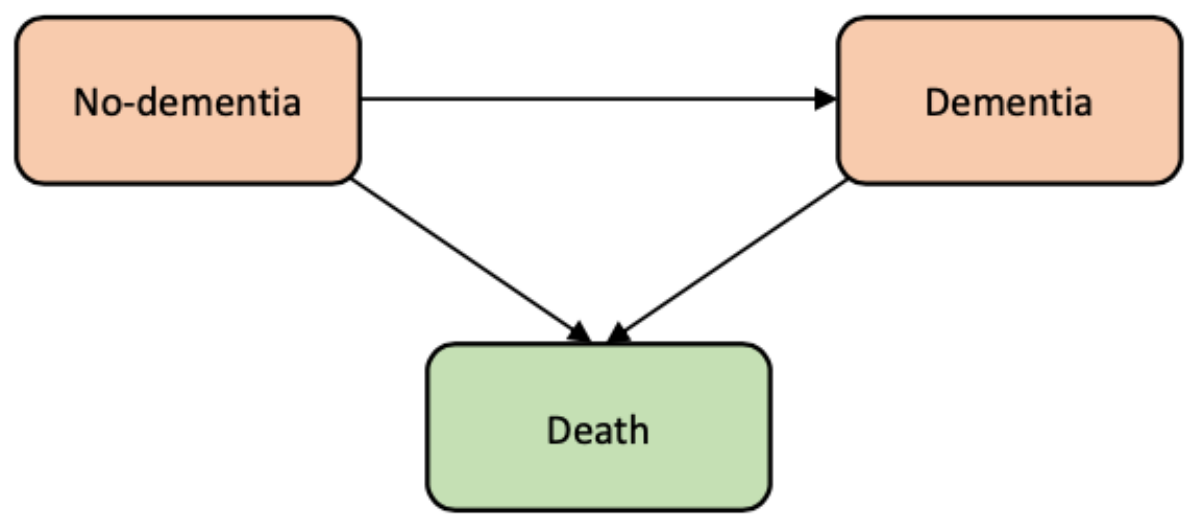}} 
\end{figure}

We model mortality and dementia incidence jointly using a multi-state
model \citep{cox1965theory,jackson2011multi} where demographics and
time affect transitions. The state space $S=\left\{ 1,2,3\right\} $,
where the transition of interest is from no-dementia $\left(1\right)$
to having dementia $\left(2\right)$ while also allowing transitions
to death $(3)$. Figure~\ref{fig:multi-state} shows the state space.

Individuals may develop dementia and die between survey waves leading
to interval censoring. Failure to account for censoring will underestimate
dementia incidence rates since those succumbing to dementia between
waves are more likely to die and thus drop out of the sample between
waves. Even if dementia incidence is constant, changing mortality
rates can produce spurious upward or downward trends if unmodeled.
We account for censoring by following \citet{leffondre2013interval,binder2014missing,CHEN2023e859}
and model mortality and health transitions explicitly.

The transition intensity or hazard from state $r$ to $s$ at time
$t$ is the instantaneous probability 
\begin{equation}
q_{rs}\left(\vect z(t),t\right)\equiv\lim_{w\downarrow0}\frac{\Prob\left\{ S\left(t+w\right)=s|S\left(t\right)=r,\vect z(t),t\right\} }{w}.\label{eq:intensity}
\end{equation}
Transition-specific intensities depend on $t$ and the demographic
variables age and sex $(\vect z_{i}\left(t\right),t)$ via 
\begin{equation}
q_{rs}\left(\vect z_{i}\left(t\right),t\right)=q_{rs0}\exp\left\{ f_{rs}\left(\vect z_{i}\left(t\right)\right)+g_{rs}\left(t\right)\right\} \,\text{for}\,r,s\in S.\label{eq:transition-probability}
\end{equation}
The parameters $q_{rs0}$ are baseline hazards and $g_{rs}\left(t\right)$
is a hazard-shifting time series. Time dependence of the transition
intensity obtains exclusively via the time-dependent covariates. Equation
(\ref{eq:transition-probability}) does not include interactions between
demographics and time; \citet{CHEN2023e859} test for these interactions
and found them to be insignificant.

Time is continuous although mortality and dementia are observed only
at discrete times. We have nine waves, giving us transitions for each
individual at points $t=1,\dots,8$ . Using wave indicator variables,
the process driving transitions from no dementia to dementia is $g_{12}\left(t\right)$
: 
\begin{equation}
g_{12}\left(t\right)=\sum_{k=1}^{T}\beta_{k}\indic\left\{ t=k\right\} ,\label{eq:trend-proc}
\end{equation}
whereas the other transitions $g_{13}\left(t\right)$ and $g_{23}\left(t\right)$
are modeled as linear time trends. Tables \ref{tab:def-f} and \ref{tab:def-f-1}
in Appendix \ref{app:functional-form} detail the functions $f_{rs}\left(.\right)$
and $g_{rs}(t)$. The object of our study is $g_{12}(t)$, which is
why we model it with maximal flexibility; the functions $g_{13}(t)$
and $g_{23}(t)$ are more parsimonious. Appendix \ref{app:matrix-notation}
discusses $q_{rs}$ and resulting probabilities in matrix notation.
For individual $i$ observed at time $t_{ik-1}$, the entries of $P\left(w,\vect z_{i}\left(t\right),t\right)$
become 
\begin{equation}
p_{rs}\left(\vect z_{i}\left(t_{ik-1}\right),t_{ik-1}\right)\equiv\Prob\left\{ S_{t_{ik}}=r\mid S_{t_{ik-1}}=s,\vect z_{i}(t_{ik-1}),t_{ik-1}\right\} \label{eq:entry-correspondence}
\end{equation}

\noindent Equation (\ref{eq:entry-correspondence}) shows the probability
of transitioning from $s$ to $r$ between $t_{ik-1}$ and $t_{ik}$,
and $S_{t_{ik}}\in S$ is individual $i$'s health state at time $t_{ik}$
for $k=1,\dots,m_{i}$ and $m_{i}$ is the number of $i$'s observed
transitions.\footnote{Based on eight two-year waves of follow-up, if an individual is observed
throughout the entire study period, they will contribute maximally
eight observations.} \citet{jackson2011multi} contains explicit expressions of $p_{rs}\left(\vect z_{i}\left(t_{ik-1}\right),t_{ik-1}\right)$
in terms of $q_{rs}\left(\vect z_{i}\left(t_{ik-1}\right),t_{ik-1}\right)$,
which shows that the relationship between the two quantities accounts
for unobserved transitions between waves. For example, it accounts
for the fact that between any two waves an individual may develop
dementia and then die, contributing to both dementia and death probabilities,
although we would only observe the latter.

We now write the individual contribution to the likelihood in terms
of transition probabilities for individual $i$, who is observed at
times $t_{i1},...t_{im_{i}}$. The likelihood for one individual obtains
from multiplying together individual transition probabilities along
all possible transitions a person can undergo. The joint distribution
for health states in all periods for an individual over the sample
period, conditional on the initial state is 
\begin{equation}
\Prob\left\{ S_{t_{i2}},\dots,S_{t_{im_{i}}}\mid S_{t_{i1}}\right\} =\Prob\left\{ S_{t_{im_{i}}}\mid S_{t_{im_{i}-1}}\right\} \dots\Prob\left\{ S_{t_{i3}}\mid S_{t_{i2}}\right\} \Prob\left\{ S_{t_{i2}}\mid S_{t_{i1}}\right\} .\label{eq:joint-dist}
\end{equation}
where we have suppressed dependence on $(\vect z_{i},w)$ for readability.
The right-hand side of (\ref{eq:joint-dist}) is a series of the conditional
probabilities shown in (\ref{eq:entry-correspondence}).

\subsubsection{Misclassification model}

Our definition of dementia uses an algorithm as well as a doctor diagnosis
to reduce bias. Inevitably, there is misclassification, which we account
for via adjusting the likelihood as follows. Define measured health
$S_{t_{ij}}^{*}$ as different from actual health. By assumption,
misclassification probabilities at time $t_{ij}$ (the $j$th period
$i$ is observed) only depend on the immediately preceding state,
i.e. $\Prob\{S_{t_{ij}}^{*}|S_{t_{i1}},\dots,S_{t_{ij}}\}=\Prob\{S_{t_{ij}}^{*}|S_{t_{ij}}\}$.
The probability of being wrongly classified as having dementia is
$\Prob\{S_{t_{ij}}^{*}=2|S_{t_{ij}}=1\}$ and the converse probability
is $\Prob\{S_{t_{ij}}^{*}=1|S_{t_{ij}}=2\}$. Death is measured with
certainty so that $\Prob\{S_{t_{ij}}^{*}=3|S_{t_{ij}}=3\}=1$ and
$\Prob\{S_{t_{ij}}^{*}=3|S_{t_{ij}}\in\{1,2\}\}=0$. Parameter estimates
of the transition intensity matrix \eqref{eq:trans-int} in Appendix
\ref{app:matrix-notation}, the misclassification matrix in Table
\ref{tab:misclassification} in Appendix \ref{app:sample_statistics},
and of the initial state distributions are estimated by maximizing
the log-likelihood based on \citet{jackson2002hidden}. Appendix \ref{app:likelihood-deriv}
derives an expression for the full log-likelihood.

The relevant sample size obtains from the number of transitions for
each $i$, $m_{i}-1$, so that $I$ individuals a total of $\sum_{i=1}^{I}(m_{i}-1)=n$
individual-transition observations. For each $i$, $L_{i}$ \eqref{eq:indiv-loglik}
contributes $m_{i}-1$ transitions. Maximizing $\ell(\gamma)=\sum_{i=1}^{I}\log L_{i}$
yields vector $\gamma$ which includes the parameters of the expressions
in (\ref{eq:transition-probability}): the dummy variables for the
time trend for dementia incidence $\{\{\beta_{k}\}_{k=1}^{T}\}$ as
defined in (\ref{eq:trend-proc}), the time trends for mortality as
part of $g_{13}(.)$ and $g_{23}(.)$, and the parameters of the function
$f_{rs}(.)$, which capture the role of demographics. The full vector
satisfies $\sqrt{n}\left(\hat{\gamma}-\gamma\right)\overset{\text{a}}{\distrib}N\left(0,\Sigma\right)$
under regularity conditions \citep{vandervaart1998}. Define the portion
of $\Sigma$ holding covariances of the estimates $\{\hat{\beta}_{k}\}_{1}^{T}$
as $\Sigma_{TT}$. Normality implies that $\hat{\beta}_{k}$ can be
written as 
\begin{equation}
\hat{\beta}_{k}=\beta_{k}+\varepsilon_{k},\label{eq:msrmt-model-scalar}
\end{equation}
where $\left\{ \varepsilon_{k}\right\} _{1}^{T}\overset{\text{a}}{\distrib}N\left(0,\Sigma_{TT}\right)$
and $\varepsilon_{k}\overset{\text{a}}{\distrib}N\left(0,\sigma_{kk}^{2}\right)$
and \eqref{eq:msrmt-model-scalar} implies that 
\[
\var\left(\{\hat{\beta}_{k}\}_{k=1}^{T}|\left\{ \beta_{k}\right\} _{k=1}^{T}\right)=\Sigma_{TT},
\]
i.e. conditional on knowing the true realization of the dementia process
$\left\{ \beta_{k}\right\} _{k=1}^{T}$, all variability in $\{\hat{\beta}_{k}\}_{1}^{T}$
is due to cross-sectional sampling uncertainty. This insight is critical
for the estimation of the time series models below.

\subsection{Time series models}

\label{sec:ts-mod}

\subsubsection{Derivation of constrained Kalman filter}

\label{subsec:Derivation-of-constrained}

This section uses the estimated dementia incidence trend $\{\hat{\beta}_{k}\}_{k=1}^{T}$
and its variance $\hat{\Sigma}_{TT}$ to estimate the underlying stochastic
process for $\{\beta_{k}\}_{k=1}^{T}$. For our main specification,
we model the process for $\{\beta_{k}\}_{k=1}^{T}$ as a random walk:
\begin{equation}
\beta_{k}=\beta_{k-1}+\eta_{k},\label{eq:state-1}
\end{equation}
where the per period shock $\eta_{k}\distrib N(0,\sigma_{\eta}^{2})$.
Our goals in this section are to recover the time series $\{\beta_{k}\}$
and the shock variance $\sigma_{\eta}^{2}$. However, what we have
is $\{\hat{\beta}_{k}\}_{k=1}^{T}$ and the time-varying measurement
variance $\sigma_{\varepsilon}^{2}\left(k\right)\equiv\sigma_{kk}^{2}.$

The Kalman filter is a popular tool for estimating the time series
properties of a variable that is measured with error \citep{harvey1990forecasting}.
The error-prone measurement in our case is the sequence of estimates
$\hat{\beta}_{k}$ from which we estimate the stochastic process $\beta_{k}$
and its dynamic variance; measurement error in $\hat{\beta}_{k}$
arises from cross-sectional sampling uncertainty. Different than most
filters where the measurement error variance is estimated jointly
with remaining model parameters, we obtain measurement error variances
in the first stage. In the second stage, variances $\sigma_{kk}^{2}$
are given, which is why we refer to our filter as ``constrained.''

Using information until time $k-1$, the forecast $\beta_{k|k-1}$
is called the prior estimate at time $k$ and is available before
$\hat{\beta}_{k}$ is. Once it is, the update $\beta_{k|k}$ is called
the posterior estimate of $\beta_{k}$ at time $k$.

Given the random walk assumption, the formula of the Kalman filter
for any period $k$ consists of the forecasting equations for $\beta$
and its variance: 
\begin{align}
\beta_{k|k-1} & =\beta_{k-1|k-1}\,\text{and}\label{eq:Kal1}\\
P_{k|k-1} & =P_{k-1|k-1}+\sigma_{\eta}^{2}.
\end{align}
The Kalman gain is 
\begin{equation}
K_{k}=\frac{P_{k|k-1}}{P_{k|k-1}+\hat{\sigma}_{kk}^{2}}.\label{eq:gen-gain}
\end{equation}
As the next $\hat{\beta}_{k}$ becomes available, the forecast error
$v_{k}\equiv\hat{\beta}_{k}-\beta_{k|k-1}$ wherefrom the posterior
$\beta_{k\mid k}$ and its variance obtain via the updating equations
\begin{align}
\beta_{k|k} & =\beta_{k|k-1}+K_{k}v_{k}\label{eq:state-update}
\end{align}
for the posterior and $P_{k|k}=\left(1-K_{k}\right)P_{k|k-1}$ for
the variance. The cross-sectional sampling uncertainty $\hat{\sigma}_{kk}^{2}$
enters the filter via the gain in (\ref{eq:gen-gain}). A large $\hat{\sigma}_{kk}^{2}$
decreases $K_{k}$ so that the model tends to attach more weight to
its prior $\beta_{k|k-1}$ in (\ref{eq:state-update}) and less weight
to the correction suggested by the observation $\hat{\beta}_{k}$.
This downweighting is intuitive since a large cross-sectional sampling
uncertainty implies the forecast error likely reflects this uncertainty
and is thus decreased. Conversely, a small cross-sectional uncertainty
$\sigma_{kk}^{2}$ implies a Kalman gain tending towards unity so
that the observation $\hat{\beta}_{k}$ receives more weight than
the prior $\beta_{k\mid k-1}$.

The solution to the Kalman filter depends on the parameters $\hat{\sigma}_{kk}^{2}$
and $\sigma_{\eta}^{2}$ with $\hat{\sigma}_{kk}^{2},k\in\{1,...,T\}$
given as explained in Section \ref{sec:msm}. The parameter $\sigma_{\eta}^{2}$
must still be estimated by running the filter on the $\{\hat{\beta}_{k}\}$
and maximizing the likelihood. Denoting the variance of the forecast
error by $F_{k}=P_{k|k-1}+\hat{\sigma}_{kk}^{2}$ , estimating $\sigma_{\eta}^{2}$
by ML is equivalent to finding the value that minimizes $\frac{1}{T}\sum_{k=1}^{T}\log(F_{k})+\frac{1}{T}\sum_{k=1}^{T}\frac{v_{k}^{2}}{F_{k}}$,
\citep{harvey1990forecasting}.

Initially, $\beta_{0|0}=0$ with posterior variance $P_{0|0}=\infty$
corresponding to a diffuse prior distribution of $\beta_{0|0}$ and
Appendix \ref{app:Kalman} contains a treatment that derives successive
updating steps explicitly. While the random walk model in (\ref{eq:state-1})
is parsimonious, it implies no drift in dementia incidence since in
that model $E[\beta_{k}|\beta_{k-1}]=\beta_{k-1}$. To allow for more
varied dynamics in dementia incidence, we augment the model with a
drift term $\nu_{k}$, 
\begin{equation}
\beta_{k}=\beta_{k-1}+\nu_{k}+\eta_{k}.\label{eq:rand-walk-drift}
\end{equation}
The drift term $\nu_{k}$ may be stochastic or constant according
to 
\begin{eqnarray}
\nu_{k}=\left\{ \begin{array}{ll}
\nu_{k-1}+\xi_{k} & {\mbox{if stochastic drift}}\\
\nu & {\mbox{if constant drift}}
\end{array}
\right.,\label{eq:drift-term}
\end{eqnarray}
where $\xi_{k}\distrib\nulls\left(0,\sigma_{\xi}^{2}\right)$. The
specification based on (\ref{eq:rand-walk-drift})-(\ref{eq:drift-term})
becomes a random walk model with time-constant drift if $\sigma_{\xi}^{2}=0$
implying that $\nu_{k}=\nu_{k-1}=\nu$. Furthermore, if $\nu_{k}=0$,
then (\ref{eq:rand-walk-drift}) reduces to (\ref{eq:state-1}), implying
no drift in dementia incidence. Empirically, a zero drift corresponds
to a flat trend which is equivalent to no change in incidence. Not
restricting $\hat{\sigma}_{kk}^{2}$ in (\ref{eq:gen-gain}) yields
the standard Kalman filter whose results we report, too.

\subsubsection{Non-parametric tests of trend}

\label{subsec:nonpar-tests-trend}

While previous studies have presented evidence on whether dementia
incidence has changed over time, they often do not show whether these
are significant. Here, we test whether we can reject the hypothesis
of no drift in dementia incidence; and, if there is a drift, whether
we can reject whether this drift is constant.

We use three approaches to study the nature of the trend. First, an
$F$-test checks whether to reject that all coefficients on the time
dummies are identically zero. Second, a $t$-test checks for the presence
of a non-zero deterministic drift. Third, another $t$-test checks
for stochastic drifts.

Under $H_{0}:\beta_{k}=0\,\forall k$, the $F$-statistic asymptotically
follows 
\begin{equation}
\hat{\beta}^{\prime}\hat{\Sigma}_{TT}^{-1}\hat{\beta}/T\distrib F\left(T-1,n-T-2\right)\overset{a}{\distrib}\chi_{T-1}^{2}.\label{eq:F-stat}
\end{equation}

\noindent In Section \ref{subsec:Derivation-of-constrained}, we assumed
the covariance matrix $\Sigma_{TT}$ to be diagonal; here, we make
use of all entries and thus do not need to adjust for auto-correlation.

To test whether the dementia incidence trend has changed over time,
we formulate $H_{0}:\nu_{k}=0,\,\sigma_{\xi}^{2}=0$, i.e. a test
of the hypothesis of zero deterministic drift against the alternative
that $\nu_{k}=\nu\neq0$ and $\sigma_{\xi}^{2}=0$, which amounts
to a zero deterministic drift. Appendix~\ref{app:t-stats-trend}
defines the test statistics and processes for the $t$-statistics
and explains covariance estimation.

\section{Results}

\label{sec:results}

\subsection{Estimates from the multi-state model}

\begin{figure}
\begin{centering}
\caption{Dementia incidence by age and time. Error bars show a $90\%$ confidence
interval.}
\centering\subfloat[\label{fig:age-effect-neat}Average dementia incidence by age.]{\centering{}\includegraphics[scale=0.335]{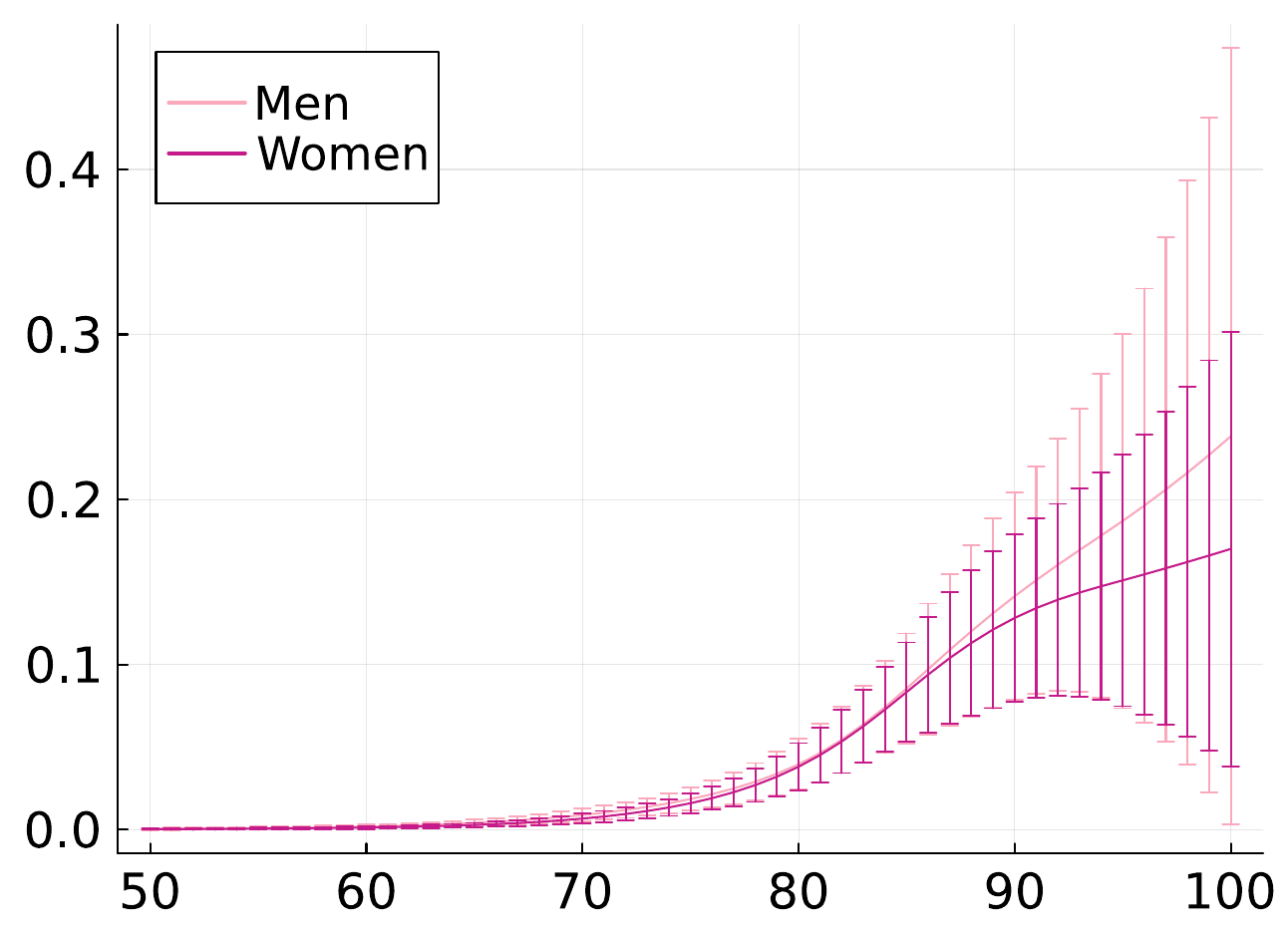}}\subfloat[\label{fig:incidence} Annual dementia incidence.]{\centering{}\includegraphics[scale=0.3]{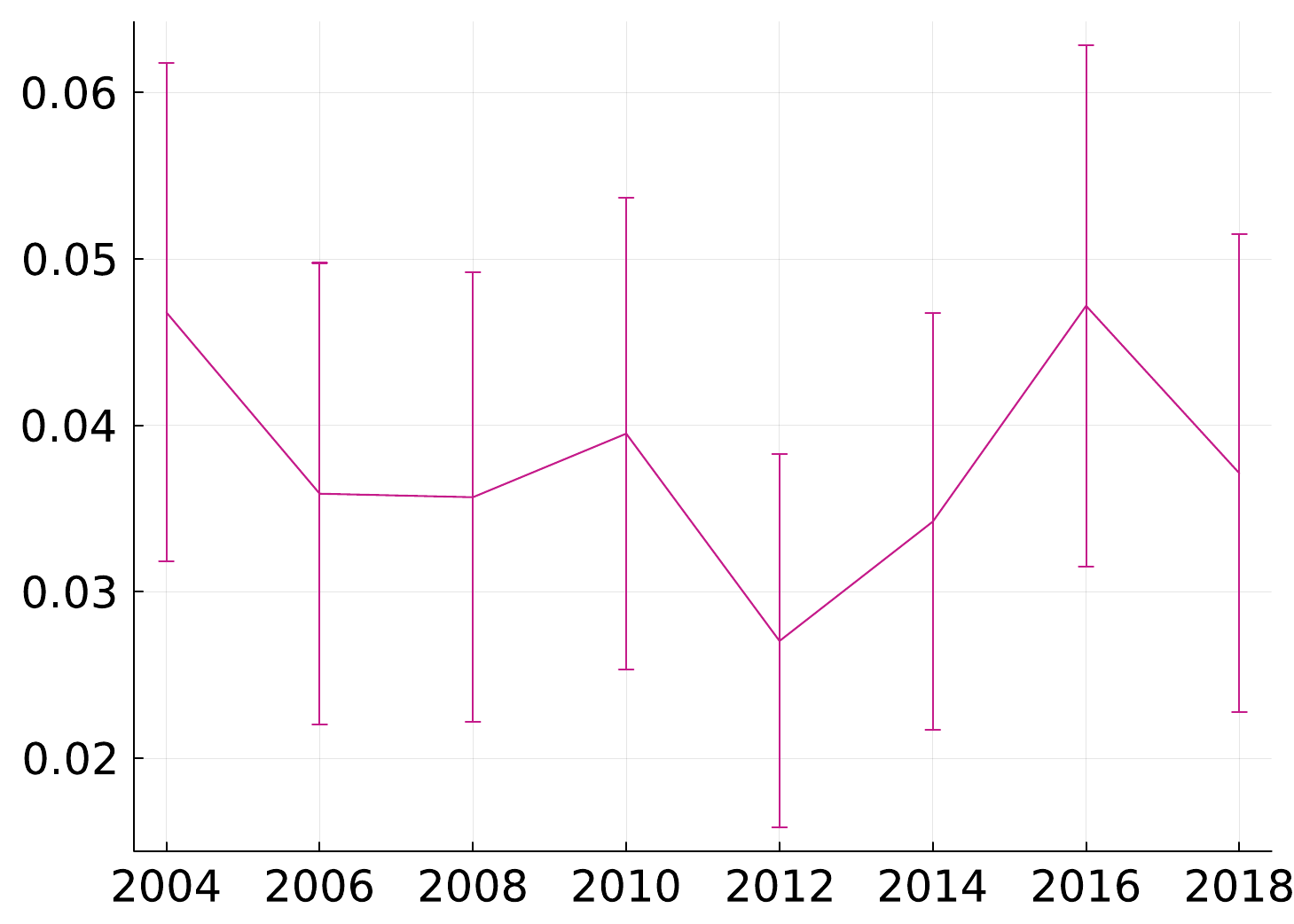}}
\par\end{centering}
{\scriptsize\textit{Notes}}{\scriptsize : The plot (a) displays estimated
dementia incidence for the sample period for each age group using
the procedures described in Section \ref{sec:msm}. For each age and
gender, we predict dementia incidence in each year of our sample period,
then take the average over all years. For higher age brackets, we
see that error bars widen significantly based on the lower sample
sizes. Consistent with the results in \citet{CHEN2023e859}, dementia
incidence is higher for men across all age groups with the split becoming
more pronounced with age. The plot (b) displays estimated annual dementia
incidence for women aged 80 obtained from the multi-state model of
Section~\ref{sec:msm}. Because the estimated model is separable
between demographics and the time trend, estimated incidence rates
of males are shifted by a constant amount where male incidence is
strictly above that of women. The error bars representing 90\% confidence
intervals are obtained from variances on the main diagonal of $\hat{\Sigma}_{TT}$
and thus reflect cross-sectional variation. The large error bars stemming
from cross-sectional variation indicate that the trend underlying
the incidence process is actually flat for both sexes.}{\scriptsize\par}

\end{figure}
Figure~\ref{fig:age-effect-neat} shows the annual dementia incidence
by age and sex using the multi-state model described in Section \ref{sec:msm}.
Dementia incidence increases rapidly with age and is higher for men.
Rates at age 80 are $3.7\%$ and $3.8\%$ for women and men, respectively.
By $90$, the rate rises to $12.3\%$ and $13.6\%$. The $90\%$ confidence
intervals grow rapidly with age, especially for men, as mortality
decreases sample size.\begin{landscape}
\begin{table}
\caption{Time series model estimates}
\vspace{0.1cm}
 \begin{tabular}
{r| ccc | ccc}
    \hline
    \hline
    &  \multicolumn{3}{c|}{{\textbf{Variance of measurement error }}} & \multicolumn{3}{c}{ {\textbf{Variance of measurement error }}}   \\
    &  \multicolumn{3}{c|}
    {{\textbf{ estimated using multi-state model}}} & \multicolumn{3}{c}{ {\textbf{ estimated using Kalman filter}}}   \\
    \hline
                        & Random walk  & Random walk  & Random walk  & Random walk  & Random walk  & Random walk  \\                     
                        &  \& zero drift &  \& const. drift &  \& stoch. drift & \& zero drift & \& const. drift &  \& stoch. drift \\ 
                                 \hline
                 $\sigma_{\eta}$ &       0.148 &                      0.137 &                       0.183 &  0.000 &                 0.000 &                   0.000 \\
                             &     (0.063,\,0.349) &                    (0.052,\,0.361) &                     (0.077,\,0.434) &      &                     &                      \\
                       $\nu$ &             &                     -0.028 &                             &             &                     -0.047 &                             \\
                             &             &                    (-0.117,\, 0.061) &                             &             &                    (-0.077,\,-0.017) &                             \\
                 $\sigma_{\xi}$ &             &                            &                 0.000 &             &                            &                       0.064 \\
                             &             &                            &                      &             &                            &                     (0.009,\,0.447) \\
          $\sigma_{\varepsilon}$ &   0.133          &  0.133                          &     0.133                        &       0.179 &                      0.210 &                       0.172 \\
                             &             &                            &                             &                         (0.115,\,0.278) &             (0.135,\,0.326) &                     (0.010,\,0.297) \\ \hline
   
                Log-likelihood &       3.606 &                      3.732 &                      -2.150 &       5.970 &                      4.308 &                      -2.054  \\
                            BIC &      -5.132 &                     -3.305 &                       8.458 &      -7.781 &                     -2.378 &                      10.346             \\
                        Q(4) &       2.854 &                      2.631 &                       6.690 &       1.944 &                      1.666 &                       6.473 \\
                          BS &       2.561 &                      2.748 &                       0.166 &       2.487 &                      1.991 &                       0.158 \\
                        r(1) &      -0.018 &                      0.013 &                      -0.281 &       0.141 &                      0.267 &                      -0.268 \\
                        r(2) &      -0.267 &                     -0.250 &                      -0.467 &      -0.303 &                     -0.232 &                      -0.474 \\
    \hline
    \hline
    \end{tabular} \label{tab:THE_ONE}
\\
 {\scriptsize\textit{Notes}}{\scriptsize : Summary of the models that
do (left three columns) and do not (right three columns) incorporate
the estimated cross-sectional variance $\hat{\Sigma}_{TT}$ estimated
using the multi-state model described in Section \ref{sec:msm}. Standard
deviations $\sigma_{\eta}$, $\sigma_{\xi}$ $\sigma_{\varepsilon}$,
of dementia incidence shocks, drift process shocks, and cross-sectional
measurement error, respectively, are presented with 90\% confidence
intervals in parentheses. The values on the left-hand columns for
$\sigma_{\varepsilon}$ show the mean sampling error $\sqrt{\frac{1}{7}\sum_{k=2}^{8}\sigma_{kk}^{2}}$
where $\sigma_{kk}^{2}$ are estimated using the multi-state model.
The parameter $\nu$ denotes the constant drift term. The random walk
model without drift in column (1) sets $\nu=\sigma_{\xi}=0$. The
random walk model with constant drift in column (2) sets $\sigma_{\xi}=0$
but with $\nu$ estimated. The model with a time-varying drift in
column (3) estimates both parameters. The next three columns represent
the same specifications, but with the cross-sectional measurement
error variance $\sigma_{\varepsilon}^{2}$ left as a free parameter.
The Bayesian Information Criterion (BIC)= $\big(-2(\widehat{\text{Log-likelihood}})+((\#\text{ of parameters})\cdot\ln T)\big)$
weighs the number of parameters against the log-likelihood and provides
a parameter-weighted metric of comparison between models. The $Q\left(4\right)$
statistic is for the Ljung-Box test which tests the null hypothesis
of no serial correlation and corresponds to a critical value taken
from a $\chi_{4}^{2}$ distribution for which $\Prob\left\{ \chi_{4}^{2}<.95\right\} \approx9.49$.
The lag length of $4$ for the Ljung-Box test was chosen based on
the rule of thumb given at \url{https://robjhyndman.com/hyndsight/ljung-box-test/}.
The statistics $r\left(\cdot\right)$ refer to auto-correlations of
residuals at the specified lag length.}
\end{table}
\end{landscape}

Figure~\ref{fig:incidence} shows the estimated annual rate over
time. For 80-year-old women, it falls from 4.7\% to 4.0\% between
2004 and 2010 before falling further to 3.7\% in 2018. 
These estimates are similar to those of \citet{CHEN2023e859}, who
use the same data and a similar model. They find that the annual rate
for 80-year-old men decreases from 4.7\% to 3.3\% from 2004 to 2010
before rebounding to 4.2\% in 2018. We adjust the years reported by
\citet{CHEN2023e859} by two since we plot end of wave estimates.
The key differences between our and their results arise from their
use of polynomials for time trend estimates while we use wave dummies
capturing more fine-grained variation. Although these results suggest
declining population level dementia incidence, the large confidence
intervals beg the question of whether the observed variation arises
from sampling variability.

\subsection{Estimates using the Kalman filter}

Next, we discuss results from the models using the filter explained
in Section~\ref{sec:ts-mod}. Table~\ref{tab:THE_ONE} presents
estimated coefficients of two model classes. The first (left panel)
is where we constrain the measurement error variance $\sigma_{\varepsilon}^{2}=\sigma_{kk}^{2}$.
Therefore, the estimated cross-sectional sampling variance from the
multi-state model enters the time series model as measurement error
variance. In the second class of models (right panel), measurement
error variance is estimated jointly with other parameters.

The top panel shows parameter estimates. The first column presents
estimates when the incidence process $\{\beta_{k}\}_{k=1}^{T}$ follows
a random walk with zero drift. The estimate of $\hat{\sigma}_{\eta}=0.148$
is significant and its confidence interval shows that it is bounded
away from zero and explains a substantial amount of variation.

Two reasons affect dementia incidence over time: (i) the underlying
\emph{population level} dementia incidence $\{\beta_{k}\}_{k=1}^{T}$
and (ii) \emph{sampling variability} in the ELSA data producing measurement
error. We reject that (ii) alone is responsible for changes displayed
in Figure~\ref{fig:incidence} as the CI for $\hat{\sigma}_{\eta}$
excludes zero.

To see how a shock to the dementia process translates to a shift in
incidence, recall that the model is estimated in logs: column (1)
implies a one-SD shock increases incidence by $14.8\%$ in the RW
model with zero drift. Allowing a drift (column 2), a one-SD shock
increases dementia incidence by $13.7\%$, while the estimated drift
implies a statistically insignificant $2.8\%$ average yearly reduction
in incidence. Column 3 presents a RW model where the drift is itself
a stochastic process. This least parsimonious model allows for shocks
to $\{\beta_{k}\}$ as well as those to the stochastic drift process
$\{\nu_{k}\}$. We do not find evidence of a stochastic drift as $\hat{\sigma}_{\xi}=0$.
Moreover, this estimate occurs on the boundary of the parameter space
and produced a zero eigenvalue in the Hessian. Switching to a model
with a deterministic drift (column 2) we find that the 90\% confidence
interval contains zero. We thus find no evidence for any drift in
dementia incidence over our sample period.

The bottom panel shows model fit and diagnostic tests. The first row
displays the log-likelihood which is found based on one-step ahead
prediction errors generated by the Kalman filter. Because out-of-sample
errors determine model fit, there is no reason that adding more parameters
should increase the log-likelihood. Indeed, adding parameters by modelling
the drift as its own stochastic process decreases the likelihood,
suggesting worse out-of-sample performance. While the likelihood is
slightly higher in column 2 than in column 1, it is considerably lower
in column 3, despite a more flexible specification.

The Bayesian Information Criterion (BIC)=$\big(-2\cdot(\widehat{\text{Log-likelihood}})+((\#\text{ of parameters})\cdot\ln T)\big)$
penalizes the likelihood for the number of parameters so that a higher
BIC implies a worse model fit. Such accounting for the number of parameters
shows the fit always deteriorating with more parameters.

The Ljung-Box ($\text{Q}\left(4\right)$) statistic measures the degree
of autocorrelation in the residuals, here at lag four. Under the null
of 0-autocorrelation, this statistic is a $\chi^{2}(4)$ variable.
While no model exceeds the $95$th percentile ($9.49$ ), the random
walk with constant drift model (column 2) minimizes this statistic.
The Bowman-Shenton ($\text{BS}$) tests measure the prediction errors'
deviations from normality via the skewness and excess kurtosis. Under
the null of normal prediction errors, the $\text{BS}$ test statistic
follows a $\chi^{2}(2)$ distribution. While again no model exceeds
the $95$th percentile ($5.99$), this time the random walk with stochastic
drift model minimizes this statistic. Finally, we include the estimated
first and second order residual autocorrelations. The random walk
model with stochastic drift induces significant negative autocorrelations,
again highlighting its overfitting problems.

The right three columns show estimates when $\sigma_{\varepsilon}^{2}$
is freely-varying and estimated as a parameter within a Kalman filter
procedure as opposed to a multi-state model one. Doing so produces
a modestly larger estimate of $\sigma_{\varepsilon}^{2}$ than when
using an estimated $\sigma_{\varepsilon}^{2}$ from the multi-state
model. In the random walk with zero drift specification, $\hat{\sigma}_{\varepsilon}=0.179$.
As a result, this approach interprets all estimated time series variation
in $\hat{\beta}_{k}$ as measurement error arising from cross-sectional
sampling variability and none of it as variability in the underlying
dementia process, so that $\hat{\sigma}_{\eta}=0$. We do not report
confidence intervals for the estimates of $\sigma_{\eta}$ in those
columns because the Hessian is singular in the direction of this parameter
for columns 4-6 making them difficult to justify or interpret.

The left panel of Table \ref{tab:THE_ONE}, where $\sigma_{\varepsilon}$
is estimated using the multi-state model, is our preferred approach,
because it uses additional information to discipline estimates of
$\sigma_{\varepsilon}$ revealing that measurement error from sampling
variability alone cannot fully explain variability in the estimated
dementia series in Figure~\ref{fig:incidence}. Instead, some of
the variability in estimated dementia incidence represents time series
variation in \emph{population level} dementia incidence.

Both the model fit indicated by the likelihoods in Table~\ref{tab:THE_ONE}
and the evidence in Table~\ref{tab:critvals} point to our preference
for a random walk model with zero drift. Although we reject the models
with non-zero constant and stochastic drifts, they nevertheless help
elucidate some of the possible dynamics.\begin{adjustwidth}{-2cm}{-2cm}\begin{landscape}
\begin{figure}
\centering \caption{Estimated dementia process and forecasts for models presented in three
left-hand side columns of Table~\ref{tab:THE_ONE}.}
\label{fig:single-plot} \subfloat[Random walk, zero drift.]{\includegraphics[scale=0.4]{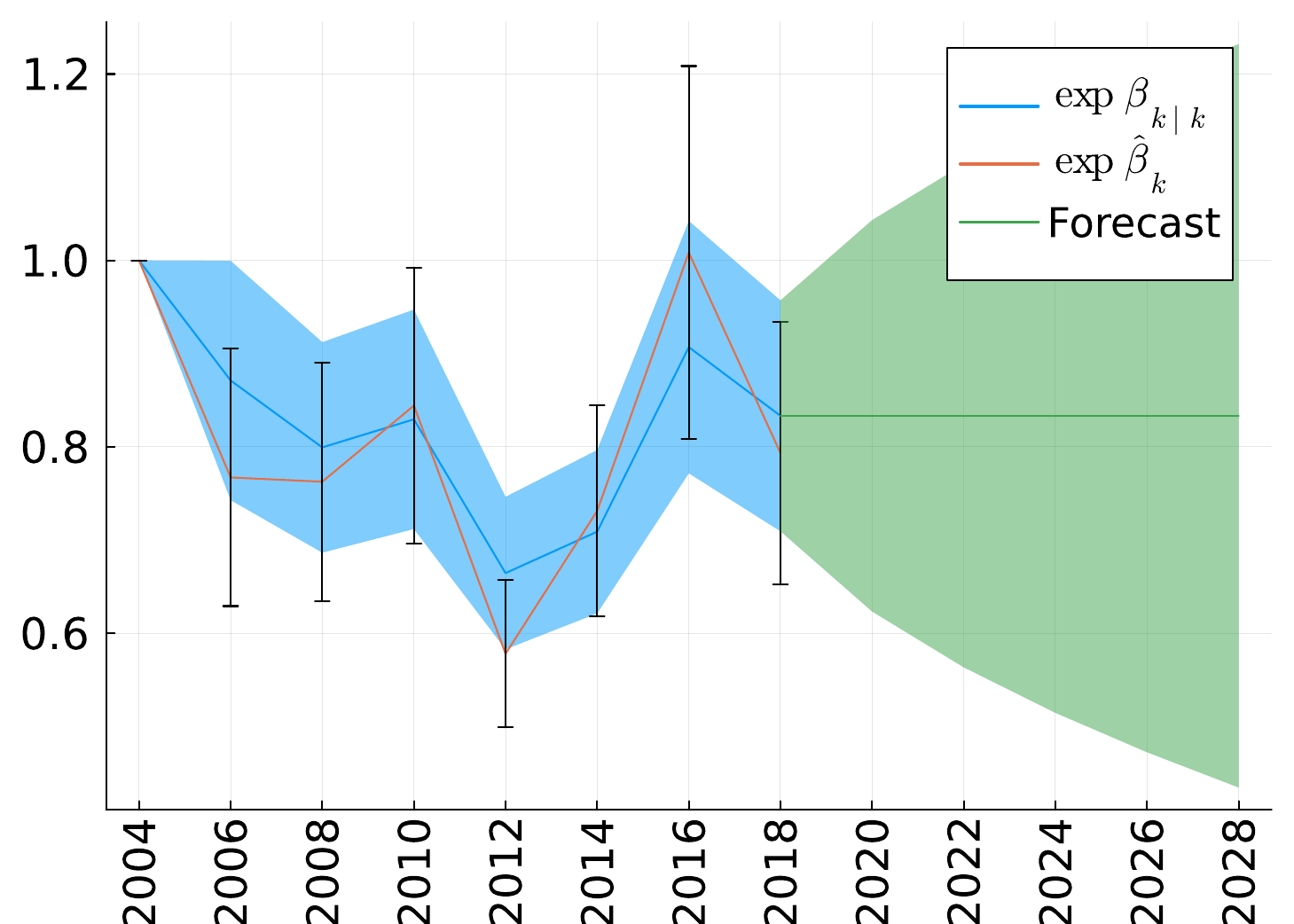}

}\hfill{}\subfloat[Random walk, fixed drift.]{\includegraphics[scale=0.4]{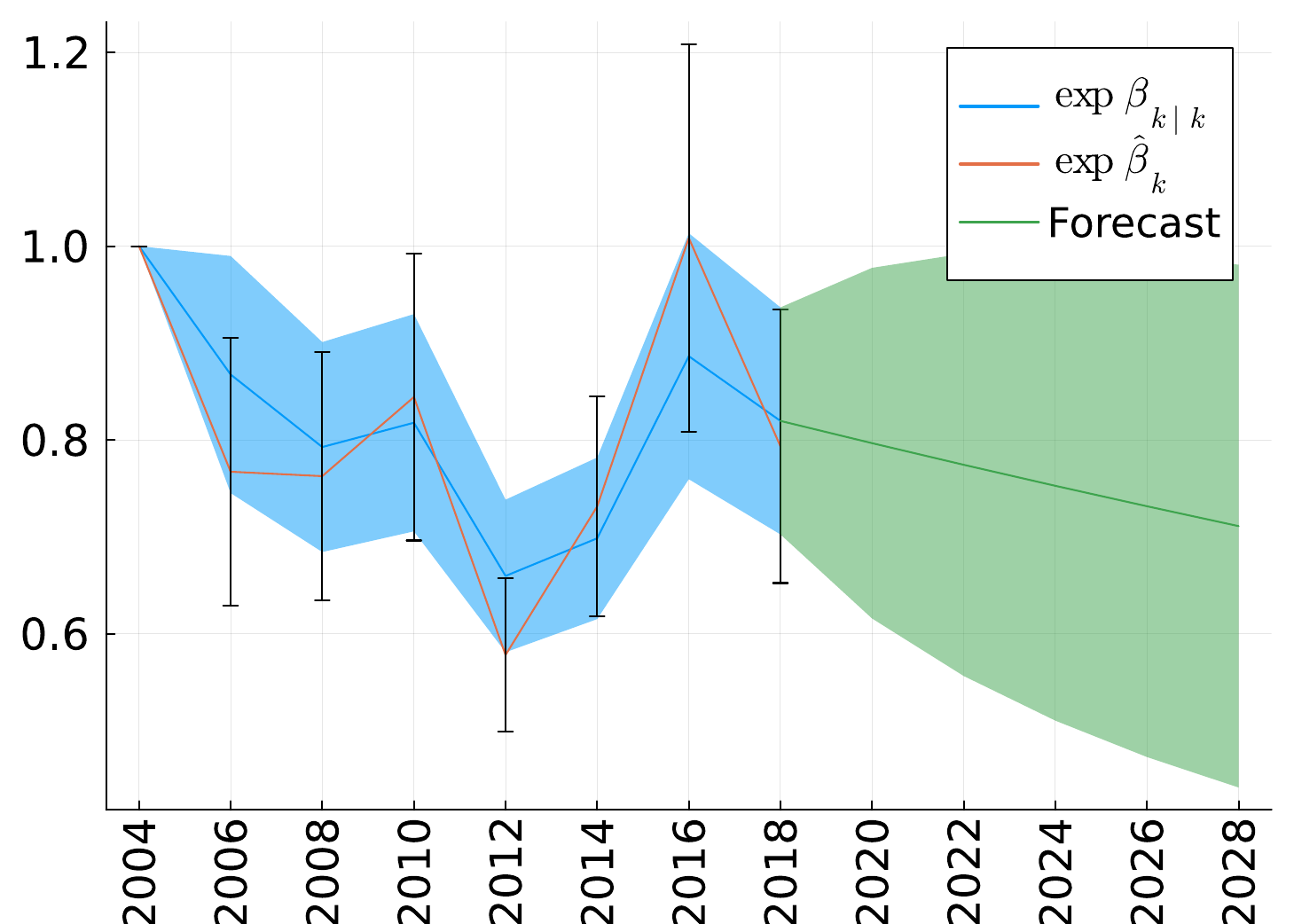}

}\vfill{}
 \subfloat[Random walk, stochastic drift process.]{\includegraphics[scale=0.4]{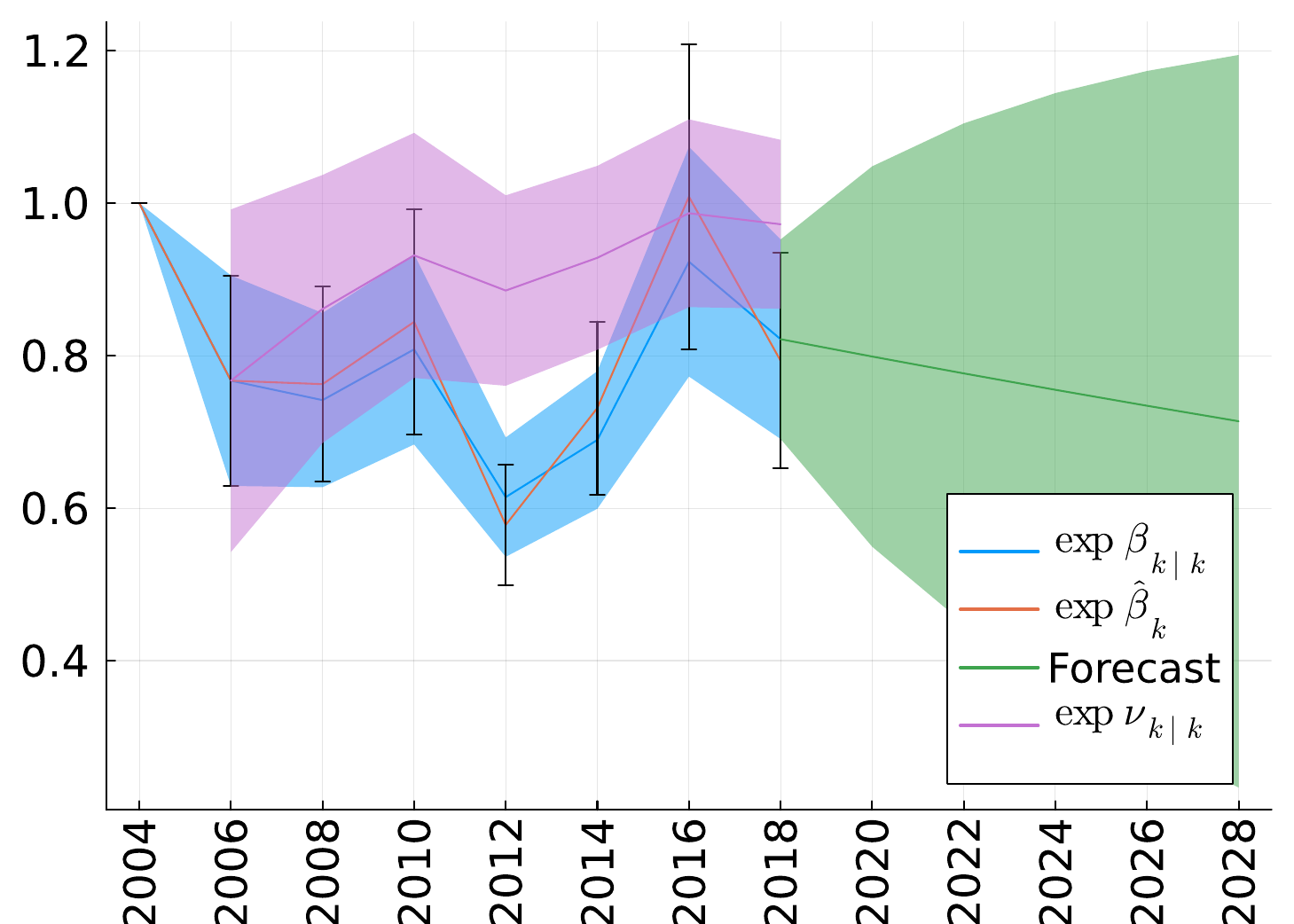}

}\justifying

\begin{singlespace}
\noindent{\tiny\emph{Notes}}{\tiny : These figures show the estimated
state from the Kalman filter $(\exp{\beta}_{k|k})$ with a blue shaded
region representing its $90\%$ confidence interval. The estimated
process $(\exp\hat{\beta}_{k})$ from the multi-state model is the
red line, with associated error bars in black. The green line is the
forecast, with the green shaded region showing the $90\%$ confidence
interval of the forecast. Column 2 of Table~\ref{tab:THE_ONE}. The
estimated process $(\exp\hat{\beta}_{k})$ from the multi-state model
is the red line, with associated error bars in black. The green line
is the forecast, with the green shaded region showing the $90\%$
confidence interval of the forecast. This model introduces a constant
drift parameter $\nu$ into the random walk model. The estimated drift
is negative $(-0.028)$ corresponding to a $3\%$ reduction of the
hazard in each two-year period. The plot (c) shows the estimated state
from the Kalman filter $(\exp{\beta}_{k|k})$ with a blue shaded region
representing its $90\%$ confidence interval. The estimated process
$(\exp\hat{\beta}_{k})$ from the multi-state model is the red line,
with associated error bars in black. The green line is the forecast,
with the green shaded region showing the $90\%$ confidence interval
of the forecast.}{\tiny\par}
\end{singlespace}
\end{figure}

\end{landscape}\end{adjustwidth}

\subsection{Forecasting dementia incidence}

Figure~\ref{fig:single-plot} presents estimates of the time series
of dementia incidence. Each panel shows the hazard shifter $\exp\{\hat{\beta}_{k}\}$
(red line) from \eqref{eq:transition-probability} with black error
bars showing $90\%$ confidence intervals. Posterior process estimates
from the Kalman filter, $\exp\left\{ \beta_{k|k}\right\} $ appear
in blue with the ribbon showing a 90\% confidence interval based on
$P_{k\mid k}$. Forecasts along with a 90\% confidence interval are
green.

Panel (a) presents forecasts from the random walk model with no drift:
Table~\ref{tab:THE_ONE}, col. 1 shows estimates of this preferred
model. The graph shows that series from the filter and the multi-state
model are close although filtered estimates are less variable than
those from the multi-state model; our modified filter downweights
observations $\hat{\beta}_{k}$ estimated from the multi-state model
if they are either extreme and/or come with large confidence intervals.
The 90\% confidence intervals from the two models are also similar,
corresponding to a roughly even split between cross-sectional and
process noise variance. The model highlights that estimates for $\hat{\beta}_{k}$
for 2012 and 2016 are outliers. In 2012, the black error bar just
about touches the Kalman estimate $\exp\left\{ \beta_{k\mid k}\right\} $
with its upper end and falls far outside the blue ribbon for its lower
end, likely making this an outlier. For the year 2016, we see that
the upward correction relative to 2014 is only moderate due to the
large amount of uncertainty communicated by the black error bar, which
per \eqref{eq:gen-gain} will result in a much lower gain thus downweighting
that observation. This specification assumes zero drift implying a
flat forecast.

Panel (b) shows estimates from the random walk with constant but potentially
non-zero drift model. The corresponding Kalman filter parameter estimates
are shown in column two of Table~\ref{tab:THE_ONE}. An estimated
non-zero negative drift parameter implies that this model predicts
a decline in dementia incidence. Relative to panel (a), the updated
posterior process estimates track the estimated $\hat{\beta}_{k}$
slightly less well than the model with zero drift implying that the
downward shift caused by $\nu<0$ is not justified. This worsening
of model fit is even greater for the one-step ahead forecasts on which
the likelihood is based but which are not pictured here.

Figure~\ref{fig:single-plot} (c) graphs the model estimates corresponding
to column three in Table~\ref{tab:THE_ONE}. This model includes
a whole drift process in purple. The process $\exp\left\{ \nu_{k\mid k}\right\} $
approaches one throughout the sample period implying that the drift
parameter in log incidence approaches zero, commensurate with a flattening
trend process. However, based on Table \ref{tab:THE_ONE}, this model
performs even worse than the previous one depicted in panel (b). Figures~\ref{fig:single-plot}
(b) and \ref{fig:single-plot} (c) show some evidence of a negative
drift but are outperformed by the zero drift model.

\subsection{Further tests for time trends}

We formally test whether the wave-to-wave variation in measured dementia
incidence is due to sampling variability, i.e. the hypothesis that
$\beta_{k}=0$ for all $k$. To this end, we evaluate the $F$-test
statistic shown in (\ref{eq:F-stat}): $\hat{\beta}^{\prime}\hat{\Sigma}^{-1}\hat{\beta}/T\approx3.298,$
which is distributed $\chi_{7}^{2}$ under the null that $\beta_{k}=0$
for all $k$. The test statistic has a $p$-value of $77.1\%$, meaning
that we cannot reject that all $\beta_{k}$ are zero and thus cannot
reject the hypothesis of no drift in dementia incidence.

Table~\ref{tab:critvals} presents specific tests for the nature
of the dementia trend. In particular, we test for whether a non-zero
drift in dementia incidence may be present. A zero drift parameter
in (\ref{eq:drift-term}) corresponds to a model with no drift. The
top row presents the test for whether a zero, deterministic drift,
$\nu=0,\,\sigma_{\xi}^{2}=0$, can be rejected against a non-zero,
deterministic drift, i.e. $\nu_{k}\neq0,\,\,\sigma_{\xi}^{2}=0$.
The test statistic, shown in equation (\ref{eq:zero-trend-test}),
has a $t-$distribution (or normal distribution in large samples)
under the null hypothesis. The test statistic is $-0.685$ with a
$p-$value of 49.5\%, meaning that we cannot reject a zero, deterministic
drift. The second row presents a test that discriminates between a
stochastic and deterministic drift $\nu_{k}$. In other words, we
test whether $\nu_{k}\neq0$ and $\sigma_{\xi}^{2}=0$, i.e. if there
is a non-zero drift and whether it is non-stochastic. The associated
test statistic appears in (\ref{eq:zero-trend-test-detr}). We cannot
reject this hypothesis at the 5\% level. Finally, the third row presents
a test for whether we can reject no drift in favor of a stochastic
one where we impose $\nu_{k}=0$ in addition to $\sigma_{\xi}^{2}=0$
under the null. The relevant test statistic, defined in (\ref{eq:stoch-trend-stats-notdetr}),
assumes no drift, making it inconsistent if one is present, which
however buys more power if this assumption is correct. We note that
the test based on $t_{\nu}$ in the first row of Table~\ref{tab:critvals}
is evidence that a deterministic trend would be zero, which inspires
confidence in the assumption that $\nu_{k}=0$, which we impose under
both null and alternative hypotheses of the test in the third row.
In summary, we do not reject the null hypothesis of no trend and we
do not reject the null hypothesis of no stochastic drift. In sum,
we do not reject that the true model has a zero, deterministic drift.

However, the inference appearing in the top row of Table~\ref{tab:critvals}
originates with a normal approximation for $t_{\nu}$ in (\ref{eq:zero-trend-test}),
which is only true if the denominator $\sigma_{\text{HAC}/\text{L}}$
in (\ref{eq:long-run-var}) or (\ref{eq:hac-var}) is known and not
estimated. Because our test statistic is based on $\hat{\sigma}$,
estimated on a short sample, $t_{\nu}$ will follow a $t$-distribution
with $T-1$ degrees of freedom. Using such critical values, the $p$-value
rises to $51.6\%$. Therefore, combining the conclusions of no stochastic
drift in rows two and three with the evidence of $\nu=0$ in row one
of Table~\ref{tab:critvals}, we conclude that the drift parameter
is a constant not different from zero. Figure~\ref{fig:p-values}
in Appendix~\ref{subsec:Sensitivity-analyses-t-tests} presents a
sensitivity analysis and graphs $p$-values for different variance
estimators and lag lengths, which all support this conclusion. Furthermore,
the evidence of the tests for trend is consistent with those results
presented in Table~\ref{tab:THE_ONE}, which implies that a model
with no drift is preferred.

\begin{table}
\caption{Tests of absence and nature of a time trend in dementia incidence.}
\label{tab:critvals} \smallskip{}
 \begin{adjustwidth}{-1cm}{-1cm} 
\begin{centering}
\begin{tabular}{p{3.5cm}p{3cm}>{\raggedright}p{2cm}|p{1.5cm}ccc}
\hline 
$H_{0}$  & $H_{1}$  & Test for:  & Critical value  & Symbol  & Test statistic  & $p-$value \tabularnewline
\hline 
\hline 
$\nu_{k}=0\ \forall\ k,\sigma_{\xi}^{2}=0$  & $\nu_{k}=\nu\neq0,\sigma_{\xi}^{2}=0$  & No constant drift  & $\pm1.96$  & $t_{\nu}$  & $-0.685$  & $49.4\%$ \tabularnewline
\hline 
$\nu_{k}=\nu\ \forall\ k,\sigma_{\xi}^{2}=0$  & $\sigma_{\xi}^{2}>0$  & No stochastic drift  & $0.452$  & $t_{\text{s,d}}$  & $0.336$  & $8.27\%$ \tabularnewline
\hline 
$\nu_{k}=0\ \forall\ k,\,\sigma_{\xi}^{2}=0$  & $\nu_{k}=0,\sigma_{\xi}^{2}>0$ or $\nu_{k}\neq0,\sigma_{\xi}^{2}>0$  & No stochastic drift  & $1.594$  & $t_{\text{s}}$  & $0.683$  & $23.4\%$ \tabularnewline
\hline 
\end{tabular}
\par\end{centering}
\end{adjustwidth}{\scriptsize\emph{Notes}}{\scriptsize : The table
summarizes the outcomes of hypothesis tests for a zero drift (row
one, with the test statistic presented in (\ref{eq:zero-trend-test}))
and for a stochastic drift (rows two and three, with test statistics
shown in equations (\ref{eq:zero-trend-test}), (\ref{eq:zero-trend-test-detr}),
and (\ref{eq:stoch-trend-stats-notdetr})) using variance estimates
$\sigma_{\text{HAC}}$ based on $\hat{\Sigma}_{TT}$ from Section
\ref{sec:msm}. The lag length is three using the rule-of-thumb $\text{lag length}=\sqrt{T}$.
The test in the first row assumes a constant drift under both null
and alternative hypotheses and checks whether it is different from
zero. The tests in rows two and three are essentially very similar,
with the difference that the test statistic for that in row three
(\ref{eq:stoch-trend-stats-notdetr}) has not been detrended, thereby
imposing a zero drift under the null hypothesis ($\nu_{k}=0$) for
all $k$ which makes this test also consistent against a constant
non-zero drift \citep[end of page 8]{slopetest}. While $t_{\nu}$
follows a normal distribution asymptotically or a $t$-distribution
with seven degrees of freedom, we obtain critical values for $t_{\text{s,d}}$
and $t_{\text{s}}$, by simulation of the stochastic integrals with
code available on request. Experiments with coarse grids for stochastic
integrals showed no noticeable difference with $10,000$ MC repetitions. }{\scriptsize\par}

\end{table}

\section{Kalman filter convergence}

\label{sec:conv-analysis}

In this section, we show that the gain and the estimated process covariance
converge rapidly to their steady state values after only a few time
periods. Thus, the Kalman filter works well even with only eight periods
of data.

Relative to joint parameter estimation, using the Kalman filter is
inefficient because finding $\beta_{1|1},\dots,\beta_{T|T}$ involves
repeated updating of the filter. Initialization using a diffuse prior
does not use future information on dementia incidence and relies on
convergence. Joint estimation of all parameters would use all information
to inform incidence in the initial period. Hence, the gain may converge
slowly leading to imprecise estimation of $\beta_{k|k}$ in a short
sample.

Recall the gain from (\ref{eq:gen-gain}) and note that $K_{k}$ is
the share of variability in incidence attributable to the underlying
process ($P_{k|k-1}$) relative to the sum of that and cross-sectional/noise
variance ($\hat{\sigma}_{kk}^{2}$). The larger this share is, the
more weight our noisy measurement $\hat{\beta}_{k}$ receives in the
updating step.

In this section, we prove that the Kalman gain $(K_{k})$ converges
to a fixed point, building on \citet{doi:10.1137/0331041}, who proved
that the posterior variance $P_{k\mid k}$ converges to a fixed point
when $\hat{\sigma}_{kk}^{2}$ is constant. We extend the proof by
both allowing for time-varying $\hat{\sigma}_{kk}^{2}$, and by showing
explicitly that $K_{k}$ converges to a fixed point. More technical
details appear in Appendix \ref{app:appendix-for-convergence}.

In our context, the cross-sectional variance $\hat{\sigma}_{kk}^{2}$
is time-varying and given exogenously, which prevents $K_{k}$'s convergence
to a constant. However, as long as $\hat{\sigma}_{kk}^{2}$ varies
only modestly, we may still consider the filter elements $K_{k}$
and $P_{k\mid k}$ convergent. Following the assumptions in Section~\ref{subsec:Derivation-of-constrained},
start the algorithm with a diffuse prior. Formally, we have
\begin{assumption}
\label{assu:diffuse-prior}~
\begin{enumerate}
\item \label{enu:diffuse-prior}For a diffuse prior, we have $P_{0|0}=\infty$
and $K_{1}=1$.
\item \label{enu:slight-variation}For $k=2,\dots,T$, the relative variation
in $\hat{\sigma}_{kk}$ satisfies $\frac{\hat{\sigma}_{kk}^{2}}{\hat{\sigma}_{k+1k+1}^{2}}<(\frac{P_{k|k}+\hat{\sigma}_{kk}^{2}+\sigma_{\eta}^{2}}{\hat{\sigma}_{kk}^{2}})^{2}$.
\end{enumerate}
\end{assumption}

The infinite prior variance in Assumption~\ref{assu:diffuse-prior}.\ref{enu:diffuse-prior}
formalizes our ignorance about the state of the system before measurements
would begin. Its main consequence is $K_{1}=1$ meaning we give full
weight to the first measurement $\hat{\beta}_{1}$ which becomes the
process estimate for $k=1$. Assumption~\ref{assu:diffuse-prior}.\ref{enu:slight-variation}
ensures we retain a meaningful concept of filter convergence even
though a time-varying variance implies that fixed points will depend
on $k$. The intuition is that the measurement variance ratio between
successive $k$ is bounded above by the square of the ratio of total
and noise variances. The RHS is always bigger than one, implying that
this condition always implies a constant observational noise variance,
i.e. $\frac{\hat{\sigma}_{kk}^{2}}{\hat{\sigma}_{k+1k+1}^{2}}=1$,
as a special case. Generally, this condition requires that $\sigma_{\eta}^{2}$
is large enough relative to $\hat{\sigma}_{kk}^{2}$. The discussion
following Lemma \ref{lem:gain-recursion} contains alternative expressions.

We define the signal-to-noise ratio as $s_{k}\equiv\sigma_{\eta}^{2}/\sigma_{kk}^{2}$,
which measures the strength of the dementia process variance relative
to the cross-sectional or noise variance $\sigma_{kk}^{2}$. Because
$\sigma_{kk}^{2}$ varies over time, $s_{k}$ varies over time, as
well, and is only constant if we set $\sigma_{\varepsilon}^{2}\equiv\sigma_{kk}^{2}$
for all $k$. The Kalman gain is given in 
\begin{lem}
\label{lem:gain-recursion-1}The Kalman gain recursions are given
by\textup{ 
\begin{equation}
K_{k+1}=\frac{s_{k}\sum_{d=1}^{k}\prod_{i=d}^{k}\left(1-K_{i}\right)+s_{k}}{s_{k}\sum_{d=1}^{k}\prod_{i=d}^{k}\left(1-K_{i}\right)+s_{k}+1}\label{eq:single-map-gain-1}
\end{equation}
}where equation (\ref{eq:single-map-gain-1}) shows that the Kalman
gain is a function not only of the history of Kalman gains $\left\{ K_{i}\right\} _{i=1}^{k}$
but also the most recent realization of the signal to noise ratio
$s_{k}$. 
\end{lem}

A proof and extended version of Lemma~\ref{lem:gain-recursion-1}
appears in Appendix \ref{app:appendix-for-convergence}. Figure~\ref{fig:map-iterations}
shows the speed of convergence of the Kalman filter using estimated
parameter values of $s_{k}=\frac{\sigma_{\eta}^{2}}{\sigma_{kk}^{2}}$.
The $y$-axis shows the values of the gain $K_{k}$ for different
values of $s_{k}$ on the $x$-axis. To keep the analysis simple yet
account for the time-varying nature of $s_{k}$, the figure evaluates
convergence for a range of plausible values of $s$, with a CI drawn
in by the black lines. The mean $\bar{s}=$ 1.26. Using this value,
after the first iteration (i.e., after the first time period) the
gain is $K_{1}=0.56$. After the second, the gain $K_{2}=0.65$. Moving
from from the second iteration to further iterations provides little
extra gain as in the limit, $K_{\infty}=0.66$. Appendix \ref{app:appendix-for-convergence}
contains an extended figure.

A potential problem is that convergence may be slow if some realizations
of $s_{k}$ are very small. However, the Kalman filter largely converges
to its asymptotic value after two periods even at the bottom of the
CI for $\bar{s}$ (denoted $s^{l}$). Put differently, the Kalman
filter is inefficient, but after one period the additional potential
efficiency gain from other estimators is negligible.
\begin{figure}
\caption{The Kalman gain as a function of the number of iterations.}
\label{fig:map-iterations} 
\begin{centering}
\includegraphics{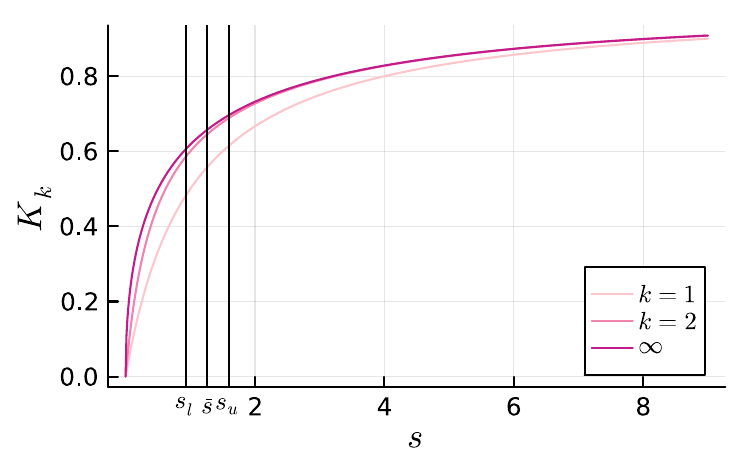} 
\par\end{centering}
{\scriptsize\emph{Notes}}{\scriptsize : The plot shows the values of
the gain map $K$ after $k$ iterations to visualize how quickly these
maps converge to their fixed points across a wide variety of signal-to-noise
ratios $s$. To account for the time-varying nature of $s$, vertical
black bars display summary statistics of the estimated $s_{k}$: $\bar{s}$
denotes the mean of all point estimates and the lower and upper ends
of the confidence intervals are denoted by $s_{l}$ and $s^{u}$,
respectively.}{\scriptsize\par}
\end{figure}

\section{Power to detect the direction of a trend}

\label{sec:power-and-size}

\citet{CHEN2023e859} show that a dementia incidence trend may have
stopped declining over time. Our findings in Section~\ref{sec:results}
lead to our preferred model being a random walk with no drift, meaning
it is not possible to predict incidence fluctuations. We further find
evidence that dementia incidence does fall in the first half of our
sample, and rises in the second, which underscores the difficulty
in predicting changes in dementia incidence.

Sampling variability possibly obscures falling dementia incidence
at the population level. While the Kalman filter addresses this problem
by down-weighting new realizations of estimated dementia incidence
when estimating population-level dementia incidence, it may still
errantly infer that population level dementia incidence has not fallen
when in fact it has. Therefore, this section evaluates the probability
that the Kalman filter correctly finds that a dementia shock $\eta_{k}$
is negative when it is in fact negative, under the assumption that
the true process is a random walk.

We consider the null hypothesis that a shock at time $k$ to the dementia
process $\eta_{k}$, defined in (\ref{eq:state-1}), is not negative
against the one-sided alternative that it is, 
\begin{equation}
H_{0}:\eta_{k}\geq0,\hspace{1cm}H_{1}:\eta_{k}<0.
\end{equation}

\noindent The false positive rate $\mathbb{P}\left\{ \text{reject}\,H_{0}|H_{0}\,\text{is true.}\right\} \equiv\alpha$
and the false negative rate $1-\mathbb{P}\left\{ \text{reject}\,H_{0}|H_{0}\,\text{is false.}\right\} \equiv1-\theta$
where $\theta$ is statistical power. Since the change in the measured
dementia incidence using the Kalman filter is $\beta_{k|k}-\beta_{k-1|k-1}$,
the power $\theta$ , or, equivalently, the probability that $\beta_{k|k}<\beta_{k-1|k-1}$
when a negative shock $\eta_{k}$ has occurred, is 
\begin{equation}
\theta=\mathbb{P}\left\{ \beta_{k|k}<\beta_{k-1|k-1}|\eta_{k}<0\right\} .\label{eq:power-prob}
\end{equation}
To express this inequality in terms of the shocks $\eta_{k}$, $\varepsilon_{k}$,
and initial conditions $\beta_{0}$ and $\beta_{1\mid0}$, we combine
(\ref{eq:Kal1}) with the state updating equation (\ref{eq:state-update})
to obtain $\beta_{k|k}-\beta_{k-1|k-1}=K_{k}v_{k}.$ Thus, (\ref{eq:power-prob})
becomes $\mathbb{P}\left\{ \beta_{k|k}<\beta_{k-1|k-1}|\eta_{k-1}<0\right\} =\mathbb{P}\left\{ K_{k}v_{k}<0|\eta_{k-1}<0\right\} $.
Lemma \ref{lem:asymptotic-coeffs} below shows that $K_{k}v_{k}$
is a function of the history of $\eta_{k}$ (the ``signal'' of the
process) as well as $\varepsilon_{k}$ (the ``noise'' of the process)
and presents an asymptotic approximation which becomes accurate beyond
$k=2$. Table \ref{tab:terms-2-and-3-simplified} shows the coefficients
to second and third order.
\begin{table}[ht]
\centering %
\begin{tabular}{cc@{\hspace{3cm}}cc}
\toprule 
\multicolumn{2}{c}{\textbf{Order 2 Coefficients}} & \multicolumn{2}{c}{\textbf{Order 3 Coefficients}}\tabularnewline
\cmidrule(lr){1-2}\cmidrule(lr){3-4}
Coefficient & Expression & Coefficient & Expression\tabularnewline
\midrule 
$c_{1}(2)$ & $K_{\infty}-K_{\infty}^{2}$ & $c_{1}(3)$ & $K_{\infty}-2K_{\infty}^{2}+K_{\infty}^{3}$\tabularnewline
$c_{2}(2)$ & $K_{\infty}$ & $c_{2}(3)$ & $K_{\infty}-K_{\infty}^{2}$\tabularnewline
 &  & $c_{3}(3)$ & $K_{\infty}$\tabularnewline
\midrule 
$d_{1}(2)$ & $-K_{\infty}^{2}$ & $d_{1}(3)$ & $-K_{\infty}^{2}+K_{\infty}^{3}$\tabularnewline
$d_{2}(2)$ & $K_{\infty}$ & $d_{2}(3)$ & $-K_{\infty}^{2}$\tabularnewline
 &  & $d_{3}(3)$ & $K_{\infty}$\tabularnewline
\bottomrule
\end{tabular}\caption{Coefficients $c_{i}(2)$, $d_{i}(2)$, $c_{i}(3)$, and $d_{i}(3)$.}
\label{tab:terms-2-and-3-simplified}
\end{table}
\begin{adjustwidth}{-2cm}{-2cm} \begin{landscape} 
\begin{figure}
\caption{Power $\theta$ and size $\alpha$ as a function of current period's
dementia shock $\eta_{k}/\sigma_{\eta}$.}
\label{fig:power}\centering\subfloat[Power as a function of standardized shock for different dates.]{\includegraphics[scale=0.45]{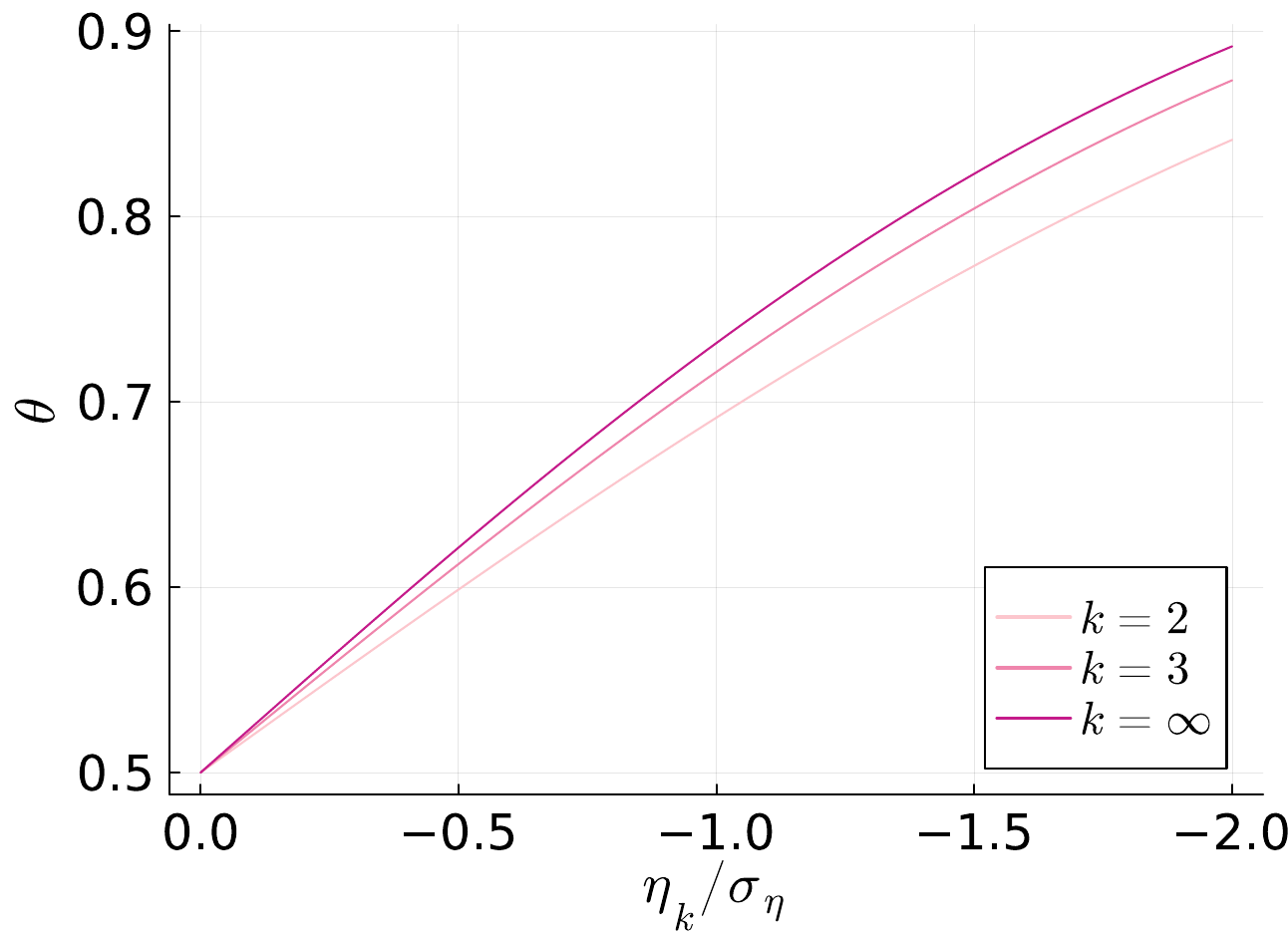}

}\hfill{}\subfloat[Power as a function of standardized shock for different $s$.]{\includegraphics[scale=0.45]{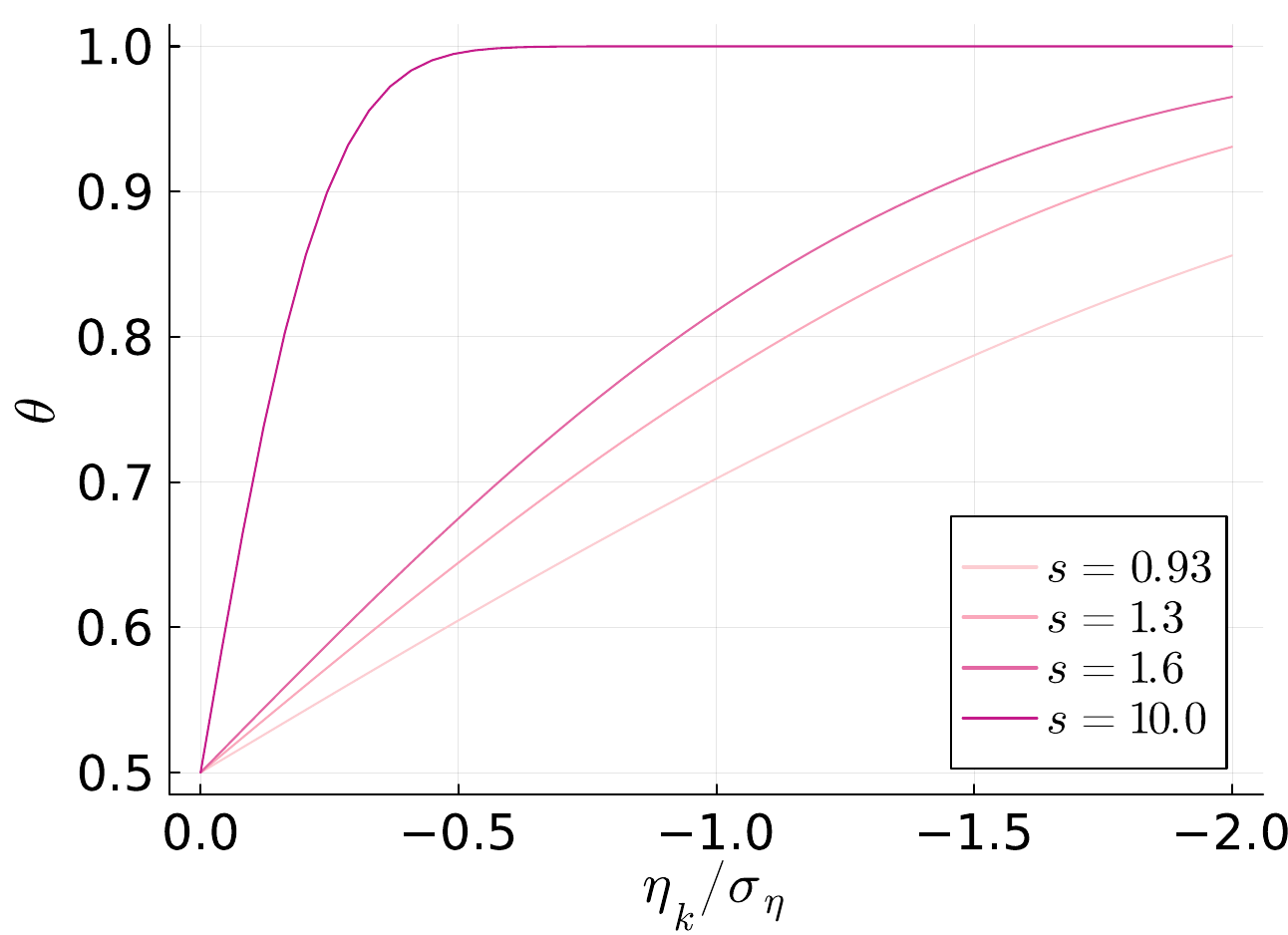}

}\vfill{}
 \subfloat[Size as a function of standardized shock for different dates.]{\includegraphics[scale=0.45]{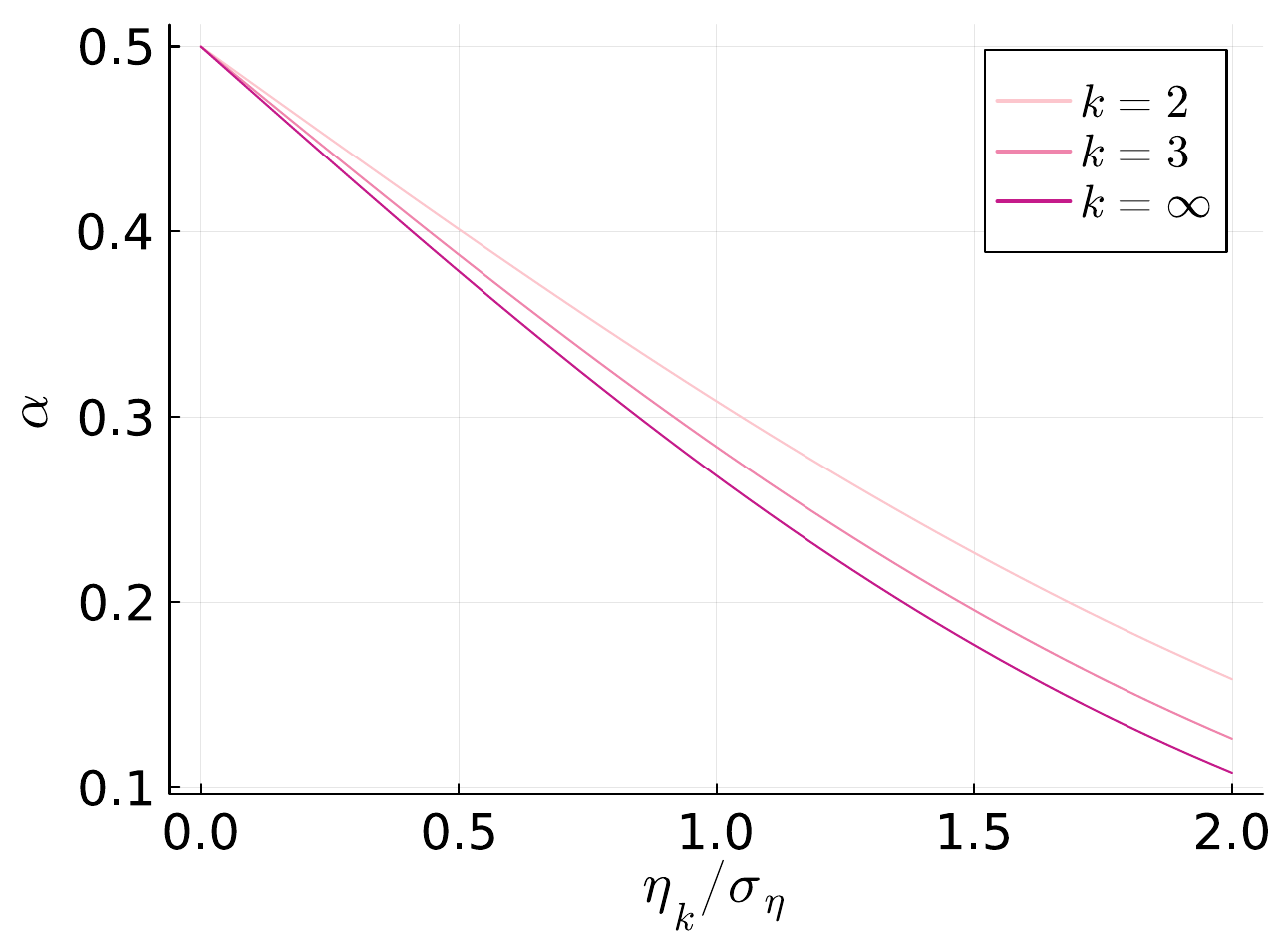}

}\hfill{}\subfloat[Size as a function of standardized shock for different $s$.]{\includegraphics[scale=0.45]{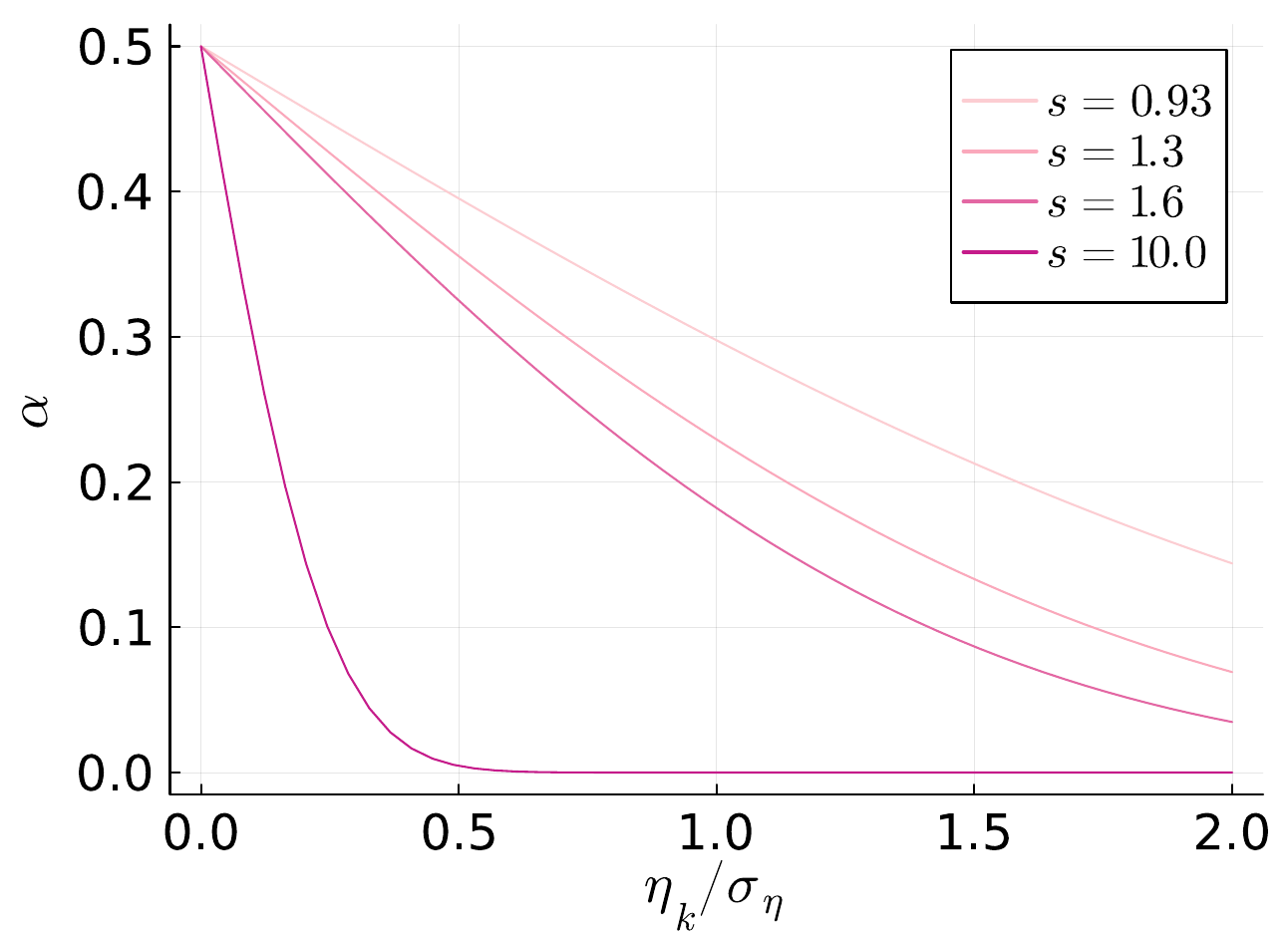}

}

\justifying {\scriptsize\textit{Notes: }}{\scriptsize The plots show
power ($\theta$) and size ($\alpha$) as a function of the current
period's shock ($\eta_{k}/\sigma_{\eta}$), normalized by its standard
deviation. The left panels keep the signal to noise ratio $s$ fixed
at the sample mean of the eight measurements and vary the number of
observations of dementia incidence to date $k$. It is evident that
with increasing $k$ power increases and size declines. The right
panels keep $k=4$ fixed and show plots of $\theta$ and $\alpha$
for the mean and $90\%$ CI of $s$ as well as the extreme value of
$s=10$. Both $\theta$ and $\alpha$ increase and decrease with $s$.}{\scriptsize\par}

\noindent{}
\end{figure}

\end{landscape} \end{adjustwidth}

Implementing the exact formulae provided by Lemma~\ref{lem:combinatorial-expansion}
in Appendix~\ref{app:power-and-size} presents a high computational
burden for sample sizes beyond $T=10$ as the length of the products
of Kalman gains $K_{1}\dots K_{k}$ doubles with each time step.\footnote{This result is established formally in Lemma~\ref{lem:combinatorial-forward}.\ref{enu:number-terms}
in Appendix \ref{app:power-and-size}.} Fortunately, because $K_{i}\in\left(0,1\right)$ by Lemma \ref{lem:gain-bound},\footnote{A similar result holds in the multi-variate case.}
the longer chains converge to zero while those of finite length approach
a stable limit.

As can be seen in Table \ref{tab:terms-2-and-3-simplified}, $c_{3}(3)\geq c_{2}(3)$
meaning that the most recent coefficients on $\eta_{i}$ outweigh
the more distant ones. Likewise, $\abs{d_{3}(3)}\geq\abs{d_{2}(3)}$.
This pattern where coefficients satisfy $\abs{d_{i}(k)}\geq\abs{d_{i-1}(k)}$
holds more generally which we show in Appendix~\ref{app:power-and-size}.
As a result, the marginal contribution by more distant $c_{i}$ and
$d_{i}$ decays to zero as $k$ grows meaning that more distant structural
and noise shocks contribute less information. The formulae are given
in
\begin{lem}[Asymptotic coefficients]
\label{lem:asymptotic-coeffs} For large $k$ such that $K_{k}$
is close to its steady state value $K_{\infty}$, then 
\begin{equation}
K_{k}v_{k}=\sum_{i=1}^{k}c_{i}(k)\eta_{i}+d_{i}(k)\varepsilon_{i},\label{eq:expansion-1}
\end{equation}
for coefficients $c_{i}(k),d_{i}(k),$ with $i\leq k$. We have the following formulae: 
\begin{equation}
c_{i}\left(k\right)\approx\sum_{m=0}^{k-1}\left(-1\right)^{m}\binom{k-i}{m}K_{\infty}^{m+1},\label{eq:eta-coefficients-1}
\end{equation}
\begin{equation}
d_{i}\left(k\right)\approx\sum_{m=0}^{k-i-1}\left(-1\right)^{m+\indic\left\{ k-i>0\right\} }\binom{k-i-1}{m}K_{\infty}^{m+1+\indic\left\{ k-i>0\right\} }.\label{eq:eps-coefficients-1}
\end{equation}
\end{lem}

\begin{figure}
\caption{Approximation of $c_{i}(k)$ and $d_{i}(k)$ for $k=3,\,4$.}
\label{fig:approximation} \centering \subfloat[Approximation of $c_{i}(k)$.]{\includegraphics[scale=0.37]{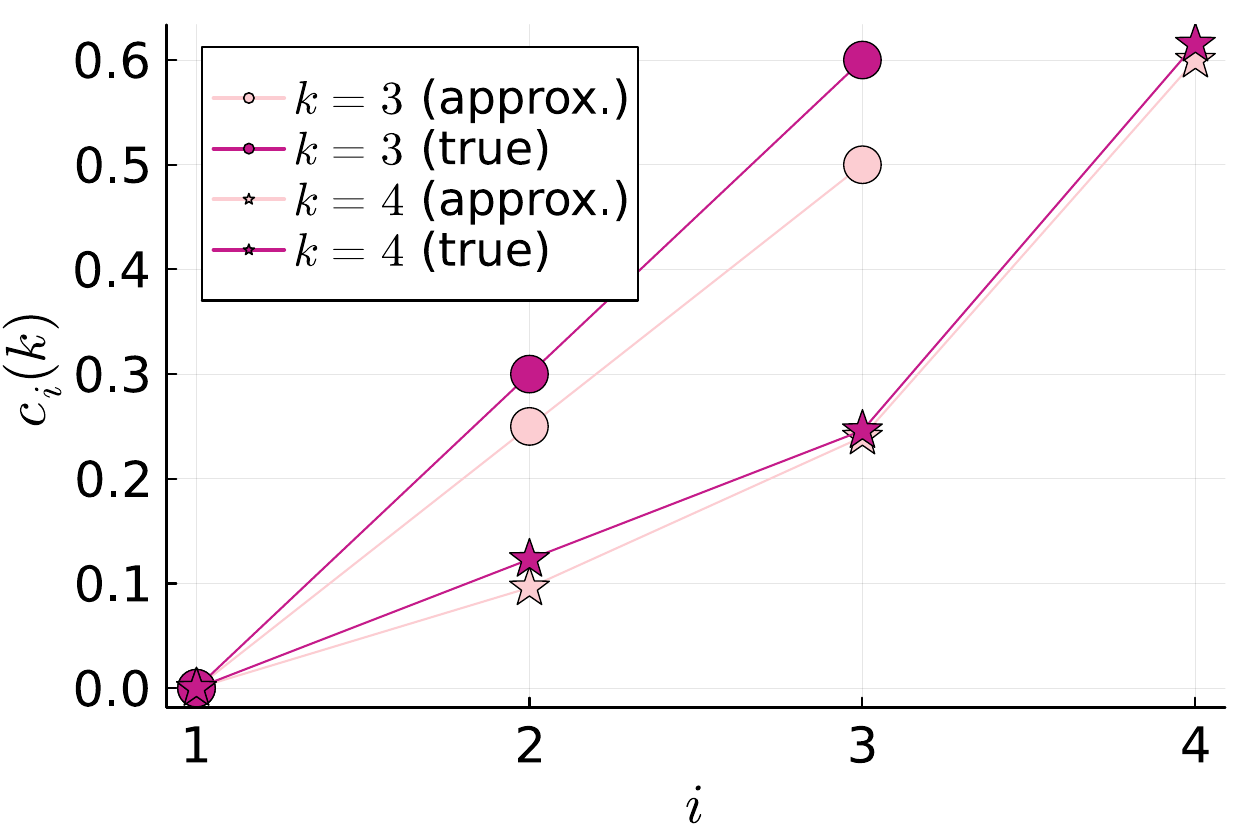}

}\hfill{}\subfloat[Approximation of $d_{i}(k)$.]{\includegraphics[scale=0.37]{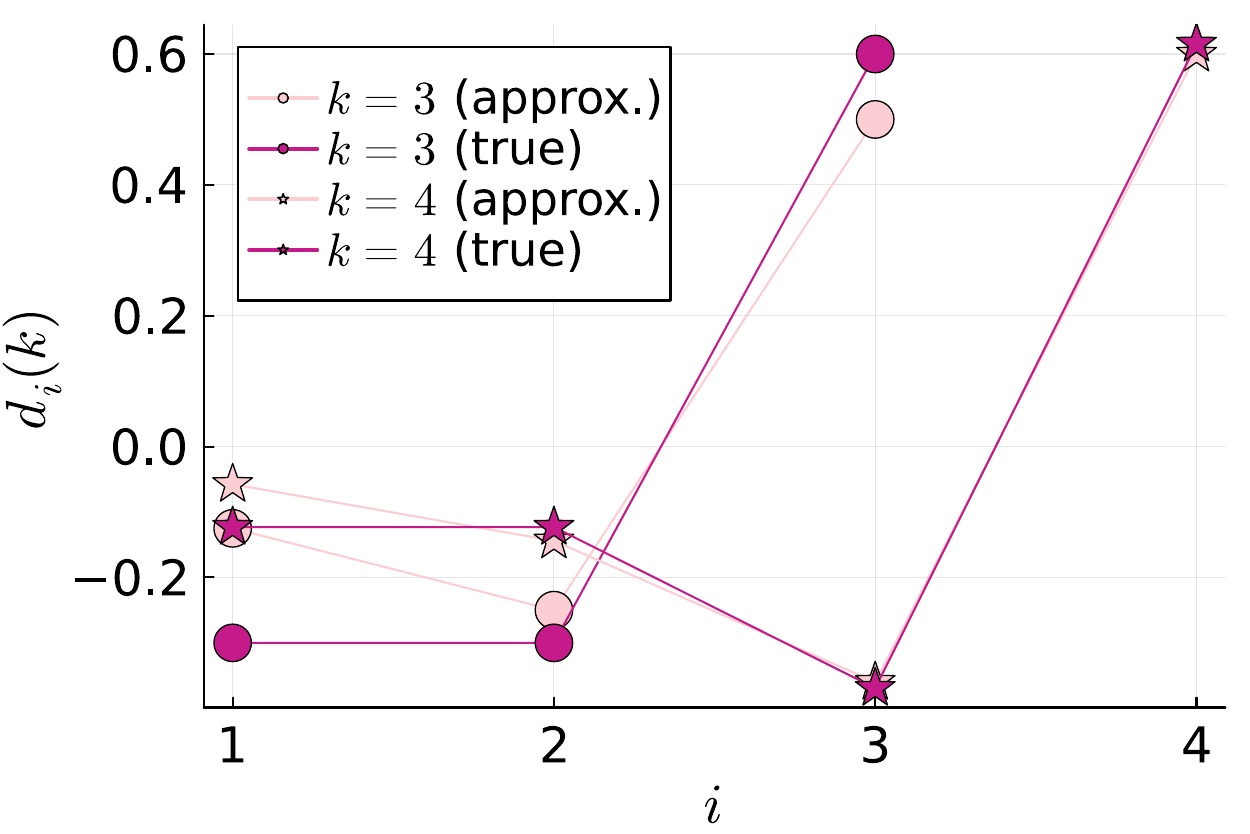}

}\justifying

\noindent{\scriptsize\textit{Notes: }}{\scriptsize The plots show
the approximation made in Lemma \ref{lem:asymptotic-coeffs} where
we replace $K_{k}$ with its limit $K_{\infty}$ for all $k$ to obtain
a simplified formula. From $k=4$ onwards, the approximation performs
very well as indicated by the starred series. We also see that in
absolute value, coefficients increase as $i$ approaches $k$. Moreover,
a general pattern is that the closer $i$ is to $k$, the larger in
abs. value both $c_{i}(k)$ and $d_{i}(k)$, i.e. more recent coefficients
are more important than more distant ones.}{\scriptsize\par}

\noindent{}
\end{figure}

The approximation becomes accurate for $k\geq3$ as shown in Figure
\ref{fig:approximation}, so that the simplified formula given in
Lemma \ref{lem:asymptotic-coeffs} should be preferred almost always.

We use (\ref{eq:pow1}) from Appendix \ref{app:power-and-size} to
derive power for $k=2$: 
\[
\theta=\Prob\left\{ K_{2}v_{2}<0\mid\eta_{2}<0\right\} =\mathbb{P}\left\{ -\varepsilon_{1}+\varepsilon_{2}<-\eta_{2}\mid\eta_{2}<0\right\} .
\]
Assuming homoskedasticity of the shocks $\big(\eta_{k}\distrib N\left(0,\sigma_{\eta}^{2}\right)$
and $\varepsilon_{k}\distrib N\left(0,\sigma_{\varepsilon}^{2}\right)\big)$,
$\frac{-\varepsilon_{1}+\varepsilon_{2}}{\sigma_{\eta}}$ has mean
0 and $\text{var}\left(\frac{-\varepsilon_{1}+\varepsilon_{2}}{\sigma_{\eta}}\right)=2\frac{\sigma_{\varepsilon}^{2}}{\sigma_{\eta}^{2}}=2/s$
so that 
\begin{equation}
\theta=\mathbb{P}\left\{ \frac{-\varepsilon_{1}+\varepsilon_{2}}{\sigma_{\eta}}\leq-\frac{\eta_{2}}{\sigma_{\eta}}\mid\eta_{2}<0\right\} \equiv F\left(-\frac{\eta_{2}}{\sigma_{\eta}}\right),\label{eq:powCDF}
\end{equation}
where $F\left(.\right)$ is the CDF of a normal random variable with
variance $2/s$. Equation (\ref{eq:powCDF}) shows that $\theta$
increases in $-\eta_{2}/\sigma_{\eta}$. Because $\text{var}\left(\frac{-\varepsilon_{1}+\varepsilon_{2}}{\sigma_{\eta}}\right)$
decreases with $\sigma_{\eta}$ (and thus $s$), and $F\left(-\frac{\eta_{2}}{\sigma_{\eta}}\right)$
(and thus $\theta$) decreases in the variance when $\eta_{2}<0$,
we have $\frac{\partial\theta}{\partial s}>0,$ i.e. power to detect
a standard deviation shock in $\eta_{k}$ increases in the signal-to-noise
ratio.

Using Lemma~\ref{lem:combinatorial-expansion} in Appendix \ref{app:power-and-size}
or Lemma \ref{lem:asymptotic-coeffs} above allows us to derive an
analytical expression for power $\theta$, which we use to construct
Figure \ref{fig:power}. The top panel of Figure~\ref{fig:power}
highlights how $\theta$ depends on the number of periods $k$ and
the signal-to-noise ratio. In particular, the top left panel displays
$\theta$ as a function of $\eta_{k}/\sigma_{\eta}$ for various values
of $k$ for a signal-to-noise ratio equal to its sample average of
$\bar{s}$=1.26 estimated from our data. Power increases with $k$
since as the number of time periods grows, more information is available
for accurate measurement. Power converges quickly: it grows only slightly
between $k=3$ and $k=\infty$. If the standardized shock to $\frac{\eta}{\sigma_{\eta}}=-1$,
then the probability of measuring a fall in $\beta$ is 0.73. If the
standardized value of the shock is $\frac{\eta}{\sigma_{\eta}}=-2$,
then the probability of measuring a fall in $\beta$ is $0.98$. The
right panel shows that power increases with $s$, holding constant
$k=3$. For the extreme value of $s=10$, we see that the filtering
algorithm is extremely sensitive.

Next we consider size, which is $\alpha=\mathbb{P}\left\{ K_{k}v_{k}<0\mid\eta_{k}\geq0\right\} $.
Inspection of this expression and noting that $\eta_{k}$ is symmetric
reveal that $\alpha=1-\theta$; we thus omit any derivations of size.
Similar to the graphs for power, we plot $\alpha$ as functions of
$\frac{\eta_{k}}{\sigma_{\eta}}$ in the bottom panels of Figure \ref{fig:power}
for various values of $k$ given a unit signal-to-noise ratio $s$
in panel (a) and for various $s$ given $k=4$ in panel (b). Panel
(a) shows that size decreases with the shock $\eta$ and converges
as fast as power does. Panel (b) shows that size decreases with $s$.

In summary, the Kalman filtering algorithm converges rapidly. As a
result, it possesses good power to reject the null after only a few
time periods.

\section{Conclusion}

\label{sec:conc}

We estimate time trends in dementia incidence in England over the
2002-2018 period and forecast future dementia incidence. While we
find some evidence that incidence falls in the early part of our sample
period, there is little evidence of a long-term trend in dementia
incidence. Our preferred model is a random walk with zero drift, meaning
that while dementia incidence changes over time, the optimal forecast
is its most recent value. We also produce confidence intervals for
these dementia forecasts and find that we cannot reject significant
increases or declines in incidence over the next decade.

Our filtering algorthim allows us to decompose the incidence process
variance into cross-sectional sampling variability and dementia incidence
variability at the population level. The advantage of our method is
that it is computationally very simple: the multi-state model is implemented
in the R msm package which produces the trend as parameter estimates
with associated variances. In the second step, we model the coefficients
using a Kalman filter for which we provide Julia code available in
the online supplement.

Data sets that contain internally consistent case definitions of dementia
over time have only recently become available, meaning that they are
relatively short in the time dimension. To assess the performance
of the Kalman filter using short samples, we perform a rigorous analysis
of its convergence properties. Its main element, the Kalman gain,
distinguishes between signal, i.e. dementia process shocks, and cross-sectional
sampling variability commonly referred to as measurement noise in
the time series literature. We find that the sequence of gains converges
so fast that only a small number of observations in the time domain
do not hinder its ability to discriminate between signal and noise.

We also study the implied power and size properties of the filter.
We define these probabilities with respect to the filter's ability
to correctly detect a negative shock (power) or falsely discover a
negative shock (size). We define our null hypothesis in terms of a
positive shock, because the major question of our study is to find
out whether the trend in dementia incidence has stopped declining.
Using the available evidence, we can reject that a positive shock
has occurred and conclude that we have a flat dementia trend.

We believe that the methods in this paper will prove useful in multiple
contexts. First, the methods in this paper can be used to forecast
dementia incidence using the newly available ELSA ``sister studies''
which are structured similarly to ELSA and contain many of the same
variables that can be used to construct measures of dementia. Second,
the problem of forecasting with samples that are of large size in
the cross-sectional dimension and short in the time dimension is very
common. Modeling and forecasting the dementia incidence process is
one example, but there are many others in the fields of epidemiology,
health, and the social sciences where the presented methodology can
be adopted.

Another interesting extension of our current approach would be to
add more health states. We model three health states and use age and
sex as explanatory variables and include a stochastic time trend.
While this approach may seem modest, it is computationally demanding.
Implementing models with more health states would be challenging,
since it would require joint modeling of many stochastic processes.
The methodology proposed by \citet{katzmaus} might be a useful way
to construct a high-dimensional model that is computationally feasible
and lets us estimate stochastic time trends. Building a richer model
would also allow us to consider prevalence of multiple health states,
i.e. the total number of people in each of those states. To achieve
this end, we could combine our transition model with the methodologies
in \citet{Bhadra01062011,Ionides03042023}, who model multiple states
and consider equilibrium impacts of transitions between those states.

\printbibliography

\newpage{} 
\begin{center}
{\large\textbf{SUPPLEMENTARY MATERIAL}}{\large\par}
\par\end{center}

\begin{center}
 
\par\end{center}
\begin{description}
\item [{Julia and R code}] We have a GitHub repository containing all
used code at \url{https://github.com/jsimons8/dementia-incidence}.
The msm package is provided by \citet{jackson2011multi} in R \citep{Rlang}.
We performed time series modeling and filter calculations in Julia,
documented in \citet{bezanson2017julia}.\footnote{\url{https://julialang.org/}} 
\end{description}

\appendix

\section{Appendix to Section \ref{sec:Data}: \nameref{sec:Data}}

\setcounter{page}{1}

\label{app:thedata}

We use the English Longitudinal Study of Ageing (ELSA), a panel survey
of individuals ages 50 or older and their spouses or partners. The
survey collects biennial measurements of both physical and cognitive
health, among many other variables. If a sample member was unable
to respond in person, a proxy respondent was asked questions in their
place. The sister studies are shown in \url{https://www.elsa-project.ac.uk/international-sister-studies}.

\subsection{Case definition of dementia}

\label{app:case-definition}

Dementia is defined using an algorithmic case definition based on
coexistence of cognitive impairment and functional impairment, or
a report of a doctor's diagnosis of dementia by the participant or
caregiver. The algorithmic case definition follows DSM-IV and other
clinical criteria in that it hinges on non-transient impairment in
two or more cognitive domains, resulting in functional impairment
\citep{ahmadi2017temporal}. With the application of stringent criteria
requiring severe cognitive and functional impairment, this definition
mainly captured moderate to severe dementia cases. Cognitive impairment
is defined as impairment in two or more domains of cognitive function
tests applied to ELSA participants (such as orientation to time, immediate
and delayed memory, verbal fluency, and numeracy function). Verbal
fluency is not assessed at wave 6 and we imputed it using information
at wave 5 and 7 \citet{authorreply}. We do not use numeracy function
as it is only measured at wave 1, 4 and later waves for those who
had not been asked before. Impairment in each domain of cognitive
function is defined as a score of 1.5 SD or more below the mean compared
with the population aged 50--80 years with the same level of education
\citep{chertkow2007mild}. To avoid the effect of transient cognitive
decline resulting from delirium or other mental disorders, participants
were considered to not have cognitive impairment if there score improved
by 1 SD or more on cognitive tests in the consecutive wave . 9.8\%
of the participants with cognitive impairment were identified as having
transient cognitive decline. For individuals unable to take the cognitive
function tests, the Informant Questionnaire on Cognitive Decline is
administered to a proxy informant (usually an immediate family member)
\citep{harrison2015informant}, and a score higher than 3.6 was used
to identify cognitive impairment. The threshold used has both high
specificity (0.84) and high sensitivity (0.82) \citep{harrison2015informant}.
Functional impairment is defined as an inability to carry out one
or more activities of daily living independently, which included getting
into or out of bed, walking across a room, bathing or showering, using
the toilet, dressing, cutting food, and eating. For more detailed
information about the definition of dementia, see \citet{ahmadi2017temporal}
and \citet{guzman2017forecasted}.

\subsection{Sample Statistics}

\label{app:sample_statistics}

The initial ELSA sample is representative of the English non-institutionalized
population aged 50 and over, but not the total population which includes
those in nursing homes and other institutions. It does, however, track
individuals if they enter a nursing home. Since our estimator is identified
using health transitions, the fact that the initial sample is not
institutionalized should not lead to bias. However, while ELSA attempts
to track sample members if they become institutionalized, ELSA sample
members face higher rates of attrition when entering nursing homes.
\citet{banks2025long} account for this problem by reweighting the
ELSA sample so that it matches the share of the age 65+ population
that is institutionalized in nursing and/or residential homes. They
find that most key results only change marginally when accounting
for this issue. Furthermore, given the ELSA design, there is likely
little trend in attrition rates, meaning that attrition is unlikely
to impact our key results.

Table \ref{tab:raw-data} describes our sample and shows the dementia
incidence rates per 1000 person-years across the nine waves. The number
of events (transitions from no dementia to dementia) is fairly stable
across the study period. The standardized rate, based on the England
and Wales 2011 Census population estimates, reflects the estimates
shown in Figure \ref{fig:incidence}.

\begin{table}
\centering \caption{Dementia incidence rates (per 1000 person-years).}
\label{tab:raw-data} \begin{tabular}{lcccccccc}
\toprule
 & 2002--04 & 2004--06 & 2006--08 & 2008--10 & 2010--12 & 2012--14 & 2014--16 & 2016--18 \\
\midrule
Events & 185 & 161 & 139 & 171 & 125 & 142 & 175 & 153 \\
Person-years & 21{,}844 & 17{,}136 & 17{,}336 & 19{,}928 & 18{,}538 & 19{,}071 & 17{,}024 & 14{,}820 \\
Crude rate & 8.5 & 9.4 & 8.0 & 8.6 & 6.7 & 7.4 & 10.3 & 10.3 \\
Standardised rate\textsuperscript{*} & 10.3 & 11.1 & 10.1 & 11.8 & 7.5 & 8.7 & 10.8 & 10.5 \\
\bottomrule
\end{tabular}
\justifying

\noindent{\scriptsize\emph{Notes}}{\scriptsize : \textsuperscript{{*}}Age-
and sex-standardized rates based on England and Wales 2011 Census
population estimates. Events are the number of transitions from no
dementia to dementia. Crude rate is defined as cases divided by total
person-years in each two-year follow-up.}{\scriptsize\par}

\end{table}

Table \ref{tab:misclassification} shows misclassification probabilities.
If a person is healthy, there is an almost one hundred percent chance
that this state is observed correctly. If a person has dementia, the
probability of observing this correctly is around 78\%.

\begin{table}
\centering \caption{Probability of correct classification.}
\label{tab:misclassification} \begin{tabular}{lcc}
\toprule
Probability & Estimate & 95\% Confidence interval \\
\midrule
$ \Prob \left \{S^{\ast} = 1 \mid S = 1 \right \}$ & 0.996 & (0.996, 0.997) \\
$ \Prob \left \{S^{\ast} = 2 \mid S = 2 \right \}$ & 0.779 & (0.753, 0.803) \\
\bottomrule
\end{tabular}

\noindent\justifying

\noindent{\scriptsize\textit{Notes}}{\scriptsize : Estimates from
misclassification model. State 1 is no dementia, State 2 is dementia.}{\scriptsize\par}

\end{table}

\section{Appendix to Section \ref{sec:main-model}: \nameref{sec:main-model}}

\subsection{Transition intensities in matrix notation}

\label{app:matrix-notation}

Following \citep[Ch. 4.5 (v)]{cox1965theory}, the transition intensity
matrix 
\begin{equation}
Q\left(\vect z_{i}\left(t\right),t\right)=\begin{pmatrix}-\left(q_{12}\left(\vect z_{i}\left(t\right),t\right)+q_{13}\left(\vect z_{i}\left(t\right),t\right)\right) & q_{12}\left(\vect z_{i}\left(t\right),t\right) & q_{13}\left(\vect z_{i}\left(t\right),t\right)\\
0 & -q_{23}\left(\vect z_{i}\left(t\right),t\right) & q_{23}\left(\vect z_{i}\left(t\right),t\right)\\
0 & 0 & 0
\end{pmatrix},\label{eq:trans-int}
\end{equation}
where rows of the intensity matrix sum to $0$. To convert the health
transition intensities in $Q\left(\vect z_{i}\left(t\right),t\right)$
to the probability of a health transition between waves we obtain
the solution to the Kolmogorov differential equation and normalize
the interval between successive waves for each individual to one.
The transition probability matrix between time $t$ and $t+w$ is
\begin{align}
P\left(w,\vect z_{i}\left(t\right),t\right) & =\exp\left(wQ\left(\vect z_{i}\left(t\right),t\right)\right)\nonumber \\
 & =\begin{pmatrix}1-p_{12}\left(\vect z_{i}\left(t\right),t\right)-p_{13}\left(\vect z_{i}\left(t\right),t\right) & p_{12}\left(\vect z_{i}\left(t\right),t\right) & p_{13}\left(\vect z_{i}\left(t\right),t\right)\\
0 & 1-p_{23}\left(\vect z_{i}\left(t\right),t\right) & p_{23}\left(\vect z_{i}\left(t\right),t\right)\\
0 & 0 & 1
\end{pmatrix}\label{eq:trans-prob}
\end{align}
where the time interval $w\equiv t_{ik-1}-t_{ik}$ is 2 years for
our study and we treat covariates $z_{i}\left(t\right)$ as constant
over this period.

\subsection{Details on functional form of time trend}

\label{app:functional-form}

Table~\ref{tab:def-f} collects all functional forms for each modeled
state transition.

\begin{table}[h]
\centering%
\begin{tabular}{|c|c|>{\centering}p{0.2\paperwidth}|}
\hline 
Origin $r$  & Destination $s$  & Functional form $f_{rs}\left(.\right)$\tabularnewline
\hline 
$1$  & $2$  & Linear in $\text{female}_{i}$, natural cubic splines in $\text{age}_{ik}$
and $\text{age}_{it}\times\text{female}_{i}$\tabularnewline
\hline 
$1$  & $3$  & Linear in $\text{female}_{i}$ and $\text{age}_{ik}$\tabularnewline
\hline 
$2$  & $3$  & Linear in $\text{female}_{i}$ and $\text{age}_{ik}$\tabularnewline
\hline 
\end{tabular}

\caption{Definitions of functional forms for covariates. We initially let $\text{age}_{ik}$
enter $f_{12}$ as a linear, additive term but found that a likelihood
ratio test provided evidence for a restricted cubic spline instead.}
\label{tab:def-f} 
\end{table}

Table~\ref{tab:def-f} summarizes all functional forms for time trends.
Therefore, we allow the most flexible functional form for the transition
of interest, while choosing more parsimonious functional forms for
the remainder.

\begin{table}[h]
\centering

\begin{tabular}{|c|c|>{\centering}p{0.25\paperwidth}|}
\hline 
Origin $r$  & Destination $s$  & Functional form of $\left\{ g_{rs,t_{k}}\right\} _{k=1}^{8}$\tabularnewline
\hline 
$1$  & $2$  & $\sum_{k=1}^{8}\beta_{k}\indic\left\{ t=t_{k}\right\} $\tabularnewline
\hline 
$1$  & $3$  & Linear in $t_{k}$\tabularnewline
\hline 
$2$  & $3$  & Linear in $t_{k}$\tabularnewline
\hline 
\end{tabular}

\caption{Definitions of functional forms for time trends. This part of the
model contributes $10$ parameters.}
\label{tab:def-f-1} 
\end{table}

\subsection{Derivation of likelihood of the misclassification model}

\label{app:likelihood-deriv}

This section explains the used indices and notation in more detail.
Recall that absent measurement error, the full likelihood is a product
of the factors defined in \eqref{eq:joint-dist} 
\begin{equation}
L=\prod_{i=1}^{I}\Prob\left\{ S_{t_{im_{i}}}\mid S_{t_{im_{i}-1}}\right\} \dots\Prob\left\{ S_{t_{i3}}\mid S_{t_{i2}}\right\} \Prob\left\{ S_{t_{i2}}\mid S_{t_{i1}}\right\} ,\label{eq:simple-loglik}
\end{equation}
where individual factors' realizations of $S_{t_{i2}},\dots,S_{t_{im_{i}}}$
depends on the observed states. Then, $\Prob\left\{ S_{t_{ik_{i}}}\mid S_{t_{ik_{i}-1}}\right\} $
is the $\left(S_{t_{ik_{i}}},S_{t_{ik_{i}-1}}\right)\text{th}$ entry
of the transition probability matrix (\ref{eq:trans-prob}) for $k_{i}=2,\dots,m_{i}$.

Using equations (\ref{eq:joint-dist}) and 
\begin{equation}
\Prob\left\{ S_{t_{ij}}^{*}|S_{t_{i1}},\dots,S_{t_{ij}}\right\} =\Prob\left\{ S_{t_{ij}}^{*}|S_{t_{ij}}\right\} ,\label{eq:markov-property-2}
\end{equation}
the joint distribution of observed health states in all periods ($S_{t_{i1}}^{\ast},\dots,S_{t_{im_{i}}}^{\ast}$),
conditional on the initial health state ($S_{t_{i1}}$) is

\begin{align}
 &  & \Prob\left\{ S_{t_{i2}}^{*},\dots,S_{t_{im_{i}}}^{*}\mid S_{t_{i1}}\right\} \nonumber \\
 & = & \Prob\left\{ S_{t_{i2}}^{*},\dots,S_{t_{im_{i}}}^{*}\mid S_{t_{i2}},\dots,S_{t_{im_{i}}}\right\} \cdot\Prob\left\{ S_{t_{i2}},\dots,S_{t_{im_{i}}}\mid S_{t_{i1}}\right\} \nonumber \\
 & = & \Prob\left\{ S_{t_{im_{i}}}^{*}\mid S_{t_{im_{i}}}\right\} \Prob\left\{ S_{t_{im_{i}}}\mid S_{t_{im_{i}-1}}\right\} \dots\nonumber \\
 &  & \Prob\left\{ S_{t_{i3}}^{*}\mid S_{t_{i3}}\right\} \Prob\left\{ S_{t_{i3}}\mid S_{t_{i2}}\right\} \Prob\left\{ S_{t_{i2}}^{\ast}\mid S_{t_{i2}}\right\} \Prob\left\{ S_{t_{i2}}\mid S_{t_{i1}}\right\} \label{eq:joint-dist-w-mis}
\end{align}
where we have applied the Markov property as in (\ref{eq:joint-dist})
and the corresponding property of conditional independence of observed
states given the true states as in \eqref{eq:markov-property-2}.
The individual likelihood contribution is obtained from summing each
individual's joint probability over all possible permutations according
to 
\begin{align}
L_{i} & =\sum_{S_{t_{i1}}\in S}\Prob\left\{ S_{t_{i1}}^{\ast}\mid S_{t_{i1}}\right\} \Prob\left\{ S_{t_{i1}}\right\} \sum_{S_{t_{i2}}\in S}\Prob\left\{ S_{t_{i2}}^{*}\mid S_{t_{i2}}\right\} \Prob\left\{ S_{t_{i2}}\mid S_{t_{i1}}\right\} \dots\label{eq:indiv-loglik}\\
 & \dots\nonumber \\
 & \sum_{S_{t_{im_{i}}}\in S}\Prob\left\{ S_{t_{im_{i}}}^{*}\mid S_{t_{im_{i}}}\right\} \Prob\left\{ S_{t_{im_{i}}}\mid S_{t_{im_{i}-1}}\right\} \nonumber 
\end{align}

\subsection{Derivation of constrained Kalman filter}

\label{app:Kalman}

This appendix derives the constrained version of the Kalman filter
that we use to estimate the parameters of the dementia process $\{\beta_{k}\}_{k=1}^{T}$
by showing the first two steps of the filter's recursions. Our filter
differs from standard Kalman filter because the estimated cross-sectional
variance that creates observational noise variance with variance $\Sigma_{TT}$
with diagonal elements $\{\sigma_{kk}^{2}\}_{k=1}^{T}$ is estimated
in a first step described in Section \ref{sec:msm}. Normally $\{\sigma_{kk}^{2}\}_{t=1}^{T}$
would be estimated as a parameter of the likelihood obtained from
running the Kalman filter. However, we are able to estimate it in
a first stage detailed in Section \ref{sec:msm}. Given this information,
we estimate $\{\beta_{k|k}\}_{k=1}^{T}$, its posterior variance $\{P_{k|k}\}_{k=1}^{T}$,
as well as the variance of the structural dementia shocks, $\sigma_{\eta}^{2}$.
Recall that the estimated series $\hat{\beta}_{k}$ differs from $\beta_{k}$
via equation \eqref{eq:msrmt-model-scalar}. Therefore, our version
of the Kalman filter has two inputs, $\hat{\beta}_{k}$ and $\sigma_{kk}^{2}$,
instead of just one.

We initialize the filter with $\beta_{0|0}=0$, which has posterior
variance $P_{0|0}=\infty$, which corresponds to a diffuse prior distribution
of $\beta_{0|0}$. Specifically, we construct estimates for the prior
distribution at $k=1$: 
\begin{align}
\beta_{1|0} & =\beta_{0|0}\label{eq:prior-state}\\
P_{1|0} & =P_{0|0}+\sigma_{\eta}^{2}.\label{eq:prior-covariance}
\end{align}
Given $\hat{\beta}_{k}$, we next find the forecast error $v_{1}=\hat{\beta}_{1}-\beta_{1|0}$
and find the constrained variance of the forecast via 
\begin{equation}
F_{1}=P_{1|0}+\hat{\sigma}_{11}.\label{eq:mse-covariance}
\end{equation}
The estimated $\hat{\sigma}_{11}$ enters \eqref{eq:mse-covariance}
which is in contrast to the usual approach that treats $\sigma_{11}$
as a parameter to be estimated. The gain 
\begin{equation}
K_{1}=\frac{P_{1|0}}{F_{1}}=\frac{P_{0|0}+\sigma_{\eta}^{2}}{P_{0|0}+\sigma_{\eta}^{2}+\hat{\sigma}_{11}^{2}}=1\label{eq:kalgain1}
\end{equation}
is obtained in the usual way as the ratio of the prior variance and
constrained forecast variance so that we update the estimated process
via 
\begin{align}
\beta_{1|1} & =\beta_{1|0}+K_{1}v_{1}\label{eq:posterior-state}\\
P_{1|1} & =\left(1-K_{1}\right)P_{1|0}.\label{eq:posterior-covariance}
\end{align}
In summary, starting at infinite prior variance $P_{0|0}$, we obtain
$K_{1}=1$, which per (\ref{eq:posterior-covariance}) results in
zero posterior variance of $\beta_{1|1}$, defined as $P_{1|1}$.
Subsequently, the prior variance rises from zero to $\sigma_{\eta}^{2}$
per 
\begin{align}
P_{2|1} & =P_{1|1}+\sigma_{\eta}^{2},
\end{align}
while the constrained variance of the forecast is $F_{2}=\sigma_{\eta}^{2}+\hat{\sigma}_{22}^{2}$,
so that the first non-trivial Kalman gain (in the second observation)
becomes 
\begin{equation}
K_{2}=\frac{\sigma_{\eta}^{2}}{\sigma_{\eta}^{2}+\hat{\sigma}_{22}^{2}},\label{eq:signal-to-total-var}
\end{equation}
i.e. the signal variance divided by the sum of the signal and measurement
error variances.

\section{Appendix to Section \ref{sec:results}: \nameref{sec:results}}

\subsection{Additional results on prevalence}

Dementia prevalence (the probability that an individual has dementia)
depends not only on dementia incidence but also on mortality incidence.
To estimate time trends in mortality incidence, both from no-dementia
and dementia, we used linear time trends in addition to functions
of age and gender. The estimated slope coefficient on the transition
from no-dementia to death is $0.950$ with a $90\%$ confidence interval
of $(0.937,0.961)$. Thus, holding age and sex constant, the incidence
of death from no dementia declines $5\%$ per two-year wave. The estimated
slope coefficient on the transition from dementia to death is $1.024$
$(1.005,1.044)$, suggesting the incidence of death from dementia
rises $2\%$ each wave. Therefore, we measure a falling trend in mortality
among those without dementia and an increasing among those with. Both
of these trends point towards a declining dementia prevalence at each
age, since those with dementia are more likely to die and those without
dementia are less likely to die over time.

\subsection{Details on $t$-statistics for tests of trend}

\label{app:t-stats-trend}

We follow \citet{slopetest} and construct $t$-test statistics defined
as 
\begin{eqnarray}
t_{\nu} & = & T^{-\frac{1}{2}}\hat{\sigma}^{-1}\left(\hat{\beta}_{T}-\hat{\beta}_{1}\right)\label{eq:zero-trend-test}\\
t_{\text{s,d}} & = & T^{-2}\hat{\sigma}^{-2}\sum_{k=1}^{T}\left(\beta_{k}-\beta_{1}-\frac{k}{T}\left(\hat{\beta}_{T}-\hat{\beta}_{1}\right)\right)^{2}\label{eq:zero-trend-test-detr}\\
t_{\text{s}} & = & T^{-2}\hat{\sigma}^{-2}\sum_{k=1}^{T}\left(\beta_{k}-\beta_{1}\right)^{2}\label{eq:stoch-trend-stats-notdetr}
\end{eqnarray}
where $\hat{\sigma}^{2}$ is the estimate of the long-run variance
of $\Delta\hat{\beta}_{k}-T^{-1}\sum_{k=2}^{T}\Delta\hat{\beta}_{k}$
for $k=2,\dots,T$. We obtain variance estimates assuming homoskedasticity
or allowing for heteroskedasticity and auto-correlation where $\hat{\sigma}=\{\hat{\sigma}_{\text{L}},\hat{\sigma}_{\text{HAC}}\}$.
Assuming homoskedasticity but allowing for auto-correlation, a variance
estimator is 
\begin{align}
\hat{\sigma}_{\text{L}}^{2}\left(m\right) & =\hat{\gamma}\left(0\right)+2\sum_{\tau=1}^{m}w\left(\tau,m\right)\hat{\gamma}\left(\tau\right),\label{eq:long-run-var}
\end{align}
for a kernel\footnote{We use the Bartlett kernel.} $w\left(\tau,m\right)$,
where $\hat{\gamma}\left(\tau\right)$ is the $\tau$th auto-covariance
of the above process.

Let $\omega_{ks}$ denote the entries of the covariance matrix of
the processes defined above. Robust covariance estimates then obtain
from 
\begin{equation}
\hat{\sigma}_{\text{HAC}}^{2}\left(m\right)=\frac{1}{T-1}\sum_{k=1}^{T}\omega_{kk}+\frac{1}{T-1}\sum_{\tau=1}^{m}\sum_{k=\tau+1}^{T}w\left(\tau,m\right)\omega_{k,k-\tau}.\label{eq:hac-var}
\end{equation}
To test whether $\nu=0$, which is constant under both null and alternative
hypotheses, we use $t_{\nu}$ and judge it against a standard normal
distribution. To test whether $\nu_{k}$ in equation \eqref{eq:drift-term}
is stochastic, we test $H_{0}:\sigma_{\xi}^{2}=0$ against $H_{1}:\sigma_{\xi}^{2}>0$
for which we have the test statistics $t_{\text{s,d}}$ and $t_{\text{s}}$
where s,d refers to the detrended version and s is the non-detrended
version.

We define the process and covariance matrix used in the t-statistics.
Let $\Psi$ be such that for $\beta_{k}$ stacked into a vector $\beta$,
the transformation $\beta\mapsto\Psi\beta$ produces a new process
with elements 
\[
\Delta\beta_{k}-\frac{1}{T-1}\sum_{k=2}^{T}\Delta\beta_{k},
\]
which demeans $\Delta\beta_{k}$ and is used as the numerator for
the $t$-statistics. Given that $\var(\hat{\beta})=\hat{\Sigma}_{TT}$,
the variance of $\Psi\hat{\beta}$ is $\Psi\hat{\Sigma}_{TT}\Psi^{\prime}$.
The variance estimates assuming homoskedasticity then follow from
using a basic variance estimator for $\Psi\hat{\beta}$. For HAC $p$-values,
we run the Newey-West Kernel-based procedure using both diagonal and
off-diagonal elements of $\Psi\hat{\Sigma}_{TT}\Psi^{\prime}$ in
lieu of covariance estimates.

The processes $B(r)$ and $W(r)$ are Brownian bridge and Wiener processes,
respectively. Then, the test statistics \eqref{eq:zero-trend-test}-\eqref{eq:stoch-trend-stats-notdetr}
follow the asymptotic distributions

\begin{eqnarray*}
t_{\text{s,d}}\overset{a}{\distrib} & \int_{0}^{1}B\left(r\right)^{2}\diff r\\
t_{\text{s}}\overset{a}{\distrib} & \int_{0}^{1}W\left(r\right)^{2}\diff r.
\end{eqnarray*}

\subsection{Sensitivity analyses}

\label{subsec:Sensitivity-analyses-t-tests}

\begin{figure}
\caption{Outcomes of $t$-tests for a zero trend.}
\label{fig:p-values} \subfloat[HAC-robust variance estimate.]{\includegraphics[scale=0.37]{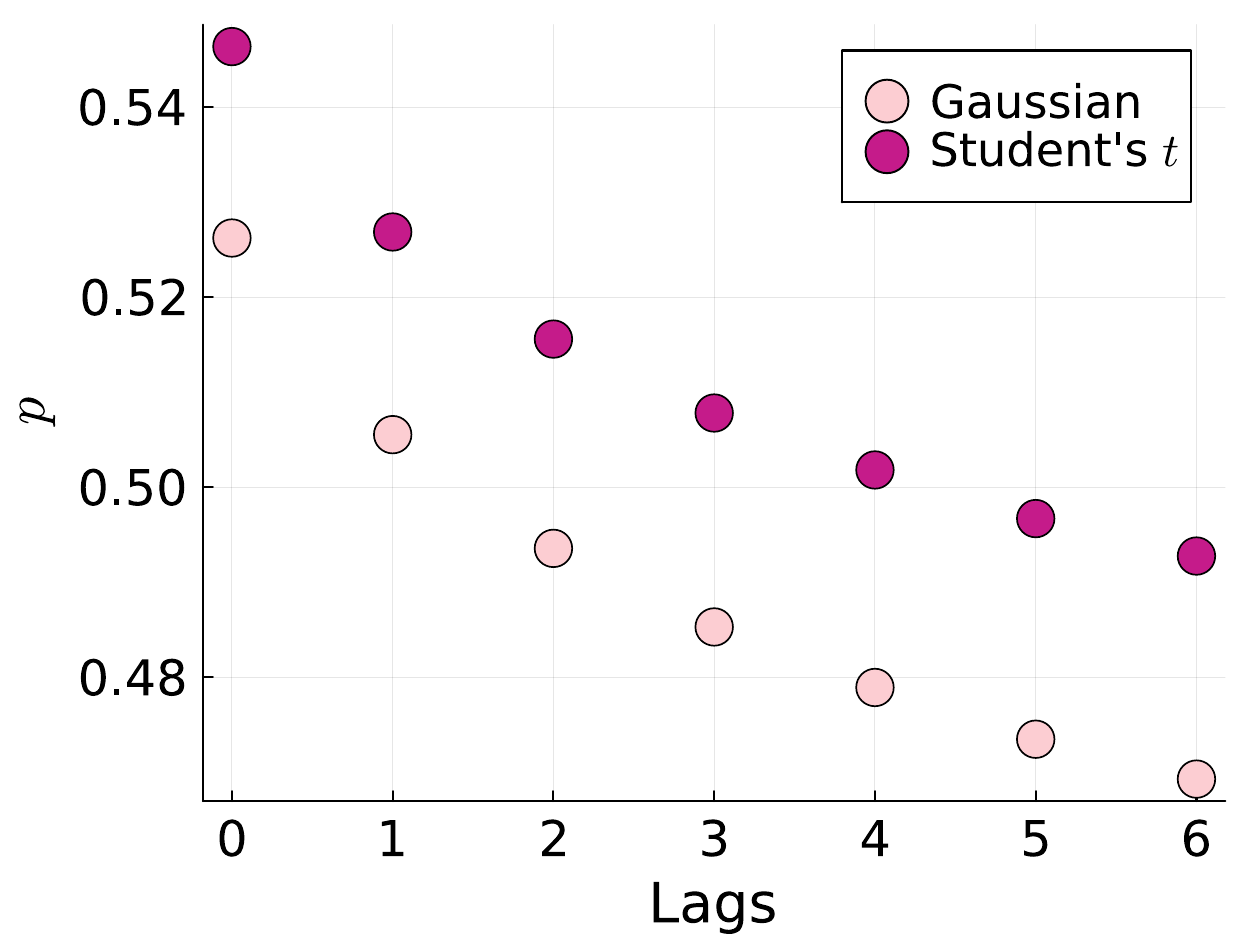}

}\hfill{}\subfloat[Homoskedastic variance estimate.]{\includegraphics[scale=0.37]{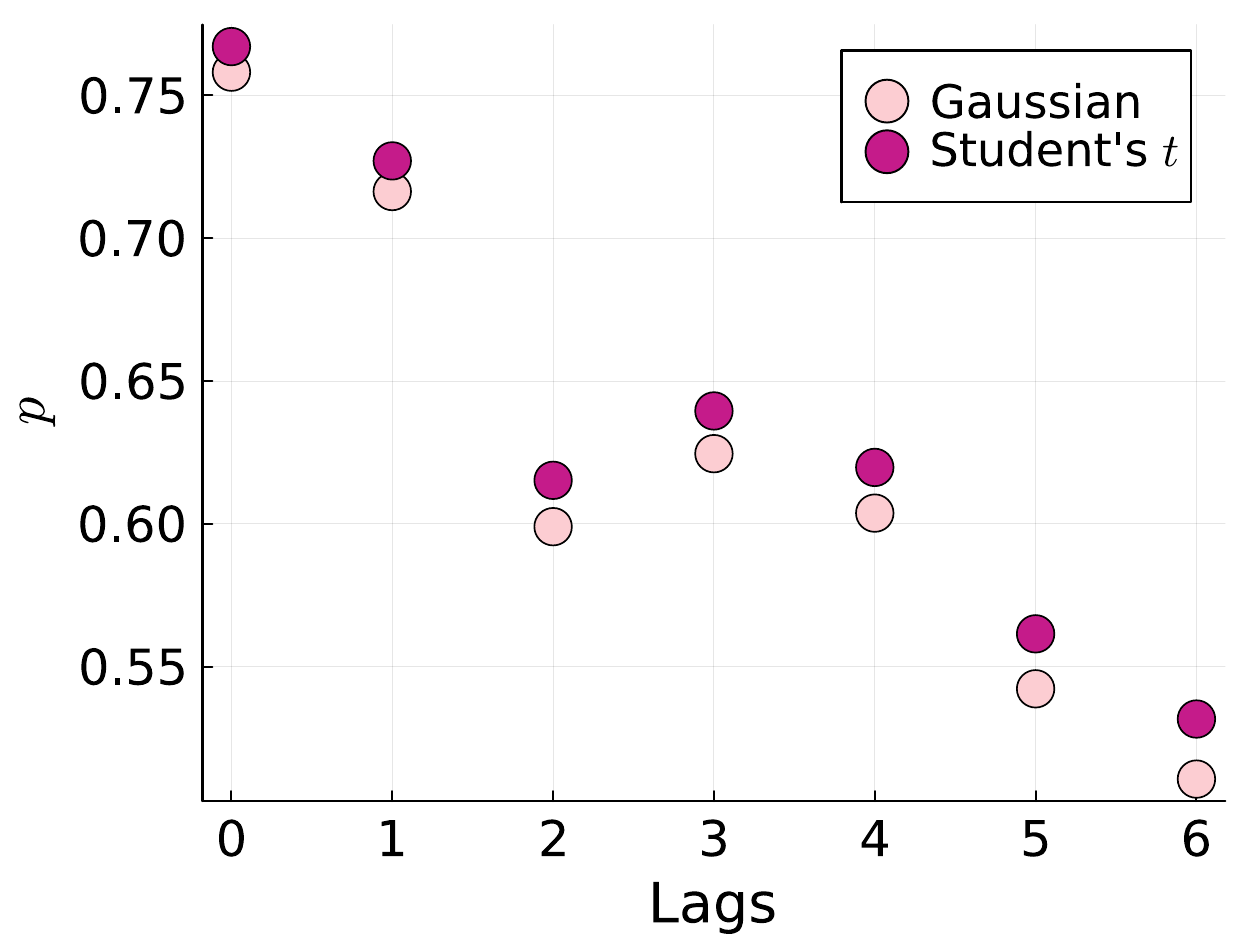}

}

\begin{singlespace}
\noindent\justifying

\noindent{\scriptsize\emph{Notes}}{\scriptsize : The figure shows
$p$-values for the non-parametric test that the trend is zero. It
corresponds to the top row of Table \ref{tab:critvals}. Assuming
it is deterministic, i.e. $\sigma_{\xi}=0$, we test $H_{0}\,:\nu=0$
against the alternative that $\nu\neq0$. Panel (a): $p$-values for
the $t$-test when the variance obtains from a HAC estimate using
all entries from the estimated covariance matrix $\hat{\Sigma}_{TT}$
from Section~\ref{sec:msm}. Panel (b): $p$-values for the $t$-test
when the variance obtains from a generic long-run variance estimate
assuming homoskedasticity.}{\scriptsize\par}
\end{singlespace}

\begin{singlespace}
 
\end{singlespace}
\end{figure}

Table \ref{tab:critvals} presents tests for a trend, where the test
statistics are defined in Section \ref{sec:ts-mod}. We test three
hypotheses in that table. The test statistic for the top row of Table
\ref{tab:critvals} is presented in equation (\ref{eq:zero-trend-test}).
This first test assumes that the trend is deterministic under both
null and alternative hypotheses, i.e. $\sigma_{\xi}=0$, and tests
$H_{0}\,:\nu=0$ against the alternative that $\nu\neq0$. The test
statistic for the middle row of the table is presented in equation
(\ref{eq:zero-trend-test-detr}). It is derived under the null hypothesis
that a deterministic slope is present and rejects against the alternative
that the slope is stochastic, formally $H_{0}\,:\nu_{k}=\nu\forall k,\,\sigma_{\xi}=0$
against $H_{1}\,:\sigma_{\xi}>0$ for any realization of the drift
parameter $\nu$ or process $\nu_{k}$. The test statistic for the
bottom row is presented in (\ref{eq:zero-trend-test-detr}). This
test is constructed for $H_{0}\,:\nu_{k}=0\ \forall\ k,\sigma_{\xi}=0$
against $H_{1}\,:\sigma_{\xi}>0$ for any $\nu$ under both null and
alternative hypotheses. This third test assumes that the drift is
zero and is thus a restricted version of the second test. If this
restriction holds true, this test has more power relative to the version
that allows for a free drift (process).

To calculate $\hat{\sigma}$, we use the default of a heteroskedasticity
and auto-correlation consistent (Newey-West) estimate, shown in equation
(\ref{eq:hac-var}) with 2 lags. When calculating critical values
for row 1, we assume normality. As an alternative, we also show estimates
that allow for auto-correlation but assume homoskedasticity in equation
(\ref{eq:long-run-var}) and also alternative numbers of lags, and
(for row 1) we compare the test statistic to the critical values implied
by Student's t distribution.

\begin{figure}
\caption{Outcomes of non-parametric tests of a stochastic trend. The dashed
line shows the 5\% line.}
\label{fig:de-and-not-de} \subfloat[HAC-robust covariance estimate.]{\includegraphics[scale=0.37]{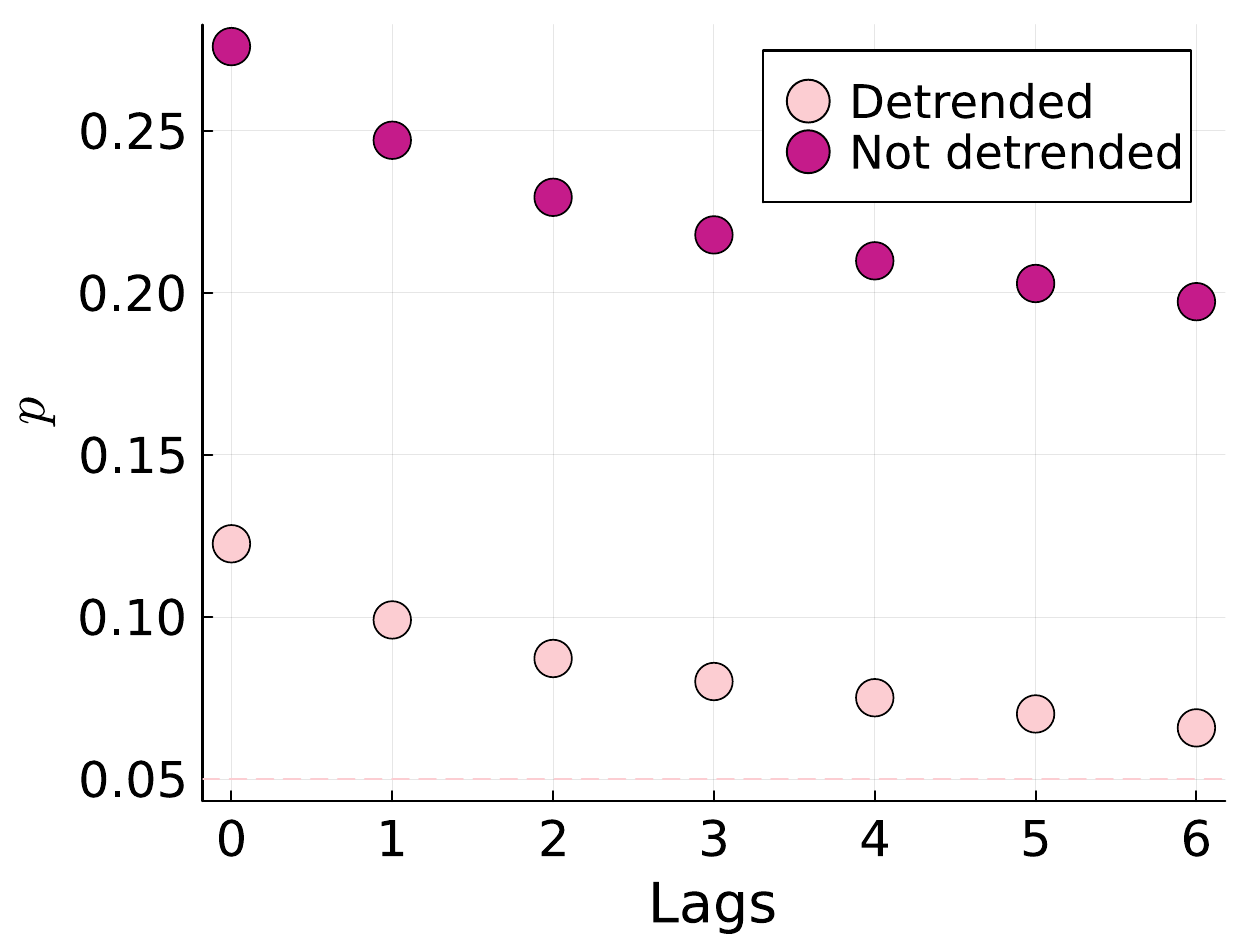}

}\hfill{}\subfloat[Homoskedastic covariance estimate.]{\includegraphics[scale=0.37]{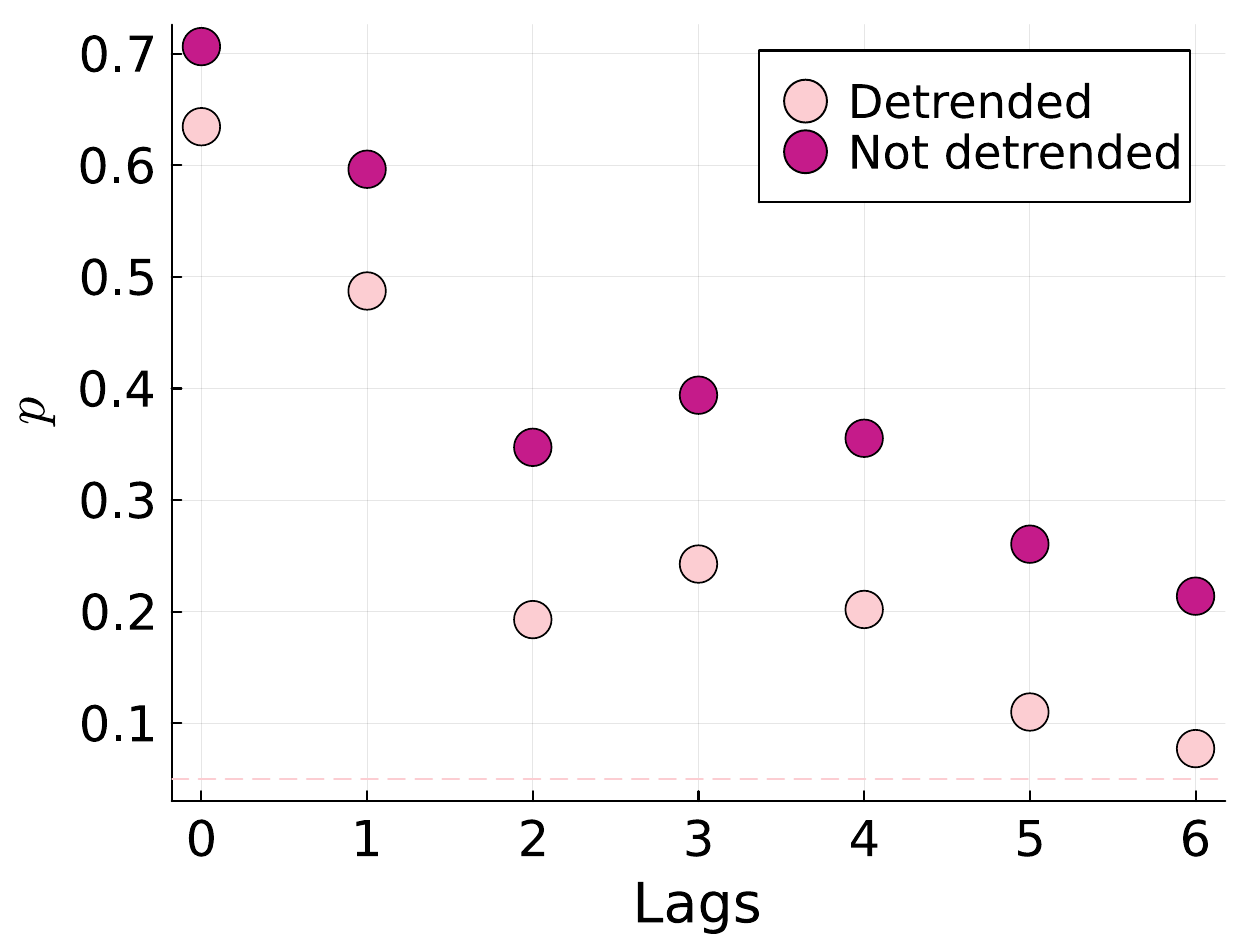}

}

\begin{singlespace}
{\scriptsize\textit{Notes}}{\scriptsize : The figure shows $p$-values
for the non-parametric tests of a stochastic trend. The detrended
versions in both panels correspond to rows tow and three in Table
\ref{tab:critvals}. Panel (a): $p$-values for the $t_{\text{s}}$-
and $t_{\text{s,d}}$ tests when the variance obtains from a HAC estimate
using all entries from the estimated covariance matrix $\hat{\Sigma}_{TT}$
from Section~\ref{sec:msm}. Panel (b): $p$-values for the $t_{\text{s}}$-
and $t_{\text{s,d}}$ tests when the variance obtains from a generic
long-run variance estimate assuming homoskedasticity. Based on the
$p$-values, we cannot reject that the trend is deterministic.}{\scriptsize\par}

\end{singlespace}
\end{figure}

Figure~\ref{fig:p-values} shows $p$-values corresponding to the
test in row 1 of Table~\ref{tab:critvals}, but varying the number
of lags, considering the additional assumption of homoskedasticity,
and comparing critical values to those implied by both a normal distribution
and a students' t distribution. It shows that the results are robust
to different assumptions on lag length, homoskedasticity, and normal
distributions. Figure \ref{fig:de-and-not-de} shows $p$-values corresponding
to the tests in rows 2 and 3 of Table~\ref{tab:critvals}, but varying
the number of lags. We also consider the additional assumption of
homoskedasticity and compare critical values to those implied by a
detrended or non-detrended test statistic. The detrended test is shown
on row 2 and the non-detrended test is shown on row 3 of Table~\ref{tab:critvals}
and the corresponding $p$-values are shown in the legends of Figure
\ref{fig:de-and-not-de}.

From the plot of p-values while varying lag lengths for both detrended
and non-detrended versions, we see that our results are not sensitive
to such variations of the testing procedure. At six lags, p-values
decline slightly but remain above five percent. Newey-West suggest
a lag length of $T^{\frac{1}{4}}$, which in our case rounds to two.

\section{Appendix to Section \ref{sec:conv-analysis}: \nameref{sec:conv-analysis}}

\label{app:appendix-for-convergence}The main maps of interest in
this Appendix are the posterior covariance and Kalman gain maps. We
compare our results with those of \citet{doi:10.1137/0331041} who
studied fixed point iterations in terms of the variance $P_{k\mid k}$
for the multi-variate case. We reparameterize the problem explicitly
in terms of the Kalman gain parameter, for two reasons. First, we
are able to measure convergence speed of the gain which helps distinguish
observational noise from signal to show that we are able to draw credible
inferences, even with relatively few years of data. Second, the Kalman
gain parameter, rather than the posterior variances alone, appear
in our analytical error probability expressions in Section \ref{sec:power-and-size}
so that we can study how fast these probabilities converge and build
an approximation formula that does not need reference to covariance
but makes use of gains directly. We believe that it is straightforward
to generalize the results below to the multi-variate case.

Our first result derives expressions for prior and posterior covariances
as functions of the history of Kalman gains and will let us substitute
out covariances. 
\begin{lem}
\label{lem:variance-recursions}For variances, we have the recursions 
\begin{enumerate}
\item \label{enu:posterior}$P_{k|k}=\sigma_{\eta}^{2}\sum_{d=1}^{t}\prod_{s=d}^{k}\left(1-K_{s}\right)$ 
\item \label{enu:prior}$P_{k+1|k}=\sigma_{\eta}^{2}\left(1+\sum_{s=2}^{k}\left(1-K_{s}\right)\right)$ 
\end{enumerate}
\end{lem}

\begin{proof}[Proof of Lemma~\ref{lem:variance-recursions}.]
We have the recursion for the prior variance 
\[
P_{k+1|k}=\left(1-K_{k}\right)P_{k|k-1}+\sigma_{\eta}^{2},
\]
which we can solve 
\[
P_{k+1|k}=\prod_{s=1}^{k}\left(1-K_{s}\right)P_{1|0}+\sigma_{\eta}^{2}\left(1+\sum_{s=2}^{k}\left(1-K_{s}\right)\right),
\]
where $\prod_{s=1}^{k}\left(1-K_{s}\right)P_{1|0}\goesto0$. For \ref{enu:prior},
we have the recursion for posterior variance 
\begin{equation}
P_{t|t}=\left(1-K_{k}\right)P_{k-1|k-1}+\left(1-K_{k}\right)\sigma_{\eta}^{2},
\end{equation}
which we can solve 
\begin{equation}
P_{k|k}=\prod_{s=1}^{k}\left(1-K_{s}\right)P_{0|0}+\sigma_{\eta}^{2}\sum_{d=1}^{k}\prod_{s=d}^{k}\left(1-K_{s}\right),
\end{equation}
where $\prod_{s=1}^{k}\left(1-K_{s}\right)P_{0|0}\goesto0$ by Assumption~\ref{assu:diffuse-prior}. 
\end{proof}
We continue our discussion with an extended version of Lemma~\ref{lem:gain-recursion-1},
which includes expressions for the normalized covariances at $k$,
$\nu_{k}\equiv\frac{P_{k|k}}{\sigma_{kk}^{2}}$. Define the relative
deviation of the measurement noise variances as $\iota_{k}\equiv1-\frac{\sigma_{kk}^{2}}{\sigma_{k+1k+1}^{2}}$.
For the case, where $\sigma_{kk}$ is time-invariant, $\iota_{k}=0$.
The following result summarizes the maps discussed in Section \ref{sec:conv-analysis}
and adds an expression for the posterior covariance recursions. 
\begin{lem}[Full version of Lemma~\ref{lem:gain-recursion-1}.]
\label{lem:gain-recursion}The following characterize the posterior
covariance and Kalman gain maps: 
\begin{enumerate}
\item \label{enu:post-cov-map}The posterior covariance map is given by
\[
\nu_{k+1}=\left(1-\iota_{k}\right)\frac{\nu_{k}+s_{k}}{\nu_{k}+s_{k}+1}.
\]
\item \label{enu:analy-fp}The fixed point of $\nu_{k}$ is 
\begin{equation}
\nu_{\infty}\left(s_{k}\right)=-\frac{s_{k}+\iota_{k}}{2}+\sqrt{s_{k}\left(1-\iota_{k}+\frac{1}{2}\iota_{k}+\frac{1}{4}s_{k}+\frac{1}{4}\frac{\iota_{k}}{s_{k}}\right)}.\label{eq:fixed-point}
\end{equation}
\item \textup{\label{enu:gain-map}The gain recursions are given by 
\begin{equation}
K_{k+1}=\frac{s_{k}\sum_{d=1}^{k}\prod_{i=d}^{k}\left(1-K_{i}\right)+s_{k}}{s_{k}\sum_{d=1}^{k}\prod_{i=d}^{k}\left(1-K_{i}\right)+s_{k}+1}.\label{eq:single-map-gain}
\end{equation}
} 
\end{enumerate}
\end{lem}

An immediate consequence of Lemma~\ref{lem:gain-recursion} is that
the covariance map is a contraction mapping as long as 
\[
\frac{\sigma_{kk}^{2}}{\sigma_{k+1k+1}^{2}}<\left(1+s_{k}+\nu_{k}\right)^{2}
\]
implied by the contraction condition $\abs{\frac{\partial\nu_{k+1}}{\partial\nu_{k}}}<1$
and assumed stipulated in Assumption \ref{assu:diffuse-prior}. The
fixed point $\nu_{\infty}$ is given by \eqref{eq:fixed-point}.

\citet{doi:10.1137/0331041} establishes that the recursions for $P_{k\mid k}$
are a contraction as well. The fact that the Kalman gain also has
that property is intuitive based on its relationship with $P_{k|k-1}$
evident in 
\begin{align*}
K_{k} & =\frac{P_{k|k-1}}{P_{k|k-1}+\hat{\sigma}_{kk}^{2}}.
\end{align*}

\begin{proof}[Proof of Lemma~\ref{lem:gain-recursion}]
Part \ref{enu:post-cov-map} is derived in \citet[Eq. (4)]{doi:10.1137/0331041},
which we re-derive here for the single-variable case. Consider the
posterior covariance 
\begin{align*}
P_{k+1|k+1} & =\left(1-K_{k+1}\right)P_{k+1|k}\\
 & =_{\left(1\right)}\left(1-K_{k+1}\right)\left(P_{k|k}+\sigma_{\eta}^{2}\right)\\
 & =_{\left(2\right)}\left(1-\frac{P_{k+1|k}}{F_{k+1}}\right)\left(P_{k|k}+\sigma_{\eta}^{2}\right)\\
 & =_{\left(3\right)}\left(1-\frac{P_{k|k}+\sigma_{\eta}^{2}}{P_{k|k}+\sigma_{\eta}^{2}+\sigma_{kk}^{2}}\right)\left(P_{k|k}+\sigma_{\eta}^{2}\right),
\end{align*}
where $=_{\left(1\right)}$ follows from the definition of prior covariance
$P_{t+1|t}$, $=_{\left(2\right)}$ follows from the definition of
the Kalman gain, and $=_{\left(3\right)}$ follows from repeated application
of the definition of prior covariance and MSE covariance. Simplifying
and dividing both sides by $\sigma_{k+1,k+1}^{2}$, we arrive at

\[
\nu_{k+1}=\frac{P_{k+1|k+1}}{\sigma_{k+1,k+1}}=\frac{\left(P_{k|k}+\sigma_{\eta}^{2}\right)}{P_{k|k}+\sigma_{\eta}^{2}+\sigma_{kk}^{2}}\cdot\frac{\sigma_{kk}^{2}}{\sigma_{k+1,k+1}},
\]
and thus 
\[
\nu_{k+1}=\left(1-\iota_{k}\right)\frac{\nu_{k}+s_{k}}{\nu_{k}+s_{k}+1},
\]
which is Part 1 of the lemma. This result is a generalization of the
results in \citet[Eq. (4)]{doi:10.1137/0331041}, who solved this
problem for the case of $\sigma_{kk}=1$. Setting $\nu_{k+1}\equiv\nu_{k}\equiv\nu_{\infty}$
delivers Part \ref{enu:analy-fp} of the lemma.

In Subsection~\ref{subsec:Derivation-of-constrained}, we went from
$k|k-1$ to $k|k$ to derive the basic filter. Here, we shall extend
these recursions all the way to $k+1|k+1$ to derive recursions from
time $k$ to $k+1$ for the sequence of Kalman gain parameters. Therefore,
we continue the recursions from (\ref{eq:prior-state}) to (\ref{eq:posterior-covariance}). 
\begin{enumerate}
\item If we come into the next period, we obtain as prior estimates 
\begin{align}
\beta_{k+1|k} & =\beta_{k|k}\nonumber \\
P_{k+1|k} & =P_{k|k}+\sigma_{\eta}^{2}\label{eq:prior-covariance-one-ahead}
\end{align}
\item Measurement becomes available $v_{k+1}=y_{k+1}-\beta_{k+1|k}$ and
we update the filter to obtain 
\begin{align}
F_{k+1} & =P_{k+1|k}+\sigma_{kk}^{2}\nonumber \\
K_{k+1} & =\frac{P_{k+1|k}}{F_{k+1}}\label{eq:gain-one-ahead}\\
\beta_{k+1|k+1} & =\beta_{k+1|k}+K_{k+1}v_{k+1}\nonumber \\
P_{k+1|k+1} & =\left(1-K_{k+1}\right)P_{k+1|k}.\nonumber 
\end{align}
\end{enumerate}
To establish part \ref{enu:gain-map}, we insert (\ref{eq:prior-covariance-one-ahead})
into (\ref{eq:gain-one-ahead}) to obtain $P_{k+1|k}=P_{k|k}+\sigma_{\eta}^{2}$.
We have 
\begin{align*}
K_{k+1} & =\frac{P_{k|k}+\sigma_{\eta}^{2}}{P_{k|k}+\sigma_{\eta}^{2}+\sigma_{kk}^{2}}\\
 & =_{\left(1\right)}\frac{\sigma_{\eta}^{2}\sum_{d=1}^{k}\prod_{s=d}^{k}\left(1-K_{s}\right)+\sigma_{\eta}^{2}}{\sigma_{\eta}^{2}\sum_{d=1}^{k}\prod_{s=d}^{k}\left(1-K_{s}\right)+\sigma_{\eta}^{2}+\sigma_{kk}^{2}}\\
 & =\frac{s_{k}\sigma_{kk}^{2}\sum_{d=1}^{k}\prod_{s=d}^{k}\left(1-K_{s}\right)+s_{k}\sigma_{kk}^{2}}{s_{k}\sigma_{kk}^{2}\sum_{d=1}^{k}\prod_{s=d}^{k}\left(1-K_{s}\right)+s_{k}\sigma_{kk}^{2}+\sigma_{kk}^{2}}\\
 & =\frac{s_{k}\sum_{d=1}^{k}\prod_{i=d}^{k}\left(1-K_{i}\right)+s_{k}}{s_{k}\sum_{d=1}^{k}\prod_{i=d}^{k}\left(1-K_{i}\right)+s_{k}+1},
\end{align*}
where $=_{\left(1\right)}$ follows from Lemma~\ref{lem:variance-recursions}.\ref{enu:posterior}. 
\end{proof}
\citet{doi:10.1137/0331041} generalizes Lemma \ref{lem:gain-recursion}.\ref{enu:post-cov-map}
to the multivariate case, assuming homoskedasticity $\sigma_{kk}=\sigma_{\varepsilon}$.
The expression \eqref{eq:single-map-gain} does not involve a normalization
by a potentially time-dependent variance as $\nu_{k}$ does above.

The following result is useful in a few contexts. 
\begin{lem}
\label{lem:gain-bound}We have $K_{1}=1$. For $k=1,\dots,T$, the
Kalman gain $K_{k+1}\in\left(0,1\right)$. 
\end{lem}

\begin{proof}[Proof of Lemma \ref{lem:gain-bound}]
We have from the definition of the Kalman gain 
\begin{align}
K_{k+1} & =\frac{F_{k+1}-\sigma_{k}^{2}}{F_{k+1}}\nonumber \\
 & =1-\frac{\sigma_{k}^{2}}{P_{k+1|k}+\sigma_{k}^{2}}\nonumber \\
 & =1-\frac{\sigma_{k}^{2}}{P_{k|k}+\sigma_{\eta}^{2}+\sigma_{kk}^{2}},\label{eq:recursion}
\end{align}
where the last line shows that $K_{k+1}\in\left(0,1\right)$ because
$P_{k|k}+\sigma_{\eta}^{2}+\sigma_{kk}^{2}>\sigma_{kk}^{2}$. 
\end{proof}
In the following, we shall express the Kalman gain sequence motivated
by (\ref{eq:single-map-gain}) as a linear map $f_{k+1}\left(x\right)=m_{k}x+b_{k}$
where $m_{k}$ and $b_{k}$ are contraction mappings. Define 
\begin{equation}
m_{k}\equiv\frac{s_{k}\sum_{d=1}^{k-1}\prod_{s=d}^{k-1}\left(1-K_{s}\right)}{s_{k}\sum_{d=1}^{k}\prod_{s=d}^{k}\left(1-K_{s}\right)+s_{k}+1}\label{eq:slope}
\end{equation}
and 
\begin{equation}
b_{k}\equiv\frac{s_{k}\left(1-K_{k}\left(1+\sum_{d=1}^{k}\prod_{s=d}^{k}\left(1-K_{s}\right)\right)\right)}{s_{k}\sum_{d=1}^{k}\prod_{s=d}^{k}\left(1-K_{s}\right)+s_{k}+1}\label{eq:intercept}
\end{equation}
Our final result establishes that the Kalman gain sequence is the
outcome of a nest of contraction mappings defined by the components
above. 
\begin{lem}
~ \label{lem:linear-map} 
\begin{enumerate}
\item The Kalman gain sequence can be expressed as $K_{1}=1$, $K_{2}=\frac{s_{k}}{s_{k}+1}$,
$K_{3}=m_{2}K_{2}+b_{2}$, $K_{4}=m_{3}K_{3}+b_{3}$, $K_{k+1}=f_{k+1}\left(K_{k}\right)$
for $k\geq2$. 
\item The slope and intercept satisfy $\abs{m_{k}}<1$ and $\abs{b_{k}}<1\,\forall\,k$. 
\item The map $f_{k+1}\left(x\right)$ defines a contraction at every $k=k^{\prime}$. 
\end{enumerate}
\end{lem}

\begin{proof}[Proof of Lemma \ref{lem:linear-map}]
For part 1, insert the expressions for $K_{1}$, $K_{2}$ etc. into
\eqref{eq:slope} and \eqref{eq:intercept} and the result follows
from (\ref{eq:single-map-gain}). For part 2, note that the denominator
is always greater than the numerator. For part 3, it follows from
part 2 that 
\[
\abs{f_{k+1}^{\prime}\left(K_{k}\right)}=\abs{m_{k}}<1\,\forall\,k
\]
so that `locking' the map at some $k=\tilde{k}$ lets us obtain $\tilde{f}\left(K_{k}\right)=m_{\tilde{k}}K_{k}+b_{\tilde{k}}$
which is a contraction map. We repeat the argument for $m_{k}$ which
for all $k$ has $\abs{m_{k}^{\prime}\left(K_{k}\right)}<1$ so that
the slope is itself a contraction. A repeated appeal to this argument
establishes that the intercept $b_{k}$ is also a contraction. Now,
let $\tilde{k}$ be such that $m_{\tilde{k}}$ and $b_{\tilde{k}}$
are very close to their fixed points. Then, iterations of $f_{k+1}\left(K_{k}\right)$
approach the fixed point $K_{\infty}$ such that $f_{\infty}\left(K_{\infty}\right)=K_{\infty}+\epsilon_{\tilde{k}}$
where $\epsilon_{\tilde{k}}$ is the error introduced by `locking'
the map rather than iterating $m_{k}$ and $b_{k}$ further. As $m_{k}\goesto m_{\infty}$
and likewise for $b_{\tilde{k}}$, $\epsilon_{\tilde{k}}\goesto0$. 
\end{proof}

\subsection{Graphical illustration}

We provide an illustration of our results. Panels (a) and (b) of Figure~\ref{fig:map-iterations-all}
show iterations of the Kalman gain maps where time stops at date $k=1,\dots,8$
while the fixed point for the covariance map corresponds to $T=\infty$.
Panels (c) and (d) show the distances $\left(\log x_{k}-\log x_{k-1}\right)^{2}$
between successive iterations $x_{k}$ of the maps in (a) and (b),
graphed for dates $k=2,\dots,8$. The $x$-axes in those panels display
the signal-to-noise ratio $s=\sigma_{\eta}^{2}/\sigma_{kk}^{2}$.\begin{adjustwidth}{-2cm}{-2cm}\begin{landscape}
\begin{figure}[h!]
\caption{Convergence of posterior $\nu\equiv\frac{P_{k\mid k}}{\sigma_{kk}}$
and Kalman gain $K_{k}$.}
\label{fig:map-iterations-all} \centering

\subfloat[\label{fig:cov_map_it}Posterior $\nu_{k}\equiv\frac{P_{k\mid k}}{\sigma_{kk}}$.]{\includegraphics[width=0.48\linewidth]{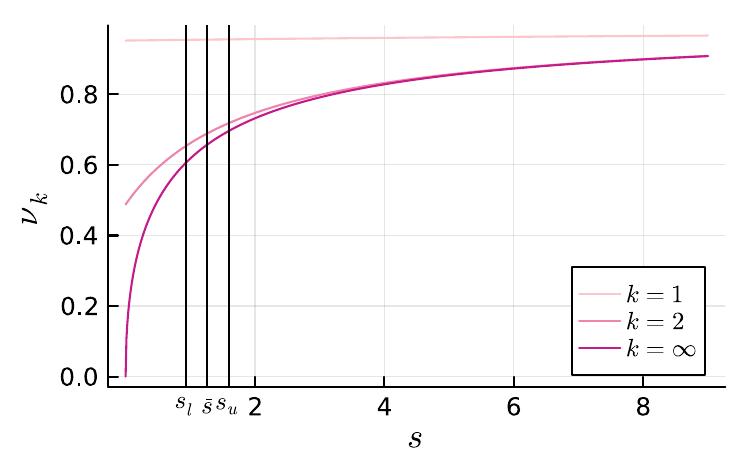}

}\hfill{}\subfloat[\label{fig:gain_map_it}Kalman gain.]{\includegraphics[width=0.48\linewidth]{gain_iterations}

}\vfill{}
 \subfloat[Distance between iterations of $\nu_{k}$.\label{fig:cov_map_dist}]{\includegraphics[width=0.48\linewidth]{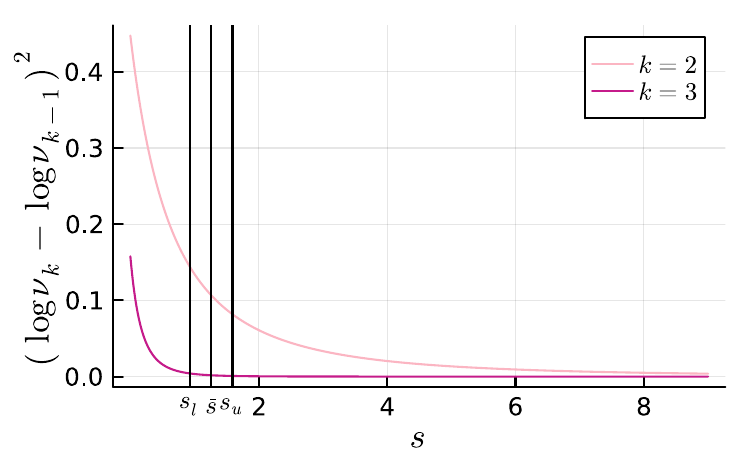}

}\hfill{}\subfloat[Distance between iterations of $K_{k}$.\label{fig:gain_map_it_dist}]{\includegraphics[width=0.48\linewidth]{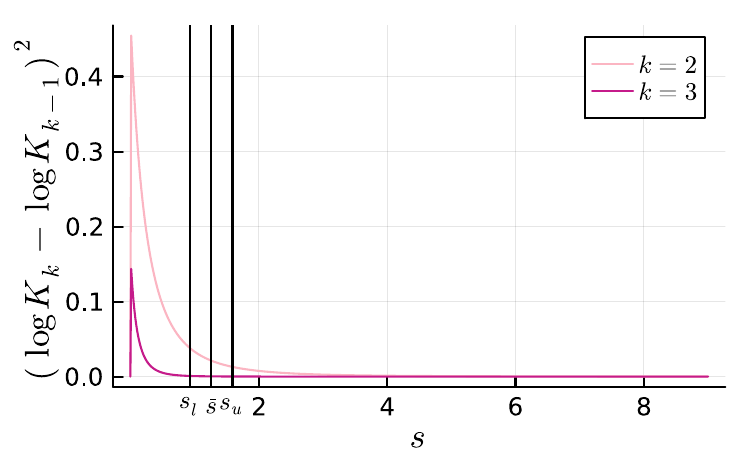}

}

\begin{singlespace}
\noindent\justifying{\scriptsize\textit{Notes: }}{\scriptsize The
plots show the normalized covariance $\varphi$ and gain maps $K$
after $k$ iterations (a,b) and the Riemann distances between successive
iterations of these maps (c,d). Top plots visualize convergence speed
to fixed points across varying signal-to-noise ratios $s$. Bottom
plots display shrinking of the distance between iterations in the
Riemann metric. The line $\varphi_{\infty}$ graphs the fixed point
of the covariance.}{\scriptsize\par}
\end{singlespace}
\end{figure}
\end{landscape}\end{adjustwidth}

What stands out from these graphs is the fact that even for modest
signal-to-noise ratios, the maps $\nu_{k}$ and $K_{k}$ collapse
very fast onto their steady state values even after only two iterations,
the value is not far from the fixed point. From panels (c) and (d),
we see the squared log differences go to zero very fast. The spikes
in the graphs correspond to the signal-to-noise ratio becoming non-zero
near the origin. For our values of $s_{k}$ graphed as $s_{k}$ along
with its confidence interval, squared log distances decay very quickly
to zero.

\section{Appendix to Section \ref{sec:power-and-size}: \nameref{sec:power-and-size}}

\label{app:power-and-size}
\begin{table}
\centering%
\begin{tabular}{cc}
\toprule 
Coefficient  & Expression\tabularnewline
\midrule 
$c_{1}\left(4\right)$  & $K_{4}-K_{4}K_{3}-K_{4}K_{2}-K_{4}K_{1}+K_{4}K_{3}K_{2}+K_{4}K_{3}K_{1}+K_{4}K_{2}K_{1}-K_{4}K_{3}K_{2}K_{1}$\tabularnewline
\midrule 
$c_{2}\left(4\right)$  & $K_{4}-K_{4}K_{3}-K_{4}K_{2}+K_{4}K_{3}K_{2}$\tabularnewline
\midrule 
$c_{3}\left(4\right)$  & $K_{4}-K_{4}K_{3}$\tabularnewline
\midrule 
$c_{4}\left(4\right)$  & $K_{4}$\tabularnewline
\midrule 
$d_{1}\left(4\right)$  & $-K_{4}K_{1}+K_{4}K_{3}K_{1}+K_{4}K_{2}K_{1}-K_{4}K_{3}K_{2}K_{1}$\tabularnewline
\midrule 
$d_{2}\left(4\right)$  & $-K_{4}K_{2}+K_{4}K_{3}K_{2}$\tabularnewline
\midrule 
$d_{3}\left(4\right)$  & $-K_{4}K_{3}$\tabularnewline
\midrule 
$d_{4}\left(4\right)$  & $K_{4}$\tabularnewline
\bottomrule
\end{tabular}\caption{Coefficients $c_{i}(4)$ and $d_{i}(4)$.}
\label{tab:terms-4} 
\end{table}
\begin{table}
\centering%
\begin{tabular}{>{\centering}p{1.5cm}>{\centering}m{10cm}}
\toprule 
Coefficient  & Expression\tabularnewline
\midrule 
$c_{1}\left(5\right)$  & $K_{5}-K_{5}K_{4}-K_{5}K_{3}-K_{5}K_{2}-K_{5}K_{1}+K_{5}K_{4}K_{3}+K_{5}K_{4}K_{2}+K_{5}K_{4}K_{1}+K_{5}K_{3}K_{2}+K_{5}K_{3}K_{1}+K_{5}K_{2}K_{1}-K_{5}K_{4}K_{3}K_{2}-K_{5}K_{4}K_{3}K_{1}-K_{5}K_{3}K_{2}K_{1}-K_{5}K_{4}K_{2}K_{1}+K_{5}K_{4}K_{3}K_{2}K_{1}$\tabularnewline
\midrule 
$c_{2}\left(5\right)$  & $K_{5}-K_{5}K_{4}-K_{5}K_{3}-K_{5}K_{2}+K_{5}K_{4}K_{3}+K_{5}K_{4}K_{2}+K_{5}K_{3}K_{2}-K_{5}K_{4}K_{3}K_{2}$\tabularnewline
\midrule 
$c_{3}\left(5\right)$  & $K_{5}-K_{5}K_{4}-K_{5}K_{3}+K_{5}K_{4}K_{3}$\tabularnewline
\midrule 
$c_{4}\left(5\right)$  & $K_{5}-K_{5}K_{4}$\tabularnewline
\midrule 
$c_{5}\left(5\right)$  & $K_{5}$\tabularnewline
\midrule 
$d_{1}\left(5\right)$  & $-K_{5}K_{1}+K_{5}K_{4}K_{1}+K_{5}K_{3}K_{1}+K_{5}K_{2}K_{1}-K_{5}K_{4}K_{2}K_{1}-K_{5}K_{4}K_{3}K_{1}-K_{5}K_{3}K_{2}K_{1}+K_{5}K_{4}K_{3}K_{2}K_{1}$\tabularnewline
\midrule 
$d_{2}\left(5\right)$  & $-K_{5}K_{2}+K_{5}K_{3}K_{2}+K_{5}K_{4}K_{2}-K_{5}K_{4}K_{3}K_{2}$\tabularnewline
\midrule 
$d_{3}\left(5\right)$  & $-K_{5}K_{3}+K_{5}K_{4}K_{3}$\tabularnewline
\midrule 
$d_{4}\left(5\right)$  & $-K_{5}K_{4}$\tabularnewline
\midrule 
$d_{5}\left(5\right)$  & $K_{5}$\tabularnewline
\bottomrule
\end{tabular}\caption{Coefficients $c_{i}(5)$ and $d_{i}(5)$.}
\label{tab:terms-5} 
\end{table}

We provide here additional results and some more explanation on technical
results as well as proofs for Lemmas\,\ref{lem:gen-formula-Kv} and
\ref{lem:combinatorial-forward}.
\begin{lem}
\label{lem:combinatorial-expansion}The Kalman update has the expansion
\begin{equation}
K_{k}v_{k}=\left(\beta_{0}-\beta_{1|0}\right)c_{1}(k)+\sum_{i=1}^{k}c_{i}(k)\eta_{i}+d_{i}(k)\varepsilon_{i},\label{eq:expansion}
\end{equation}
for coefficients $c_{i}(k),d_{i}(k),$ with $i\leq k$. 
\begin{equation}
c_{i}\left(k\right)=\sum_{\sigma\left(i\dots k\right)\backslash\sigma\left(i\dots s\right),\,s<k}\left(-1\right)^{l_{\sigma}+1}K_{i}\dots K_{k},\label{eq:eta-coefficients}
\end{equation}
\begin{equation}
d_{i}\left(k\right)=\sum_{\sigma\left(i\dots k\right)\backslash\sigma\left(n\dots s\right),\,i<n,\,s<k}\left(-1\right)^{l_{\sigma}+1}K_{i}\dots K_{k},\label{eq:eps-coefficients}
\end{equation}
where $K_{i}\dots K_{k}$ is the product $K_{i}K_{i+1}\cdot\cdot\cdot K_{k-1}K_{k}$
and $l_{\sigma}$ is the length of the combination, i.e., the number.

For coefficients $c_{i}(k)$, the expression $\sigma\left(i\dots k\right)\backslash\sigma\left(i\dots s\right),\,s<k$
denotes the sum over all combinations without replacement of all lengths
from $i$ to $k$ that include $k$ as the largest element. We denote
these combinations by listing all combinations of all lengths without
replacement and removing those whose largest element is $s<k$ in
them. For coefficients $d_{i}(k)$, $\sigma\left(i\dots k\right)\backslash\sigma\left(n\dots s\right),\,i<n,\,s<k$
denotes the sum over all combinations without replacement of all lengths
that have $i$ and $k$ as their smallest and largest elements, respectively.
We denote these combinations here by listing all combinations of all
lengths without replacement and removing those that do not have both
$i$ and $k$ in them.

A diffuse prior implies $K_{1}=1$ and thus $c_{1}\left(k\right)=0$
for $k\geq2$, which implies that the terms \textup{$\beta_{0}-\beta_{1|0}$}
and $\eta_{1}$ disappear beyond $k=2$.\textup{ } 
\end{lem}

Table~\ref{tab:terms-2-and-3} presents coefficients for $k=2$ and
3, while those for $k=4,5$ appear in Appendix \ref{app:power-and-size}
in Tables~\ref{tab:terms-4} and \ref{tab:terms-5}.\footnote{When we have $k\ge2$ observations, $c_{1}\left(k\right)=0$, which
implies that neither $\eta_{1}$ nor our choice of $\beta_{1|0}$
affect power, which corresponds to the reasoning in \citet[Example 3.2.1]{harvey1990forecasting}.}

\begin{table}[ht]
\centering %
\begin{tabular}{cc@{\hspace{3cm}}cc}
\toprule 
\multicolumn{2}{c}{\textbf{Order 2 Coefficients}} & \multicolumn{2}{c}{\textbf{Order 3 Coefficients}}\tabularnewline
\cmidrule(lr){1-2}\cmidrule(lr){3-4}
Coefficient & Expression & Coefficient & Expression\tabularnewline
\midrule 
$c_{1}(2)$ & $K_{2}-K_{2}K_{1}$ & $c_{1}(3)$ & $K_{3}-K_{3}K_{1}-K_{3}K_{2}+K_{3}K_{2}K_{1}$\tabularnewline
$c_{2}(2)$ & $K_{2}$ & $c_{2}(3)$ & $K_{3}-K_{3}K_{2}$\tabularnewline
 &  & $c_{3}(3)$ & $K_{3}$\tabularnewline
\midrule 
$d_{1}(2)$ & $-K_{2}K_{1}$ & $d_{1}(3)$ & $-K_{3}K_{1}+K_{3}K_{2}K_{1}$\tabularnewline
$d_{2}(2)$ & $K_{2}$ & $d_{2}(3)$ & $-K_{3}K_{2}$\tabularnewline
 &  & $d_{3}(3)$ & $K_{3}$\tabularnewline
\bottomrule
\end{tabular}\caption{Coefficients $c_{i}(2)$, $d_{i}(2)$, $c_{i}(3)$, and $d_{i}(3)$.}
\label{tab:terms-2-and-3}
\end{table}

To build intuition for the expansion in Lemma \ref{lem:combinatorial-expansion},
we derive coefficients for $k=2$. The full proof is in Appendix \ref{app:power-and-size}.
To this end, we first find a recursion that relates the forecast error
in period $k$, $v_{k}$, with that of period $k-1$. Substitute equation
(\ref{eq:msrmt-model-scalar}) into 
\begin{equation}
v_{k}=\hat{\beta}_{k}-\beta_{k|k-1}\label{eq:forecast-error}
\end{equation}
to obtain 
\begin{equation}
v_{k}=\beta_{k}+\varepsilon_{k}-\beta_{k|k-1}
\end{equation}
and make use of the fact that in the random walk model, this period's
prior is equal to last period's posterior $\beta_{k\mid k-1}=\beta_{k-1\mid k-1}$
so that the display above becomes 
\begin{equation}
v_{k}=\beta_{k}+\varepsilon_{k}-\beta_{k-1\mid k-1}.\label{eq:basic-recursion-almostthere}
\end{equation}
To eliminate $\beta_{k-1\mid k-1}$, we roll back (\ref{eq:state-update})
by one period to obtain 
\begin{equation}
\beta_{k-1\mid k-1}=\beta_{k-1\mid k-2}+K_{k-1}v_{k-1}.
\end{equation}
and substitute this expression into (\ref{eq:basic-recursion-almostthere})
to arrive at 
\begin{equation}
v_{k}=\beta_{k}+\varepsilon_{k}-\beta_{k-1\mid k-2}-K_{k-1}v_{k-1},\label{eq:basic-prediction-error}
\end{equation}
We multiply (\ref{eq:basic-prediction-error}) by $K_{k}$, set $k=2$,
note that $K_{1}=1$ (from the assumption of a diffuse prior), and
recognize that $\beta_{2}=\beta_{0}+\eta_{1}+\eta_{2}$ (from equation
(\ref{eq:state-1})), to obtain 
\begin{eqnarray}
K_{2}v_{2} & = & K_{2}\beta_{2}+K_{2}\varepsilon_{2}-K_{2}\beta_{1\mid0}-K_{2}K_{1}\left(\beta_{0}+\eta_{1}+\varepsilon_{1}-\beta_{1\mid0}\right)\nonumber \\
 & = & K_{2}\beta_{2}+K_{2}\varepsilon_{2}-K_{2}\left(\beta_{0}+\eta_{1}+\varepsilon_{1}\right)\nonumber \\
 & = & K_{2}\left(\eta_{2}-\varepsilon_{1}+\varepsilon_{2}\right).\label{eq:pow1}
\end{eqnarray}
The derivation in the preceding display gives coefficients shown in
the left panel of Table \ref{tab:terms-2-and-3}, since it shows that
$c_{2}(2)=d_{2}(2)=K_{2},c_{1}(2)=0,d_{1}(2)=-K_{2}$.

We generate a few terms using the next result and then prove that
the observed pattern is indeed correct. 
\begin{lem}
\label{lem:gen-formula-Kv}The expression $K_{k}v_{k}$ follows the
recursion given by 
\begin{align}
K_{k}v_{k} & =\sum_{i=2}^{k}\left(-1\right)^{k-i+2}K_{i}\dots K_{k}\left(\beta_{0}+\sum_{s=1}^{i}\eta_{s}\right)\label{eq:gen-formula-Kv}\\
 & +\sum_{i=2}^{k}\left(-1\right)^{k-i+2}K_{i}\dots K_{k}\varepsilon_{i}\nonumber \\
 & +\sum_{i=3}^{k}\left(-1\right)^{k-i+1}K_{i}\dots K_{k}\left(\beta_{1\mid0}+\sum_{s=1}^{i-2}K_{s}v_{s}\right)\nonumber \\
 & +\left(-1\right)^{k-1}K_{1}\dots K_{k}\left(\beta_{0}+\eta_{1}+\varepsilon_{1}\right)\nonumber 
\end{align}
\end{lem}

Using Lemma \ref{lem:gen-formula-Kv}, we see that the terms in Tables
\ref{tab:terms-2-and-3}, \ref{tab:terms-4}, and \ref{tab:terms-5}
display the following patterns, one for coefficients on $\varepsilon_{i}$
and one for $\eta_{i}$. 
\begin{proof}[Proof of Lemma~\ref{lem:gen-formula-Kv}.]
To arrive at a recursion that relates the forecast error in period
$k$, $v_{k}$, with that of period $k-1$, we roll back (\ref{eq:basic-prediction-error})
to obtain 
\[
v_{k-1}=\beta_{k-1}+\varepsilon_{k-1}-\beta_{k-2\mid k-3}-K_{k-2}v_{k-2},
\]
and substitute 
\begin{align*}
v_{k} & =\beta_{k}+\varepsilon_{k}-\beta_{k-1\mid k-2}-K_{k-1}\left(\beta_{k-1}+\varepsilon_{k-1}-\beta_{k-2\mid k-3}-K_{k-2}v_{k-2}\right)\\
 & =\beta_{k}+\varepsilon_{k}-\beta_{k-1\mid k-2}-K_{k-1}\beta_{k-1}-K_{k-1}\varepsilon_{k-1}+K_{k-1}\beta_{k-2\mid k-3}+K_{k-1}K_{k-2}v_{k-2}
\end{align*}
where we roll back again 
\[
v_{k-2}=\beta_{k-2}+\varepsilon_{k-2}-\beta_{k-3\mid k-4}-K_{k-3}v_{k-3}
\]
to obtain 
\begin{align*}
v_{k} & =\beta_{k}+\varepsilon_{k}-\beta_{k-1\mid k-2}-K_{k-1}\beta_{k-1}-K_{k-1}\varepsilon_{k-1}+K_{k-1}\beta_{k-2\mid k-3}+K_{k-1}K_{k-2}v_{k-2}\\
 & =\beta_{k}+\varepsilon_{k}-\beta_{k-1\mid k-2}-K_{k-1}\beta_{k-1}-K_{k-1}\varepsilon_{k-1}+K_{k-1}\beta_{k-2\mid k-3}\\
 & +K_{k-1}K_{k-2}\beta_{k-2}+K_{k-1}K_{k-2}\varepsilon_{k-2}-K_{k-1}K_{k-2}\beta_{k-3\mid k-4}-K_{k-1}K_{k-2}K_{k-3}v_{k-3},
\end{align*}
which we rearrange to obtain 
\begin{align*}
v_{k} & =\beta_{k}-K_{k-1}\beta_{k-1}+K_{k-1}K_{k-2}\beta_{k-2}\\
 & +\varepsilon_{k}-K_{k-1}\varepsilon_{k-1}+K_{k-1}K_{k-2}\varepsilon_{k-2}\\
 & -\beta_{k-1\mid k-2}+K_{k-1}\beta_{k-2\mid k-3}-K_{k-1}K_{k-2}\beta_{k-3\mid k-4}\\
 & -K_{k-1}K_{k-2}K_{k-3}v_{k-3}.
\end{align*}
The previous display contains four types of terms. We carry out $k-2$
such substitutions to arrive at 
\begin{align}
K_{k}v_{k} & =\sum_{i=2}^{k}\left(-1\right)^{k-i+1+1}K_{i}\dots K_{k}\beta_{i}\label{eq:gen-formula-almost}\\
 & +\sum_{i=2}^{k}\left(-1\right)^{k-i+1+1}K_{i}\dots K_{k}\varepsilon_{i}\nonumber \\
 & \underbrace{+\sum_{i=2}^{k}\left(-1\right)^{k-i+1}K_{i}\dots K_{k}\beta_{i-1\mid i-2}}_{\text{E}}\nonumber \\
 & \underbrace{+\left(-1\right)^{k-1}K_{1}\dots K_{k}\left(\beta_{1}+\varepsilon_{1}-\beta_{1|0}\right)}_{\text{F}}\nonumber 
\end{align}
We substitute into the above equation \eqref{eq:state-1}, i.e. $\beta_{1}=\beta_{0}+\eta_{1}$
and equation (\ref{eq:prior-state}) so that the last summand in equation
(\ref{eq:gen-formula-almost}) becomes 
\begin{equation}
\left(-1\right)^{k-1}K_{1}\dots K_{k}\left(\beta_{0}+\eta_{1}+\varepsilon_{1}-\beta_{1|0}\right)
\end{equation}

The number of substitutions is relevant because (\ref{eq:basic-prediction-error})
already has one prior term without substitution and $k-1$ after all
substitutions. Also, note that the prior term is negative when the
product has an odd number of terms contrary to the first two and the
last term is negative whenever $k$ is even.

We delay the start of the sum in summand E in equation (\ref{eq:gen-formula-almost})
to $i=3$ to uncouple the $\beta_{1\mid0}$ term and thus obtain 
\[
\text{E}=\sum_{i=3}^{k}\left(-1\right)^{k-i+1}K_{i}\dots K_{k}\beta_{i-1\mid i-2}+\left(-1\right)^{k-1}K_{2}\dots K_{k}\beta_{1\mid0}.
\]
To substitute out $\beta_{i-1\mid i-2}$, we insert \eqref{eq:Kal1}
into \eqref{eq:state-update} and employ backwards substitution to
obtain 
\begin{equation}
\beta_{i-1\mid i-2}=\beta_{1\mid0}+\sum_{s=1}^{i-2}K_{s}v_{s}.\label{eq:prior-backwards-sub}
\end{equation}
Following on, we insert \eqref{eq:prior-backwards-sub} into E and
obtain 
\[
C=\sum_{i=3}^{k}\left(-1\right)^{k-i+1}K_{i}\dots K_{k}\left(\beta_{1\mid0}+\sum_{s=1}^{i-2}K_{s}v_{s}\right)+\left(-1\right)^{k-1}K_{2}\dots K_{k}\beta_{1\mid0}.
\]
Observe that $K_{1}=1$ so that we can combine the terms $\left(-1\right)^{k-1}K_{2}\dots K_{k}\beta_{1\mid0}$
and 
\[
\left(-1\right)^{k-1}K_{2}\dots K_{k}\left(\beta_{0}+\eta_{1}+\varepsilon_{1}-\beta_{1|0}\right)
\]
so that E and F become 
\begin{equation}
\sum_{i=3}^{k}\left(-1\right)^{k-i+1}K_{i}\dots K_{k}\left(\beta_{1\mid0}+\sum_{s=1}^{i-2}K_{s}v_{s}\right)+\left(-1\right)^{k-1}K_{1}\dots K_{k}\left(\beta_{0}+\eta_{1}+\varepsilon_{1}\right).\label{eq:C-D-combo}
\end{equation}
Recall that $\beta_{i}=\beta_{0}+\sum_{s=1}^{i}\eta_{s}$. We now
insert \eqref{eq:C-D-combo} into \eqref{eq:gen-formula-almost} to
obtain the formula 
\begin{align*}
K_{k}v_{k} & =\sum_{i=2}^{k}\left(-1\right)^{k-i+2}K_{i}\dots K_{k}\left(\beta_{0}+\sum_{s=1}^{i}\eta_{s}\right)\\
 & +\sum_{i=2}^{k}\left(-1\right)^{k-i+2}K_{i}\dots K_{k}\varepsilon_{i}\\
 & +\sum_{i=3}^{k}\left(-1\right)^{k-i+1}K_{i}\dots K_{k}\left(\beta_{1\mid0}+\sum_{s=1}^{i-2}K_{s}v_{s}\right)\\
 & +\left(-1\right)^{k-1}K_{1}\dots K_{k}\left(\beta_{0}+\eta_{1}+\varepsilon_{1}\right),
\end{align*}
from which we can generate the first few $K_{1}v_{1}$, $K_{2}v_{2}$,
etc. 
\end{proof}
\begin{proof}[Proof of Lemma\,\ref{lem:combinatorial-expansion}.]
We use the general formula from Lemma\ \ref{lem:gen-formula-Kv}
to generate expansion coefficients shown in Tables\ \ref{tab:terms-2-and-3},
\ref{tab:terms-4}, and \ref{tab:terms-5} and then use induction
on $k$ to establish (\ref{eq:expansion}). Recall (\ref{eq:basic-prediction-error})
and multiply by $K_{k}$ to obtain 
\begin{equation}
K_{k}v_{k}=K_{k}\beta_{k}+K_{k}\varepsilon_{k}-K_{k}\beta_{k-1\mid k-2}-K_{k}K_{k-1}v_{k-1}.\label{eq:basic-recursion}
\end{equation}
We shall use induction on $k$ to establish (\ref{eq:expansion}).
We already checked the formula for $k=2,\dots,5$ above. We additionally
verify going from $k=2$ to $k=3$ via the recursion (\ref{eq:basic-recursion}).
Thus 
\begin{align*}
K_{3}v_{3} & =K_{3}\beta_{3}+K_{3}\varepsilon_{3}-K_{3}\beta_{3-1\mid3-2}-K_{3}K_{2}v_{2}.\\
 & =K_{3}\left(\beta_{0}+\eta_{1}+\eta_{2}+\eta_{3}\right)+K_{3}\varepsilon_{3}\\
 & -K_{3}\left(\beta_{1\mid0}+K_{1}\left(\beta_{0}+\eta_{1}+\varepsilon_{1}-\beta_{1\mid0}\right)\right)\\
 & -K_{3}\left(\eta_{1}\left(K_{2}-K_{2}K_{1}\right)+\eta_{2}K_{2}+\varepsilon_{1}\left(-K_{2}K_{1}\right)+\varepsilon_{2}K_{2}\right).
\end{align*}
We collect terms to obtain 
\begin{align*}
K_{3}v_{3} & =\varepsilon_{3}K_{3}-\varepsilon_{2}K_{2}K_{3}+\varepsilon_{1}\left(-K_{1}K_{3}+K_{1}K_{2}K_{3}\right)\\
 & +\beta_{0}\left(-K_{2}K_{3}+K_{3}K_{2}K_{1}-K_{1}K_{3}+K_{3}\right)\\
 & +\eta_{1}\left(-K_{2}K_{3}+K_{3}K_{2}K_{1}-K_{1}K_{3}+K_{3}\right)\\
 & +\eta_{2}\left(K_{3}-K_{2}K_{3}\right)\\
 & +\eta_{3}K_{3}\\
 & +\beta_{1\mid0}\left(K_{2}K_{3}-K_{3}K_{2}K_{1}+K_{1}K_{3}-K_{3}\right)
\end{align*}
where the terms involving $\beta_{0}$, $\beta_{1\mid0}$, and $\eta_{1}$
are zero when $K_{1}=1$. We note, that we recover the formula given
in Table\ \ref{tab:terms-2-and-3}.

The induction step is now to show that (\ref{eq:expansion}) is correct
for $k+1$ assuming it is correct for $k$. Therefore, we assume that
the formula is correct for the expansion of $K_{k}v_{k}$ and show
that we indeed obtain (\ref{eq:expansion}) for $k+1$ using (\ref{eq:basic-recursion})
rolled forward by one, that is 
\begin{equation}
K_{k+1}v_{k+1}=K_{k+1}\beta_{k+1}+K_{k+1}\varepsilon_{k+1}-K_{k+1}\beta_{k\mid k-1}-K_{k+1}K_{k}v_{k}.\label{eq:rolled-forw}
\end{equation}
We repeat \eqref{eq:rolled-forw} and collect terms

\begin{align*}
K_{k+1}v_{k+1} & =K_{k+1}\beta_{k+1}+K_{k+1}\varepsilon_{k+1}-K_{k+1}\beta_{k\mid k-1}-K_{k+1}K_{k}v_{k},\\
 & =\underbrace{K_{k+1}\left[\beta_{0}+\eta_{1}+\dots+\eta_{k+1}\right]+K_{k+1}\varepsilon_{k+1}}_{\text{A}}\\
 & -K_{k+1}\left[\beta_{1\mid0}+K_{1}v_{1}+\dots+K_{k-1}v_{k-1}\right]\\
 & -K_{k+1}K_{k}v_{k},
\end{align*}
where the expansion of $\beta_{k\mid k-1}$ follows from backwards
substitution to find that $\beta_{k\mid k-1}=\beta_{k-1\mid k-1}=\beta_{1|0}+K_{1}v_{1}+\dots+K_{k-1}v_{k-1}$.\footnote{In more general models, a transition matrix would appear as well.}
Introduce some shorthand for combinatorial coefficients and denote
by $c_{i}^{s}\equiv c_{i}\left(s\right)$ and $d_{i}^{s}\equiv d_{i}\left(s\right)$
defined in (\ref{eq:eta-coefficients}) and (\ref{eq:eps-coefficients})
respectively. Then, we write, using $v_{1}=\beta_{0}+\eta_{1}+\varepsilon_{1}-\beta_{1\mid0}$,

\begin{align*}
K_{k+1}v_{k+1} & =\text{A}-K_{k+1}\left[\beta_{1\mid0}+\underbrace{K_{1}\left(\beta_{0}+\eta_{1}+\varepsilon_{1}-\beta_{1\mid0}\right)}_{K_{1}v_{1}}\right]\\
 & -K_{k+1}\underbrace{\left[c_{1}^{2}\eta_{1}+c_{2}^{2}\eta_{2}+d_{1}^{2}\varepsilon_{1}+d_{2}^{2}\varepsilon_{2}\right]}_{K_{2}v_{2}}\\
 & -K_{k+1}\left[\dots+\dots\right]\\
 & -K_{k+1}\left[c_{1}^{k-1}\eta_{1}+c_{2}^{k-1}\eta_{2}+\dots+c_{k-1}^{k-1}\eta_{k-1}+d_{1}^{k-1}\varepsilon_{1}+c_{2}^{k-1}\varepsilon_{2}+\dots+d_{k-1}^{k-1}\varepsilon_{k-1}\right]\\
 & -K_{k+1}\left[c_{1}^{k}\eta_{1}+c_{2}^{k}\eta_{2}+\dots+c_{k-1}^{k}\eta_{k-1}+c_{k}^{k}\eta_{k}+d_{1}^{k}\varepsilon_{1}+c_{2}^{k}\varepsilon_{2}+\dots+d_{k}^{k}\varepsilon_{k}\right],
\end{align*}
where we have used Lemma \ref{lem:combinatorial-forward}.\ref{enu:diffuse-prior}
to establish $c_{1}^{k}=0$ for $k\geq2$ and dropped terms corresponding
to $\beta_{1\mid0}$ and $\beta_{0}$ but kept $c_{1}^{2}\eta_{1}$
for clarity in the second line of the preceding display. We collect
terms 
\begin{align*}
K_{k+1}v_{k+1} & =\varepsilon_{k+1}\left[K_{k+1}\right]+\varepsilon_{k}\left(-K_{k+1}\right)\left[d_{k}^{k}\right]+\varepsilon_{k-1}\left(-K_{k+1}\right)\left[d_{k-1}^{k}+d_{k-1}^{k-1}\right]\\
 & +\varepsilon_{k-2}\left(-K_{k+1}\right)\left[d_{k-2}^{k}+d_{k-2}^{k-1}+d_{k-2}^{k-2}\right]\\
 & +\dots\\
 & \underbrace{+\varepsilon_{i}\left(-K_{k+1}\right)\left[d_{i}^{k}+d_{i}^{k-1}+d_{i}^{k-2}+\dots+d_{i}^{i}\right]}_{\text{B}}\\
 & +\dots\\
 & +\varepsilon_{1}\left(-K_{k+1}\right)\left[d_{1}^{k}+d_{1}^{k-1}+\dots+d_{1}^{2}+\underbrace{d_{1}^{1}}_{K_{1}}\right]\\
 & +\eta_{k+1}K_{k+1}+\eta_{k}\left(-K_{k+1}\right)\left[-1+c_{k}^{k}\right]+\eta_{k-1}\left(-K_{k+1}\right)\left[-1+c_{k-1}^{k}+c_{k-1}^{k-1}\right]\\
 & +\eta_{k-2}\left(-K_{k+1}\right)\left[-1+c_{k-2}^{k}+c_{k-2}^{k-1}+c_{k-2}^{k-2}\right]\\
 & +\dots\\
 & \underbrace{+\eta_{i}\left(-K_{k+1}\right)\left[-1+c_{i}^{k}+c_{i}^{k-1}+c_{i}^{k-2}+\dots+c_{i}^{i}\right]}_{\text{C}}\\
 & +\dots\\
 & +\eta_{1}\left(-K_{k+1}\right)\left[-1+c_{1}^{k}+c_{1}^{k-1}+c_{1}^{k-2}+\dots+\underbrace{c_{1}^{1}}_{K_{1}}\right]\\
 & +\beta_{1\mid0}\underbrace{\left(K_{1}K_{k+1}-K_{k+1}\right)}_{=0}.
\end{align*}
Note that in the last line, we have made use of $K_{1}=1$ to drop
the $\beta_{1\mid0}$ term. Now, apply Lemma\ \ref{lem:combinatorial-forward}
to establish that 
\[
\left(-K_{k+1}\right)\left(d_{i}^{k}+\dots+d_{i}^{1}\right)=d_{i}^{k+1}
\]
and 
\[
\left(-K_{k+1}\right)\left(-1+c_{i}^{k}+c_{i}^{k-1}+c_{i}^{k-2}+\dots+c_{i}^{i}\right)=c_{i}^{k+1}.
\]
Therefore, we obtain $\text{B}=d_{i}^{k+1}\varepsilon_{i}$ and $C=c_{i}^{k+1}\eta_{i}$
for all $i=1,\dots,k+1$. Hence, $K_{k+1}v_{k+1}$ attains the same
expression as produced by (\ref{eq:expansion}) with $k+1$ substituting
$k$. Finally, the claim that all coefficients of $c_{1}\left(k\right)$
vanish for $k\geq2$ follows from Lemma \ref{lem:combinatorial-forward}.\ref{enu:simplification-diffuse-prior}. 
\end{proof}
The following lemma helps in the proof of Lemma\,\ref{lem:combinatorial-expansion}. 
\begin{lem}[Auxiliary formulae for combinatorial induction]
\label{lem:combinatorial-forward}Let the combinatorial coefficients
$c_{i}^{k}$ and $d_{i}^{k}$ be given as in (\ref{eq:eta-coefficients})
and (\ref{eq:eps-coefficients}) Then, 
\begin{enumerate}
\item \label{enu:forward-eta}Coefficients on $\eta_{i}$ satisfy 
\begin{equation}
\left(-K_{k+1}\right)\left(-1+c_{i}^{k}+c_{i}^{k-1}+c_{i}^{k-2}+\dots+c_{i}^{i}\right)=c_{i}^{k+1}\label{eq:forward-eta-rel}
\end{equation}
\item \label{enu:forward-eps}Coefficients on $\varepsilon_{i}$ satisfy
\begin{equation}
\left(-K_{k+1}\right)\left(d_{i}^{k}+\dots+d_{i}^{i}\right)=d_{i}^{k+1}.\label{eq:forward-eps-rel}
\end{equation}
\item \label{enu:number-terms}With each increasing order, the number of
terms in each coefficient doubles, i.e. $\abs{c_{i}^{k+1}}=2\abs{c_{i}^{k}}$
and likewise for $d_{i}^{k+1}$. 
\item \label{enu:simplification-diffuse-prior}For a diffuse prior $K_{1}=1$
implies $c_{1}^{k}=0$ for $k\geq2$. 
\end{enumerate}
\end{lem}

\begin{proof}[Proof of Lemma\,\ref{lem:combinatorial-forward}]
We start with proving part Lemma\ \ref{lem:combinatorial-forward}.\ref{enu:forward-eps}
because it is more straightforward and proceed with induction on $k$.
The subscript $i$ is not important and can be set to $1$ wlog. To
verify $k=2$, $d_{1}^{3}=K_{1}K_{2}K_{3}-K_{1}K_{3}$ whereas the
LHS fulfills $\left(-K_{3}\right)\left(d_{1}^{2}+d_{1}^{1}\right)=\left(-K_{3}\right)\left(-K_{1}K_{2}+K_{1}\right)$,
which checks out. Now, we assume the claim is true for $k$ and wish
to verify it for $k+1$. We have the induction assumption 
\[
\left(-K_{k}\right)\left(d_{i}^{k-1}+\dots+d_{i}^{i}\right)=d_{i}^{k}.
\]
To apply the induction assumption to (\ref{eq:forward-eps-rel}),
we multiply its LHS by $\left(-K_{k}\right)$ to obtain 
\begin{align}
 & \left(-K_{k+1}\right)\left(-K_{k}\right)\left(d_{i}^{k}+d_{i}^{k-1}+\dots+d_{i}^{i}\right)\nonumber \\
 & =\left(-K_{k+1}\right)\left(\left(-K_{k}\right)d_{i}^{k}+\underbrace{\left(-K_{k}\right)\left(d_{i}^{k-1}+\dots+d_{i}^{i}\right)}_{d_{i}^{k}\text{ per ind. ass.}}\right)\nonumber \\
 & =\left(-K_{k}\right)\left(\frac{K_{k+1}}{K_{k}}d_{i}^{k}-K_{k+1}d_{i}^{k}\right).\label{eq:one-forw-eps}
\end{align}
First, note that $\frac{K_{k+1}}{K_{k}}d_{i}^{k}$ means the following:
first, multiplication by $\frac{1}{K_{k}}$ makes all combinations
lose the factor $K_{k}$ so that we go from $1,\dots,k$ to $1,\dots,k-1$.
Then, multiplication by $K_{k+1}$ replaces these factors by $K_{k+1}$,
i.e. having combinations $1,\dots,k-1,k+1$ with a gap left at $k$.
No sign change is necessary because the combination length stays the
same after the swap. Now, observe that $-K_{k+1}d_{i}^{k}$ adds back
in the combinations involving $k$ so that we create those of the
form $1,\dots,k,k+1$, which we had just eliminated. Multiplication
in $-K_{k+1}d_{i}^{k}$ also increases the length of each combination
by one, so that each sign gets flipped. Hence, $\frac{K_{k+1}}{K_{k}}d_{i}^{k}-K_{k+1}d_{i}^{k}=d_{i}^{k+1}$
so that 
\[
\left(-K_{k+1}\right)\left(-K_{k}\right)\left(d_{i}^{k}+d_{i}^{k-1}+\dots+d_{i}^{i}\right)=\left(-K_{k}\right)d_{i}^{k+1}
\]
whence the claim follows. To see an example of how the transformation
$\frac{K_{k+1}}{K_{k}}d_{i}^{k}-K_{k+1}d_{i}^{k}$ adjusts the combinations,
let $k=3$ so that 
\begin{align*}
-K_{k+1}d_{1}^{k} & =-K_{4}d_{1}^{3}\\
 & =K_{4}K_{3}K_{1}-K_{4}K_{3}K_{2}K_{1}
\end{align*}
and $\frac{K_{4}}{K_{3}}d_{1}^{3}=K_{4}K_{1}-K_{4}K_{2}K_{1}$. So
$\frac{K_{4}}{K_{3}}d_{1}^{3}$ has appended $K_{4}$ everywhere and
removed $K_{3}$ everywhere while $-K_{4}d_{1}^{3}$ appends $K_{4}$
everywhere that has $K_{3}$ in it (as it has to be because all combinations
must involve $1$ and $3$) and flips the sign as those combinations
have increased in length by one.

Regarding, Lemma~\ref{lem:combinatorial-forward}.\ref{enu:forward-eta},
we use a similar argument. Set $k=2$ and note that $c_{1}^{3}=K_{3}-K_{3}K_{1}-K_{3}K_{2}+K_{1}K_{2}K_{3}$
and evaluate the LHS to obtain 
\begin{align*}
\left(-K_{3}\right)\left(-1+c_{1}^{2}+c_{1}^{1}\right) & =\left(-K_{3}\right)\left(-1+K_{2}-K_{2}K_{1}+K_{1}\right)
\end{align*}
which checks out. Now, assume that the claim is true for $k$, so
that the induction assumption is 
\[
\left(-K_{k}\right)\left(-1+c_{i}^{k-1}+\dots+c_{i}^{i}\right)=c_{i}^{k}.
\]
To apply the induction assumption to \ref{enu:forward-eta}, we multiply
the LHS of (\ref{eq:forward-eta-rel}) by $\left(-K_{k}\right)$ to
obtain 
\begin{align}
 & \left(-K_{k}\right)\left(-K_{k+1}\right)\left(-1+c_{i}^{k}+c_{i}^{k-1}+c_{i}^{k-2}+\dots+c_{i}^{i}\right)\nonumber \\
 & =\left(-K_{k+1}\right)\left(\left(-K_{k}\right)c_{i}^{k}+\underbrace{\left(-K_{k}\right)\left(-1+c_{i}^{k-1}+c_{i}^{k-2}+\dots+c_{i}^{i}\right)}_{c_{i}^{k}\text{ per ind. ass.}}\right)\nonumber \\
 & =\left(-K_{k+1}\right)\left(\left(-K_{k}\right)c_{i}^{k}+c_{i}^{k}\right)\nonumber \\
 & =\left(-K_{k}\right)\left(\frac{K_{k+1}}{K_{k}}c_{i}^{k}-K_{k+1}c_{i}^{k}\right).\label{eq:one-forw-eta}
\end{align}
Now, we witness the same phenomenon as above. Note that all chains
must end in $K_{k+1}$ as above. The term $\frac{K_{k+1}}{K_{k}}c_{i}^{k}$
swaps out combinations like $1,\dots,k$ for $1,\dots,k-1,k+1$ with
a gap for $k$ while leaving the combination length intact. The term
$-K_{k+1}c_{i}^{k}$ exchanges combinations like $1,\dots,k$ for
$1,\dots,k,k+1$, i.e. fills the gaps that the first transformation
created and also extends the length by one, thus flipping all signs.
The last line of the preceding display equals $\left(-K_{k}\right)c_{i}^{k+1}$
so that again we obtain (\ref{eq:forward-eta-rel}) multiplied by
$\left(-K_{k}\right)$.

For Lemma \ref{lem:combinatorial-forward}.\ref{enu:number-terms},
we see that rolling forward $c_{i}^{k}$ or $d_{i}^{k}$ to $k+1$
adds the same number of terms of those found for $k$ as per (\ref{eq:one-forw-eps})
and (\ref{eq:one-forw-eta}) and no terms cancel because the difference
have no terms in common. Therefore, we are doubling the number of
terms each time we increase $k$ by one.

Finally, for Lemma \ref{lem:combinatorial-forward}.\ref{enu:simplification-diffuse-prior},
we again proceed by induction. Suppose $K_{1}=1$, then equation (\ref{eq:forward-eta-rel})
for $k=2$ implies that $c_{1}^{1}=1$ and 
\[
-K_{2}\left(-1+c_{1}^{1}\right)=c_{1}^{2}=0.
\]
Now, assume that $c_{1}^{s}=0$ for $s\leq k$ except $c_{1}^{1}=1$
so that by part \ref{enu:forward-eta} 
\[
\left(-K_{k+1}\right)\left(-1+c_{1}^{k}+c_{1}^{k-1}+c_{1}^{k-2}+\dots+1\right)=c_{1}^{k+1}
\]
becomes 
\[
\left(-K_{k+1}\right)\left(-1+0+0+0+\dots+1\right)=0.
\]
Hence, we have established that $c_{1}^{k+1}=0$ if $c_{1}^{s}=0$
for $s\leq k$, thus establishing the claim for all $k\geq2$. 
\end{proof}
\begin{proof}[Proof of Lemma \ref{lem:asymptotic-coeffs}]
For coefficients $c_{i}\left(k\right)$ we sum over combinations
of length $m$ from $m=0$ onwards. First, we recognize that for any
given $k$, we actually need all combinations from $i$ to $k-1$
with combination length $\left(k-1\right)-i+1=k-i$. Because these
combinations are always one shorter than the actual combination length,
we need power $m+1$ and prefactor $\left(-1\right)^{m}$ so that
odd powers, $\left(m+1\right)$ being odd, have a positive coefficient.

For coefficients $d_{i}\left(k\right)$, we note that we want combinations
from $i+1$ to $k-1$, which have lengths $m=0,\dots,k-i-1$. Finally,
there are $\binom{k-i-1}{m}$ combinations of length $m$ and the
power accounts for the combination length $m$ plus the beginning
at $i$ adding one, and adding one more if the beginning and end are
different. Even powers are negative, hence $\left(-1\right)^{m+\indic\left\{ k-i>0\right\} }$. 
\end{proof}
\addtolength{\textheight}{.5in} \addtolength{\textheight}{-.3in}

\end{document}